\documentclass{aa}  

\usepackage{graphicx}
\usepackage{txfonts}
\usepackage{longtable}
\begin{document}

   \title{Understanding the evolution and dust formation of carbon stars in the LMC with a look at the {\itshape JWST}}


   \author{E. Marini\inst{1,2}, F. Dell'Agli\inst{2}, M. A. T. Groenewegen$^{5}$, D. A. Garc\'{\i}a--Hern\'andez$^{3,4}$, \\ 
   L. Mattsson$^{6}$, D. Kamath$^{7}$, P. Ventura\inst{2}, F. D'Antona\inst{2}, M. Tailo\inst{8}
          }

   \institute{Dipartimento di Matematica e Fisica, Università degli Studi Roma Tre, 
              via della Vasca Navale 84, 00100, Roma, Italy \and
              INAF, Observatory of Rome, Via Frascati 33, 00077 Monte Porzio Catone (RM), Italy \and
              Instituto de Astrof\'{\i}sica de Canarias (IAC), E-38200 La Laguna, Tenerife, Spain  \and
              Departamento de Astrof\'{\i}sica, Universidad de La Laguna (ULL), E-38206 La Laguna, Tenerife, Spain  \and
              Koninklijke Sterrenwacht van Belgi{\"e}, Ringlaan 3, 1180 Brussels, Belgium \and
              Nordita, KTH Royal Institute of Technology and Stockholm University, Roslagstullsbacken 23, SE-106 91 Stockholm, Sweden \and
              Department of Physics and Astronomy, Macquarie University, Sydney, NSW 2109, Australia \and
              Dipartimento di Fisica e Astronomia 'Galileo Galilei', Univ. di Padova, Vicolo dell?Osservatorio 3, I-35122 Padova, Italy
             }

   \date{Received September 15, 1996; accepted March 16, 1997}


  \abstract
   {Carbon stars have been and are extensively studied, given their complex internal
   structure and their peculiar chemical composition, which make them living laboratories
   to test stellar structure and evolution theories of evolved stars. Furthermore, they
   are the most relevant dust manufacturers, thus playing a crucial role in the evolution
   of galaxies.}
   {We study the dust mineralogy of circumstellar envelope of carbon stars in the Large Magellanic Cloud (LMC),
   to achieve a better understanding of the dust formation process in the outflow of these
   objects. We investigate the expected distribution of carbon stars in the
   observational planes built with the filters of the Mid-Infrared
   Instrument (MIRI) mounted
   onboard the James Webb Space Telescope ({\itshape JWST}), to select the best planes allowing
   an exhaustive characterisation of the stars.}
   {We compare the synthetic spectral energy distributions, obtained by the modelling
   of asymptotic giant branch stars and of the dust formation process in the wind, with the
   spectra of carbon stars in the LMC, taken with the Infrared Spectrograph (IRS)
   onboard the Spitzer Space Telescope. From the detailed comparison between 
   synthetic modelling and observation we characterise the individual sources and derive
   the detailed mineralogy of the dust in the circumstellar envelope.}
   {The sample of the stars considered here is composed by stars of diverse mass, formation
   epoch, degree of obscuration and metallicity. We find that precipitation of MgS on
   SiC seeds is common to all non metal-poor carbon stars. Solid carbon is the dominant
   dust component, with percentages above $80\%$ in all cases; a percentage between 
   $10\%$ and $20\%$ of carbon dust is under the form of graphite, the remaining being
   amorphous carbon. Regarding the observational planes based on the MIRI filters,
   the colour-magnitude ([F770W]-[F1800W], [F1800W]) plane allows the best understanding of the
   degree of obscuration of the stars, while the ([F1800W]-[F2550W], [F1800W]) diagram allows
   a better discrimination among stars of different metallicity.}
   {}


   \titlerunning{Evolution and dust formation in LMC carbon stars}
   \authorrunning{Marini et al.}
   \maketitle
%

\section{Introduction}
Asymptotic giant branch (AGB) stars are important dust manufacturers in the Universe. Contrary to early studies, that suggest a dominant role of supernovae, it is now generally recognized that the contribution from AGB stars to dust production in the Universe cannot be neglected, even in early epochs \citep{valiante09, valiante17}. Consequently, the knowledge of the AGB phase and the modality with which these stars form dust are important to investigate the physics of the interstellar medium and more generally the galaxy evolution, considering 
the critical role played by dust in influencing the star formation process 
\citep{hollenbach71, mathis90, draine03, gong17}.

The recent years have witnessed significant steps forward in the description of the AGB phase, with the implementation of the description of dust formation in the winds departing from the photosphere of the central star \citep{ventura12, nanni13, nanni14, ventura14, flavia19a}. These models have been successfully applied to study the evolved stellar populations of the Magellanic Clouds \citep{flavia14b, flavia15a, flavia15b, nanni16, nanni18, nanni19b} and other galaxies in the Local Group \citep{flavia16, flavia18, flavia19a}. Further fields of application are the modelling of dust evolution in the Milky Way \citep{ginolfi18}, in local and high-redshift galaxies \citep{raffa14, nanni20a} and in the Universe \citep{gioannini17}.

In the near future, the studies aimed at improving our understanding of dust production in the envelope of AGB stars will benefit from the launch of the {\itshape JWST}, that will allow an unprecedented exploration of the evolved stellar populations in the Local Universe. The large aperture ($6.5$ m) and the subarsecond spatial resolution will allow the study of resolved dusty stellar populations at moderate and large distances, up to $\sim 4$ Mpc \citep{jones17}.  {\itshape JWST} will provide spectroscopy in the $5-28.5~\mu$m range \citep{bouchet15}, thus offering a unique opportunity to study the evolution of AGB stars and the dust formation process in their expanding wind, in a large variety of environments.
The full exploitation of the {\itshape JWST} potentialities will allow the characterisation of the evolved stellar populations observed in the galaxies of the Local Group, and possibly beyond, provided that we know how AGB stars evolve in the observational planes built with the {\itshape JWST} filters.

A first step in this direction was achieved by \citet{jones17}, who considered the {\itshape Spitzer} spectroscopic data obtained with the IRS instrument of $\sim 1000$ objects in the Magellanic Clouds, and obtained the magnitudes in the various
MIRI filters, via convolution with the corresponding transmission curves. The aim of this work was fixing reliable classification schemes to disentangle different classes of objects. Regarding the evolved stellar populations (AGB and RSG stars) the analysis by \citet{jones17} was based on existing libraries of spectra, spanning a wide range of photospheric parameters and dust composition.

In this context we have recently started a research aimed at exploring how the {\itshape JWST} space mission will allow the characterisation of the evolved stellar populations of galaxies. We base our investigation on the IRS data of LMC stars, because only a detailed analysis of the spectral energy distribution (SED) allows a correct understanding of the dust composition in the circumstellar envelope. An example of this approach is reported in the work by \citet{jones14}, aimed at deriving the amount of alumina dust in the wind of oxygen-rich stars in the LMC.

Differently from \citet{jones17}, we use evolutionary sequences of AGB stars that include the description of dust production, which allows to predict the SED evolution, consisting in a sequence of synthetic spectra, each representing a specific evolutionary stage of the AGB phase.
In \citet{ester20} we studied the sample of M-type AGB stars in the LMC. By comparing IRS and synthetic SED we characterised the individual sources and we investigated the obscuration sequences of oxygen-rich AGB stars in the planes built with the {\itshape {\itshape JWST}} filters.

Here we extend the analysis done in \citet{ester20} to the C-rich stars sample in the LMC. This step is important in the general context of dust production by stars in galaxies, because the overall dust production rate from AGB sources is dominated by carbonaceous species in the Magellanic Clouds \citep{raffa14, nanni18, nanni19b, nanni19c} and even more in metal poor environments \citep{cioni03, boyer13}. For these reasons carbon stars have been deeply investigated, in terms of the capability of carbon dust in driving an AGB wind \citep{lars10, lars11, bladh19, sandin20} and in studies aimed at constraining the physical parameters and the dynamical properties of the winds of individual carbon stars \citep{rau17, brummer18, maercker18, rau19}.

To date, the most exhaustive studies focused on the carbon stars population of the LMC, based on the results from AGB+dust modelling, are those by \citet{flavia15a} and \citet{nanni19b}. The former used results from the InfraRed Array Camera (IRAC) and the Multi-band Imaging Photometer
(MIPS) photometry to characterise the stars, according to their position in the observational planes. \citet{nanni19b} considered different observations in the infrared and optical data from Gaia, and derived important trends connecting mass loss rate, dust production rate, dust-to-gas ratio of the stars with the infrared colours, properly taken as indicators of the degree of obscuration. 

This work, similarly to \citet{flavia15a} and \citet{nanni19b}, is based on results from
stellar evolution and dust formation modelling, calculated on purpose for the present study, 
with the evolutionary sequences extended until the beginning of the post-AGB phase. 
However, following the same approach as in \citet{ester20}, we mostly focus on the results 
from IRS spectroscopy to reach a better understanding of dust formation in the 
wind of carbon stars; furthermore, we project the results into the {\itshape JWST} perspective, 
to provide the scientific community with the appropriate tools to analyse the carbon stars population of galaxies.

The present investigation is developed along two lines. We first derive the luminosities and the optical depths of the individual sources, by comparing the IRS and the synthetic spectral energy distributions, the latter being obtained by the modelling of the AGB evolution and of the dust formation process. These estimates are rather robust, because the extinction of carbon stars is mostly due to solid carbon particles and is substantially independent of the presence of other dust species. This step allows the characterisation of the stars considered, in terms of mass, formation epoch and chemical composition of the progenitors.

After the identification of the sources we focus on the dust mineralogy,
to determine the percentages of the different dust species in the circumstellar
envelope. This demands a detailed analysis of the IRS SED, considering that the
morphology of the different spectral features is extremely sensitive to the
type and the quantity of the dust grains formed in the stellar wind. 
Our aim is to improve our knowledge of the dynamical and chemical structure of the 
wind of carbon stars, to shed new light on the dust formation process, to understand
the conditions that lead to the formation of specific dust compounds: these information
will prove crucial to the characterisation of the carbon stars samples in galaxies 
different from the Magellanic Clouds, which will be observed by the {\itshape JWST}.

In this investigation we focus on the distribution of the stars in the observational
planes built with the MIRI filters, because the IRS spectra cover the $5-37 ~ \mu$m
spectral region. Our goal is to select the planes that allow the best characterisation
of the sources observed and to identify those planes where the bifurcation of the
various sequences can be associated with some peculiar properties of the star, e.g.
the presence of a specific dust particles or the metallicity.

The paper is organised as follows: in section \ref{sample} we present the sample of stars used for the analysis; the numerical codes used to calculate the evolutionary sequences, to model dust formation and to build the synthetic spectra are described in section \ref{modinput}; the evolution and the dust formation mechanism in the models used are discussed in section \ref{cstars}; in section \ref{lmcc} we present the characterisation of the sources observed, whereas in section \ref{spectra} we address the determination of the dust mineralogy and the information obtained regarding the structure of the winds; the expected distribution of carbon stars in the observational planes built with the
MIRI filters is discussed in section \ref{JWST}.

\section{The selected sample}
\label{sample}
In this paper we use the LMC sample of 147 C-rich AGB spectra from the SAGE-Spec database at the NASA/IPAC Infrared Science Archive, reduced by \citet{jones17}, using the method discussed in \citet{kemper10}. This database includes all staring-mode observations with the {\itshape Spitzer} IRS taken in the area of the sky covered by the SAGE survey \citep{meixner06}, which entirely covers the LMC. \citet{jones17} classified the sources of the sample by using the decision-tree classification method by \citet{woods11}: the primary criterion for the classification was the {\itshape Spitzer} spectrum itself, but it also relied on the additional information provided by the SED, on the estimate of the bolometric luminosity and supplementary information from the literature.  C-AGB stars, in particular, are usually identified by molecular absorption features, particularly C$_2$H$_2$ (acetylene) at 7.5 and 13.7 $\mu$m, and by the dust features at $11.3 ~ \mu$m and $30~\mu$m.

To study the expected distribution of the stars in the observational planes built
with the magnitudes in the MIRI filters, we used the mid-IR magnitudes ([F770W], [F1000W], [F1130W], [F1280W], [F1500W], [F1800W], [F2100W], [F2550W]) calculated by \citet{jones17}, who integrated the IRS spectra of each source over the MIRI spectral response. 
In some cases the IRS data do not match IRAC and MIPS photometry, which 
could be an effect of variability, considering that IRS, IRAC and MIPS data were collected in different epochs. On this regard, we used the criterion proposed
by \citet{martin18}, that variability can account for up to a $20\%$ difference
between the fluxes derived from the analysis of the IRS SED and those found
via the IRAC and MIPS magnitudes. In these cases we left the MIRI magnitudes
found by \citet{jones17} unchanged. In the cases where the difference is
above $20\%$ we scaled the IRS spectra to match the corresponding IRAC and MIPS photometry, and recalculated the corresponding MIRI magnitudes via convolution over the MIRI transmission curves.

The majority of these objects covers the $5-37~\mu$m range of the low-resolution 
modules of the IRS, so the fluxes for the F560W  filter are not available. For 27 out of these  objects the spectral coverage by the {\itshape Spitzer}-IRS is limited to the $5-14~\mu$m range, therefore the mid-IR photometry beyond this wavelength is not available, leading us to exclude them from our analysis.
In addition, there are a paucity of sources which show a steep rise starting from $\sim 20~\mu$m and no decline up to
the end of the spectra. We believe that these peculiar SEDs are affected by nearby background emission, as also stated by \citet{30micron}, and for this reason are not considered in the analysed sample.

\section{Physical and numerical input}
\label{modinput}
\subsection{Stellar evolution modelling}
\label{agbinput}
The evolutionary sequences were calculated by means of the ATON code for stellar 
evolution \citep{ventura98}.
An exhaustive description of the numerical details of the code and the most recent updates 
can be found in \citet{ventura13}. 

\subsubsection{The chemical composition}
Most of the analysis is based on models of metallicity
$Z=0.008$, helium mass fraction $Y=0.26$, mixture taken from \citet{gs98}, with
$\alpha-$enhancement $[\alpha/$Fe$]=+0.2$. This is the same chemical composition used in 
\citet{ventura14}; we preferred to recalculate the evolutionary sequences from the
beginning, to follow in detail the very final phases, which, as we will see 
in the following sections, are those with the largest degree of obscuration, and that
can account for the presence of the objects with the largest infrared emission. 

The models not undergoing the helium flash were evolved from the pre-main sequence until the almost total consumption of the envelope. Low-mass models ($M \leq 2~M_{\odot}$) experiencing the helium flash were evolved from the horizontal branch, starting from the core mass and surface chemical composition calculated until the tip of the red giant branch (RGB). For the objects whose spectral energy distribution suggests a metal-poor composition we used $Z=0.001$ and $Z=0.002$
tracks with initial helium $Y=0.25$ and $[\alpha/$Fe$]=+0.4$ mixture from \citet{gs98}. Finally, models of metallicity $Z=0.004$, $Y=0.26$, $[\alpha/$Fe$]=+0.2$ were also considered. The $Z=0.001$, $Z=0.002$ and $Z=0.004$ models were taken from \citet{ventura14}.

\subsubsection{Convection}
The temperature gradient within regions unstable to convection is calculated via the Full Spectrum of Turbulence (FST) model \citep{cm91}. Overshoot of convective eddies within radiatively stable regions is modelled by assuming that the velocity of convective elements decays exponentially beyond the neutrality point, fixed via the Schwartzschild criterion. The e-folding distance of the velocity decays during the core (hydrogen and helium) burning phases and during the AGB phase is taken as $0.02H_{\rm P}$ and $0.002H_{\rm P}$, respectively. The latter values reflect the calibrations discussed, respectively, in \citet{ventura98} and \citet{ventura14}.

\subsubsection{Mass loss}
The mass loss rate during the phases when the star is oxygen-rich was determined via the \citet{blocker95} treatment, with the parameter entering the \citet{blocker95}'s formula set to $\eta=0.02$; this choice follows the calibration given in \citet{ventura00}. For carbon stars we used the mass loss description from the Berlin group \citep{wachter02, wachter08}. 

\subsubsection{Opacities}
The radiative opacities are calculated according to the OPAL release, in the version documented by \citet{opal}. The molecular opacities in the low-temperature regime ($T < 10^4$ K) are calculated with the AESOPUS tool \citep{marigo09}. The opacities are constructed self-consistently, by following the changes in the chemical composition of the envelope, particularly of the individual abundances of carbon, nitrogen, and oxygen.
 
\subsection{Dust production}
\label{dustmod}
The interpretation of the IR spectra of obscured carbon stars requires the knowledge of the dust present in the circumstellar envelope, namely the density, the radial distribution and the percentages of the different dust species formed. To this aim we modelled the formation and growth of dust particles in the wind of AGB stars according to  the schematization proposed by the Heidelberg group \citep{fg06}, similarly to previous works by our team \citep{ventura12, ventura14, ventura15, ventura16} and used in a series of papers by the Padua group \citep{nanni13, nanni14, nanni16, nanni18, nanni19a, nanni20b}.  The interested reader can find all the relevant equations in \citet{ventura12}. Here we provide a brief description of the methodology used.

In the present framework dust particles are assumed to form and grow in the wind,
which expands isotropically from the photosphere of the star. Each dust species starts to form in the so called condensation point, where the growth rate exceeds the vaporisation rate; the latter is dependent on the thermodynamic properties of the solid compounds, mainly on the formation enthalpies of the solid compounds and of the gaseous molecules involved in the formation reaction \citep{fg06}.

The dynamics of the wind is described by the momentum equation, where the acceleration is determined by the  competition between gravity and radiation pressure
acting on the newly formed dust grains. The coupling between grain growth and wind dynamics is given by the extinction coefficients, describing absorption and scattering of the radiation by dust particles. The evolution with time of dust grains is determined by the difference between the growth and the vaporization rate. The former is given by the gas molecules hitting the already formed grains, the latter is related to the vapour pressure of gaseous molecules over the solid compounds. 

In the basic configuration we assume that the dust species formed in the winds of
carbon stars are silicon carbide (SiC) and amorphous carbon (C). This assumption is suitable to determine the main properties of the wind, namely the terminal velocity and the degree of obscuration, which we quantify via the optical depth at the wavelength $\lambda = 10 \ \mu$m, $\tau_{10}$. For the detailed interpretation of the SED of the various sources we also account for the formation of graphite and of MgS; for the latter species we consider either the possibility that it forms as individual species or that MgS mantles form on SiC cores \citep{svetlana08}. The extinction coefficients were found by using the optical constants from \citet{zubko} (amorphous carbon), \citet{peg} (SiC), \citet{begemann94} (MgS) and \citet{draine84} (graphite). The optical constants of SiC+MgS compounds were calculated on the basis of the method described in \citet{svetlana08}.

The modelling of dust formation, as described above, allows the determination of the asymptotic velocity of the gas particles, the size reached by the grains of the various species and the optical depth. Furthermore, this computation provides
an estimate of the surface fraction of gaseous silicon, carbon and magnesium 
condensed into dust (see eq. 20-23 and 34-35 in \citet{fg06}) and the dust production rate for each dust species, which depends on the gas mass-loss
rate, the surface mass fractions of the afore mentioned chemical elements, and the fraction of the latter species condensed into dust (see Section 5.2 in \citet{fg06}).

\subsection{Spectral energy distribution}
The interpretation of the IRS SEDs demands the computation for each stellar mass 
of a sequence of theoretical SEDs, which correspond to the evolutionary stages of
carbon stars and to the dust formed in the circumstellar envelope, modelled according to the methods described earlier in this section.

To characterise the individual AGB sources and derive the mass and
formation epoch of the progenitor stars, we confront the IRS data with
the sequences of synthetic spectra obtained for each model star, which represent
the evolution of the SED of the star during the AGB phase.
To build these sequences, we first select a few points along the evolutionary track, distributed among the various interpulse phases. Typically we pick one 
evolutionary stage corresponding to the largest luminosity during the inter-pulse, before the ignition of the following TP. Towards the final evolutionary phases, when the mass-loss rate rises and a significant fraction of the envelope is lost during a single inter-pulse, we select one or two points more, in order to follow in more details the variation of the stellar properties as mass is lost from the envelope. For completeness, we also considered
a couple of stages following the extinction of each TP, when the contribution
of the CNO burning shell accounts to $10\%$ and $30\%$ of the overall energy release of the star. Based on the physical parameters attained by the star
during these phases, particularly luminosity, effective temperature, current mass,
mass loss rate and surface chemical composition, we model dust formation, which
allows the determination of the amount of dust formed, the dust mineralogy, the
size of the dust particles and $\tau_{10}$.

The last step consists of the calculation of the synthetic SED, which is done by means of the code DUSTY \citep{nenkova99}. 
As input radiation, we used the spectral energy distribution found by interpolation in effective temperature, surface gravity and C/O ratios among the COMARCS model atmospheres \citep{aringer09} of the appropriate metallicity.

\begin{figure*}
\begin{minipage}{0.48\textwidth}
\resizebox{1.\hsize}{!}{\includegraphics{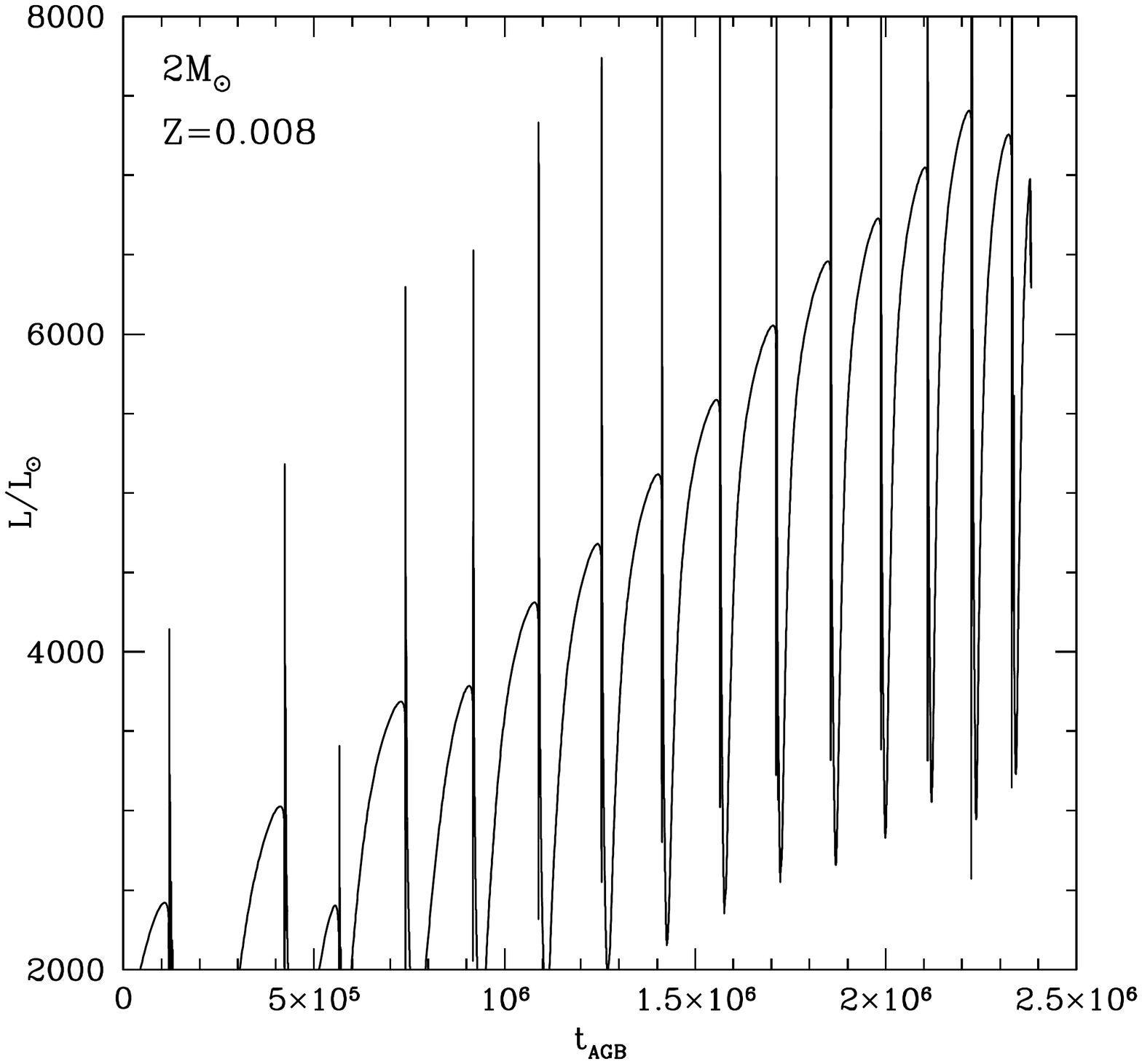}}
\end{minipage}
\begin{minipage}{0.48\textwidth}
\resizebox{1.\hsize}{!}{\includegraphics{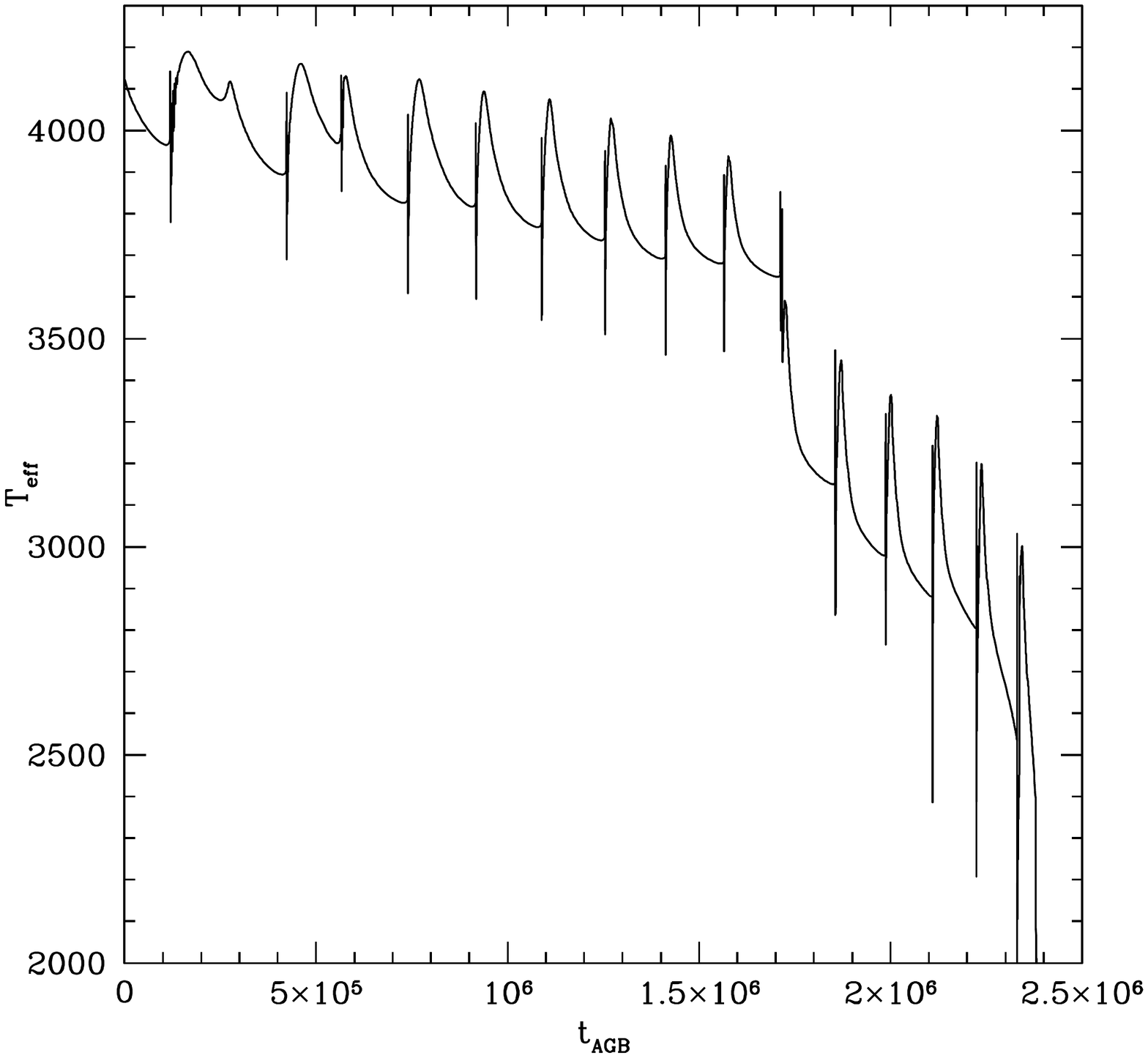}}
\end{minipage}
\vskip-70pt
\begin{minipage}{0.48\textwidth}
\resizebox{1.\hsize}{!}{\includegraphics{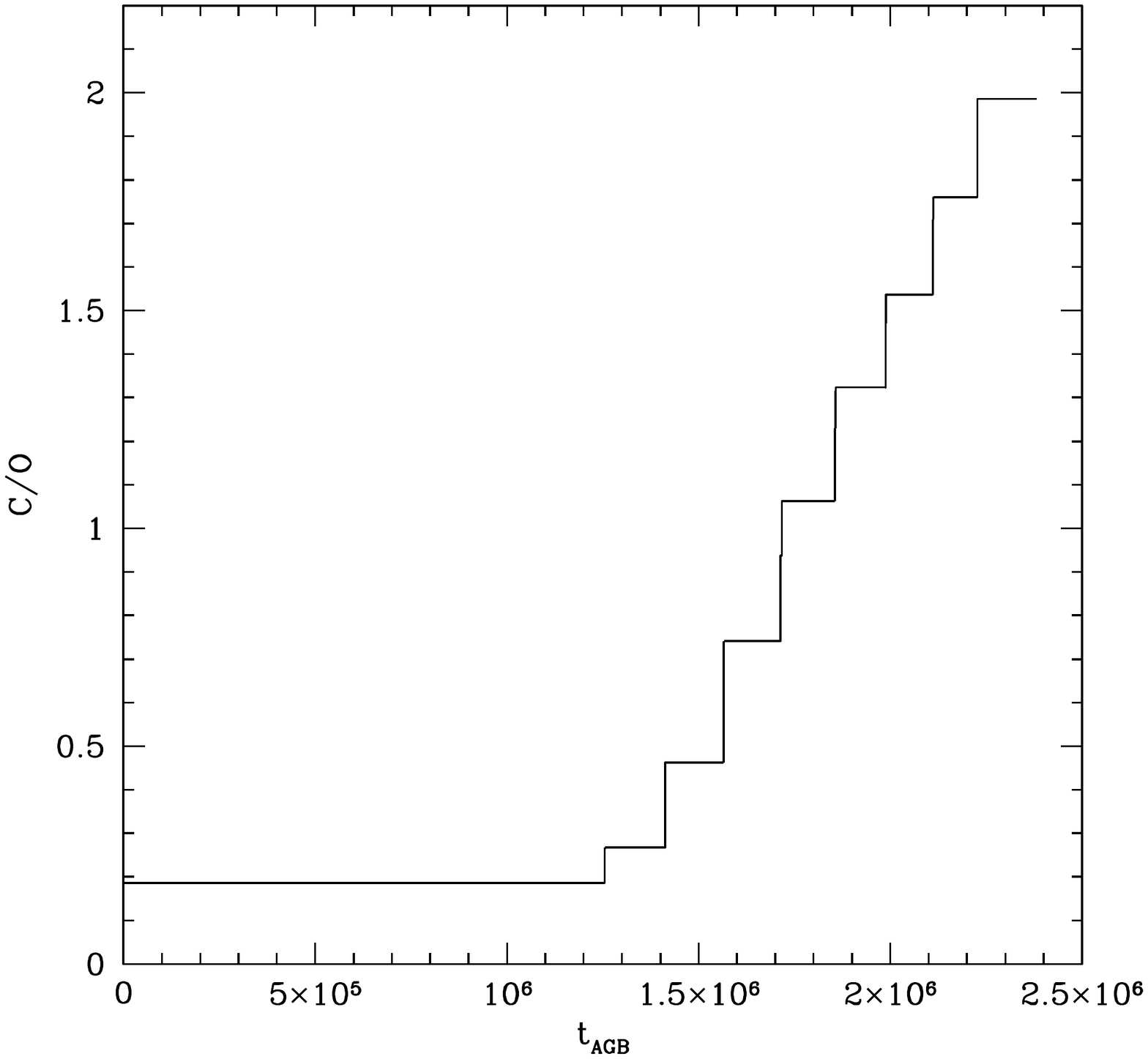}}
\end{minipage}
\begin{minipage}{0.48\textwidth}
\resizebox{1.\hsize}{!}{\includegraphics{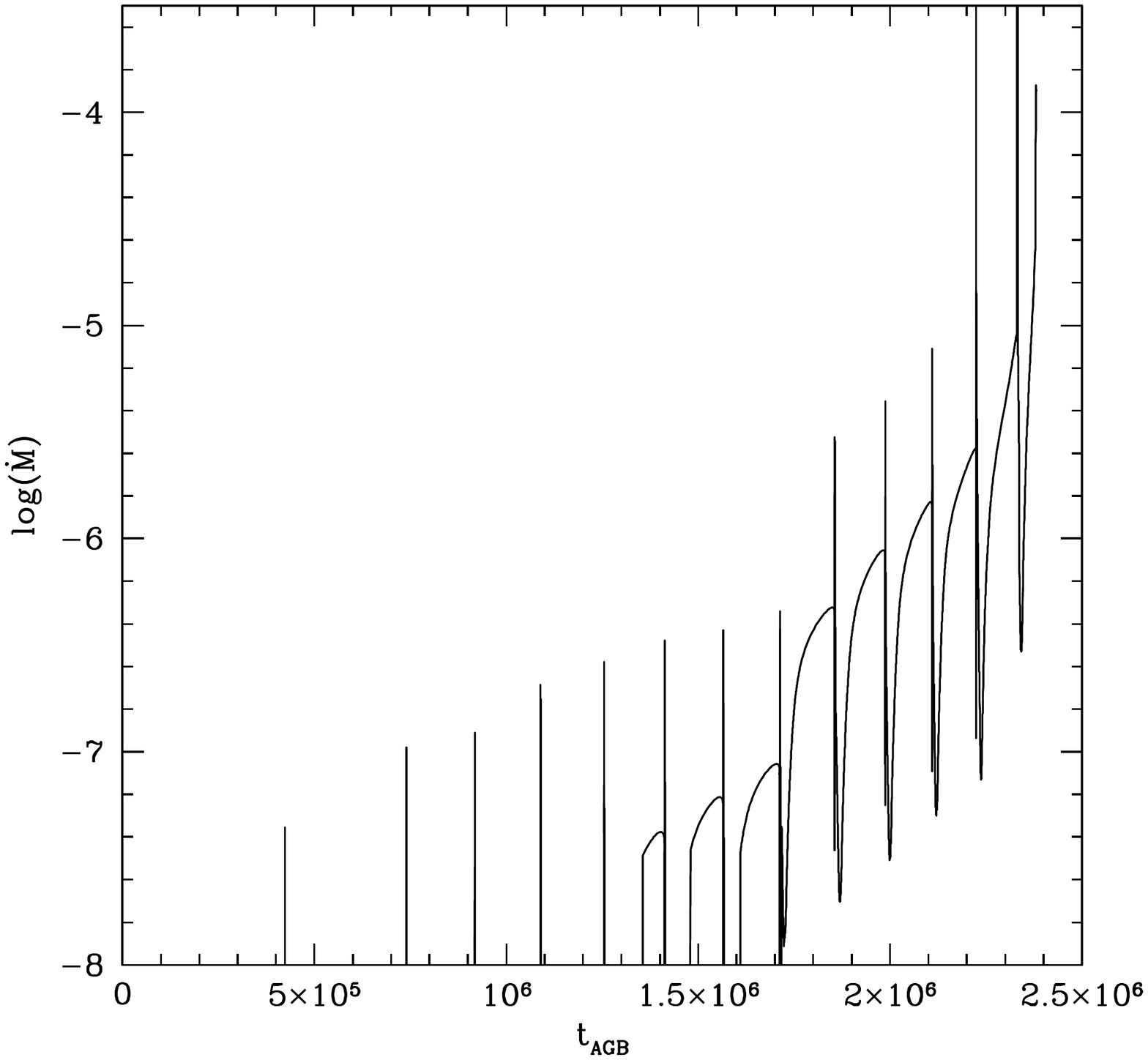}}
\end{minipage}
\vskip-40pt
\caption{The evolution as a function of time (counted from the beginning of the TP phase) of the luminosity (top, left panel), effective temperature (top, right), surface C$/$O (bottom, left) and mass loss rate (bottom, right) of a model of initial mass $2~M_{\odot}$ and metallicity $Z=0.008$.}
\label{fevol20}
\end{figure*}

\section{Carbon stars: evolution and dust production}
\label{cstars}
The carbon star stage is reached as a consequence of repeated third dredge-up (TDU)
episodes, which take place during the AGB phase, after the occurrence of thermal pulses
(TP). During these events the surface convective layer penetrates inwards,
down to regions of the star previously exposed to helium burning nucleosynthesis
\citep{iben74, iben75, lattanzio87, lattanzio93, boothroyd88}. The consequence of TDU
is the transportation of carbon nuclei synthesized via 3$\alpha$ reactions to the surface 
regions, which become more and more enriched in carbon: once the number of C nuclei
overcomes the number of O, the star becomes a C-star.

Stars of mass below $\sim 1~M_{\odot}$ do not become C-star. This threshold mass decreases slightly with metallicity, because lower $Z$ stars, owing to the smaller amount of oxygen, reach more easily the C$/$O$>1$ condition \citep{karakas14a}. For the $Z=0.008$ models used in the present work we find that the minimum mass required to attain the C-star stage is $1.25~M_{\odot}$\footnote{This lower limit refers to the mass at the beginning of the AGB phase, under the hypothesis that no mass loss
occurred during the previous evolutionary phases. Should mass loss during the
RGB be considered, the threshold mass under which no carbon stars form would be $\sim 1.4~M_{\odot}$. This is slightly lower than the lower limit for the formation of Galactic carbon stars found by \citet{marigo20}, a result consistent with the lower metallicity of LMC stars compared to the Milky Way.
}. 

The mass range of carbon stars is also limited from above, because stars with  CO cores of mass above $\sim 0.8~M_{\odot}$ \citep{ventura13} experience hot bottom burning (HBB), consisting 
in the activation of a series of p-capture reactions at the base of the external envelope, which favour the destruction of the surface carbon. In this investigation we find that the  aforementioned limit in the core mass translates into a threshold initial mass of $3.3~M_{\odot}$. 
 
It is not excluded that stars of higher mass go through evolutionary phases during which they are C-stars. This may occur either during the initial AGB stages, before HBB is activated, or at the very end of the AGB life, when HBB is turned off and the occurrence of a few TDU events might rise the C$/$O ratio 
above unity \citep{frost98}. However, while we will take this into account for the interpretation of a few outliers in the sample that we are going to analyse, in the following we will focus our attention on the stars with initial mass $M \leq 3.3~M_{\odot}$, because only these objects are expected to reach a significant degree of obscuration and to provide a not negligible contribution to the dust enrichment of the interstellar medium during the C-star phase.

\subsection{The behaviour of $2~M_{\odot}$ stars}
\label{2msun}
To illustrate more clearly the key aspects of the evolution of these objects we show in Fig.~\ref{fevol20} the variation with time of luminosity, effective temperature, surface C$/$O and mass loss rate of a star of initial mass $2~M_{\odot}$. The figure refers to the evolutionary phases between the beginning of the TP phase (from which the times on the abscissa are counted) and the almost complete ejection of the external mantle.

\begin{figure*}
\begin{minipage}{0.48\textwidth}
\resizebox{1.\hsize}{!}{\includegraphics{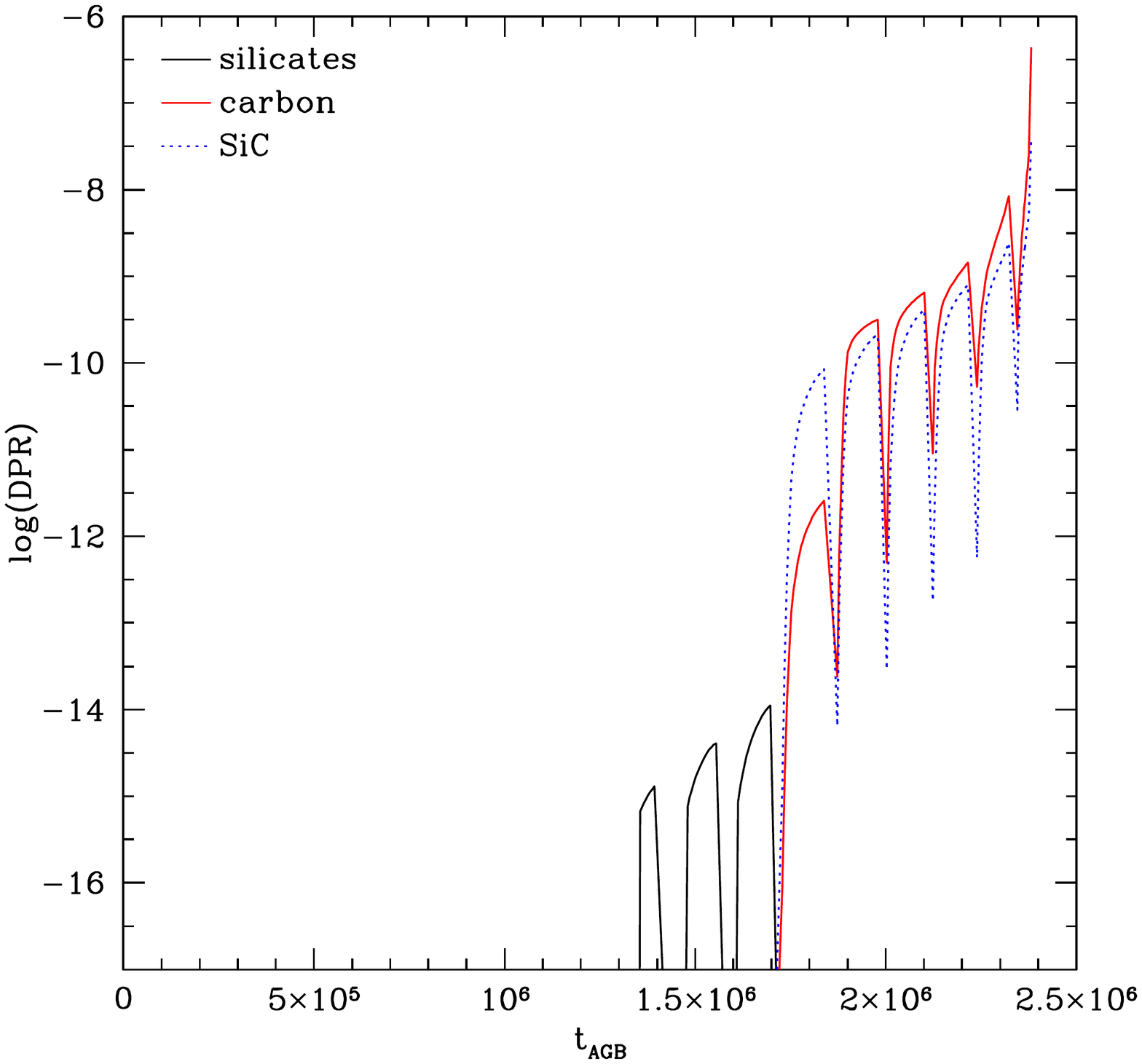}}
\end{minipage}
\begin{minipage}{0.48\textwidth}
\resizebox{1.\hsize}{!}{\includegraphics{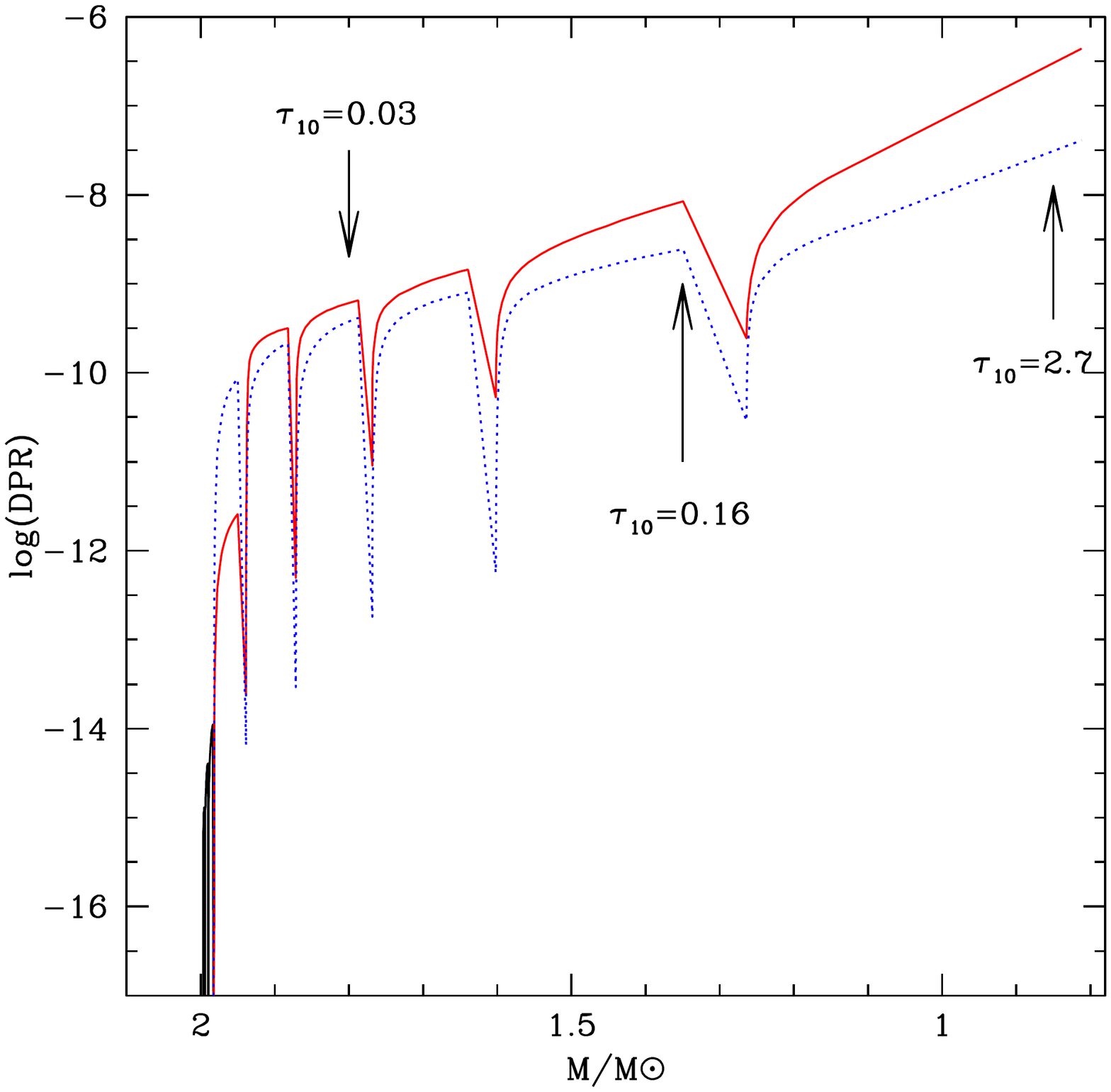}}
\end{minipage}
\vskip-70pt
\begin{minipage}{0.48\textwidth}
\resizebox{1.\hsize}{!}{\includegraphics{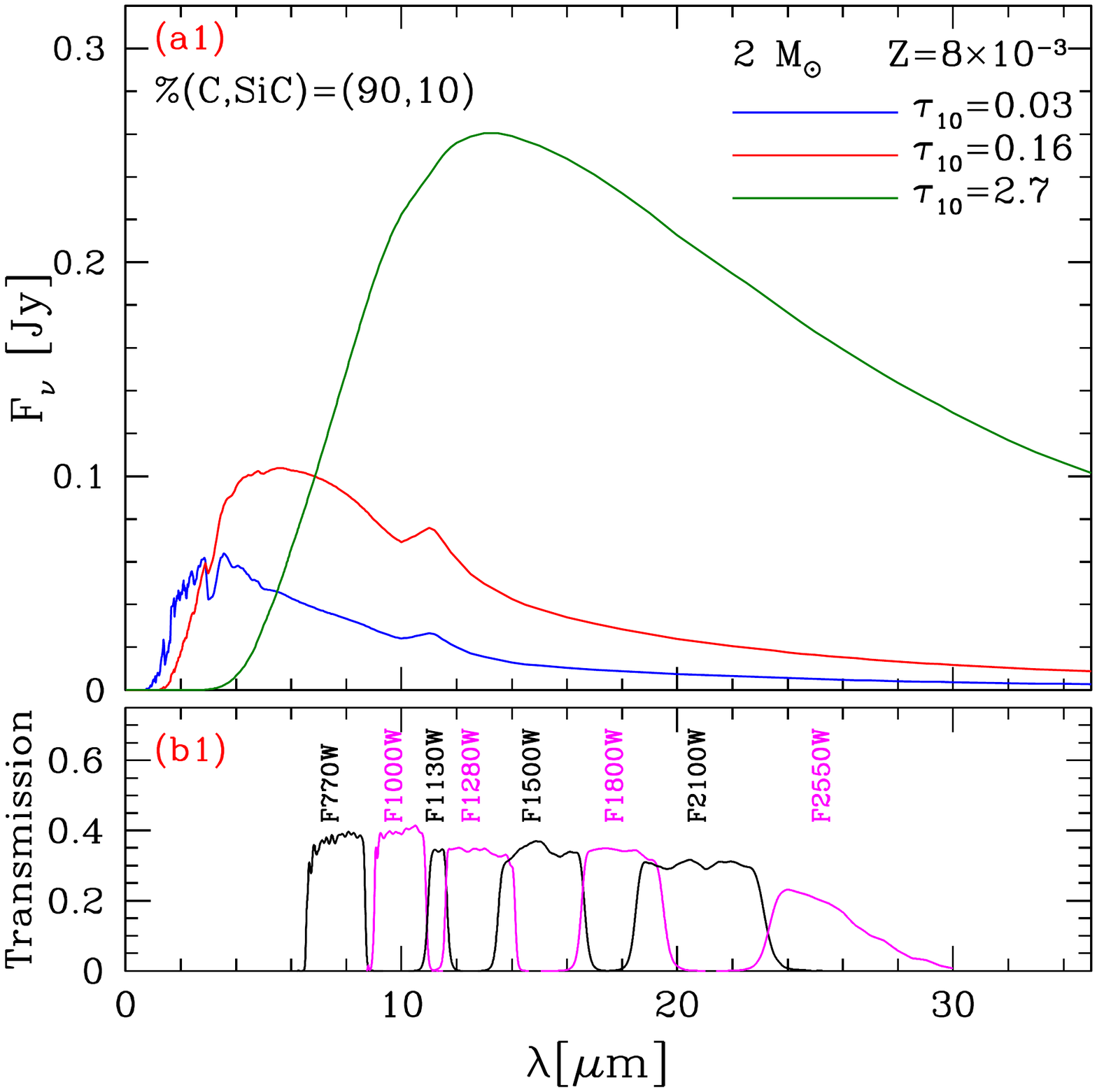}}
\end{minipage}
\begin{minipage}{0.48\textwidth}
\resizebox{1.\hsize}{!}{\includegraphics{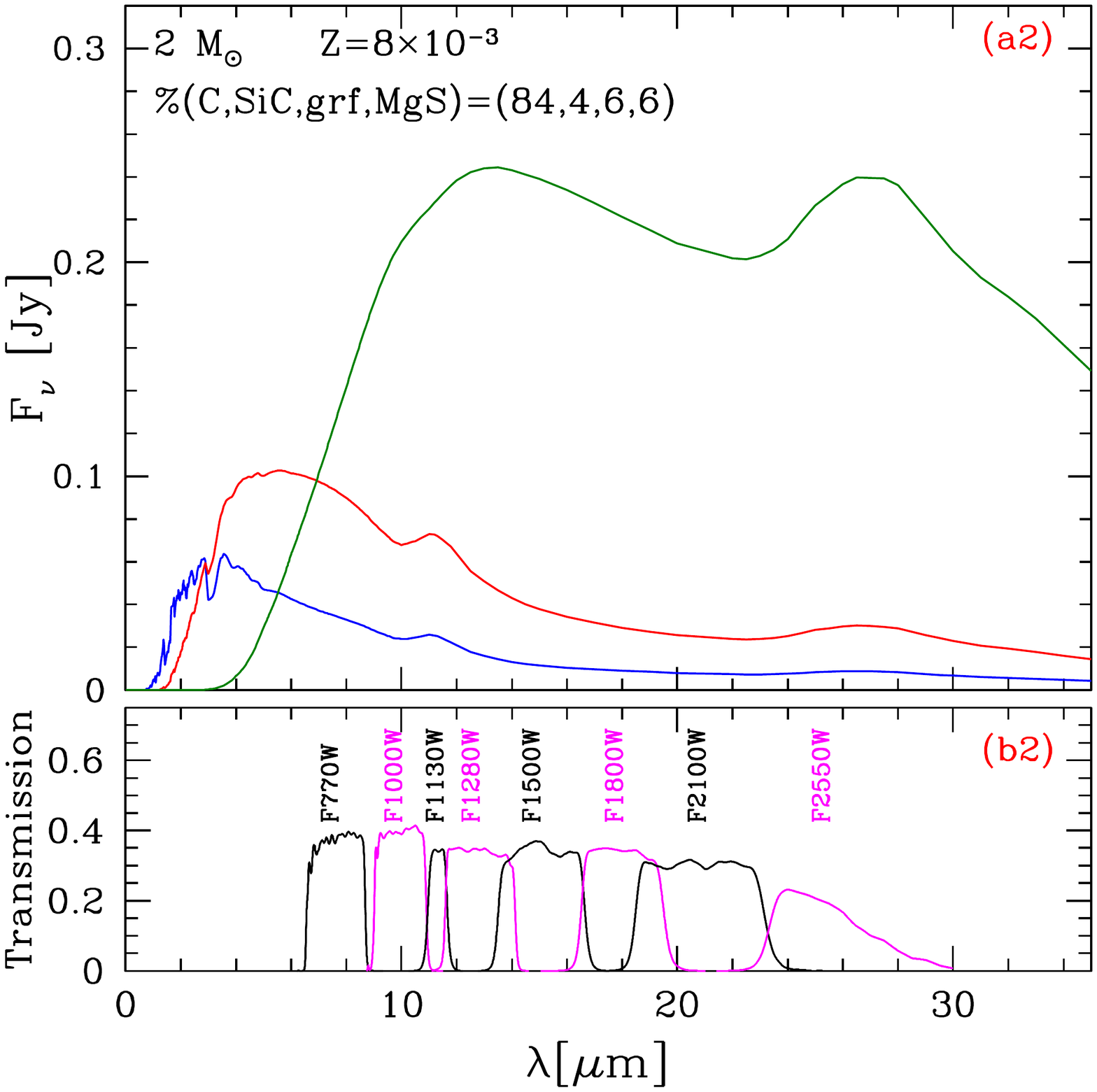}}
\end{minipage}
\vskip-40pt
\caption{Top panels: variation of the dust production rate as a function of
time (left) and of the current mass of the star (right) during the AGB evolution
of the model of initial mass $2~M_{\odot}$ and metallicity $Z=0.008$
presented in Fig.~\ref{fevol20}. The individual contributions from silicates, solid carbon and silicon carbide are shown. Bottom panels: variation of the expected SED during 4 different stages, indicated with the arrows in the top, right panel, when considering only solid carbon and SiC (left) and when also grafite and MgS are considered (right). In these examples the number
density of the seeds of amorphous carbon dust relative to hydrogen is assumed 
$10^{-13}$, whereas the density of the seeds of the other dust species are scaled
with respect to amorphous carbon according to the percentages given in the
two panels.}
\label{fdust20}
\end{figure*}

The $2~M_{\odot}$ model shown in Fig.~\ref{fevol20} experiences 14 TPs. The C-star stage is reached after 9 TPs: the duration of the C-star phase is $\sim 6\times 10^5$ Myr, which corresponds to $\sim 25 \%$ of the AGB life (2.4 Myr). The relatively short duration of the C-star phase is a common property of carbon stars \citep{karakas14a}. It is related to the transition from M- to C-star and the afore mentioned increase in the mass loss rate \citep{vm09, vm10}, which renders the remaining part of the AGB lifetime shorter. This effect can be seen in the right panels of Fig.~\ref{fevol20}, where we note that the effective temperature drops by $\sim 1000$ K during the last 4 interpulse phases, while at the same time the mass loss rate increases by almost two orders of magnitude, up to $\dot{M} \sim 10^{-4}~M_{\odot}/$yr.

 During the inter-pulse periods of the C-star phase we find that the luminosity
of the star is constant within $\sim 20\%$, spanning the $6500<L/L_{\odot}<7500$  $L_{\odot}$ range. These phases account for $90 \%$ of the total AGB lifetime.
As clear in Fig.~\ref{fevol20} the luminosity drops to $\sim 2000-3000~L_{\odot}$
after each TP, before the CNO activity is fully restored. 

The almost constant luminosity characterising the quiescent CNO burning evolution is partly explained by the afore mentioned short relative duration of the C-star phase, during which the stars experience only a few TPs. An additional reason is that after the C-star stage is reached, the TDU experienced becomes deeper and deeper, which prevents a significant growth of the core mass, hence of the luminosity. The small variation of the luminosity of the star during the C-star phase, combined with the fact that stars of different mass evolve at different luminosities, opens the way to use it as a mass indicator, which turns important to identify the progenitors of the stars.

To discuss dust production in the $2~M_{\odot}$ star, we present in the top-left panel of Fig.~\ref{fdust20} the variation of the dust production rate (DPR), i.e. the amount of dust produced by the star in a given evolutionary phase, expressed in $M_{\odot}/$yr unit. In the top-right panel of Fig.~\ref{fdust20} the DPR is shown as a function of the current mass of the star, to have an idea of the dust yields expected.

During the first part of the AGB evolution, when the star is oxygen-rich, the dust formed is under the form of alumina dust and silicates. Alumina dust forms in a more internal region of the envelope, at a distance of $\sim 2$ stellar radii ($R_*$) from the photosphere of the star, whereas the condensation region of silicates is more external, at a distance of $5-10~R_*$ from the stellar surface \citep{flavia12, flavia14a}. The extinction properties of the circumstellar envelope are mostly related to silicates, because alumina dust forms in small quantities and is extremely transparent to the radiation. During most of the O-rich phase dust production is negligible, because the mass loss rate is extremely small and the effective temperatures are close to $4000$ K (see Fig.~\ref{fevol20}): both these conditions prevent significant dust formation in the wind.

During the final part of the AGB evolution the dust formed is made up of carbonaceous dust, mostly SiC and solid carbon \citep{ventura12}. 
The latter species generally forms in higher quantities, with the exception of the first interpulse phase after the achievement of the C-star stage, during which the carbon excess is smaller than, or of the same orders of magnitude of the silicon abundance. This is shown in the top panels of Fig.~\ref{fdust20}, where see that the SiC contribution to the DPR is higher than that of solid carbon during the phases immediately following the achievement of the C-star stage.
The overall degree of obscuration of the star is almost entirely determined by
solid carbon dust; this is not only due to the fact that this is generally the most abundant dust species, but, more important, to the larger extinction coefficients of solid carbon compared to SiC (see Fig.~10 in \citet{fg02}).

Inspection of Fig.~\ref{fdust20} shows that the DPR during the inter-pulse
phase increases by 2-3 orders of magnitude when the transition from O-rich
to carbon star occurs. This behaviour, common to all the stars that eventually become carbon stars, is due to two reasons: a) the rate of mass loss is much higher during the C-star phase (see bottom-left panel of Fig.~\ref{fevol20}), which favours dust production, owing to the higher densities of the wind (see Eq.~14 in \citet{fg01}); b) the formation of carbon dust is favoured compared to silicates, because the surface carbon excess with respect to oxygen is significantly higher than the density of gaseous silicon, which is the key quantity to determine the formation of silicates during the O-rich phases.

Because of the mass is lost during the C-star phase, most of the dust produced by the stars the eventually reach the C-star stage is made up of carbon dust \citep{fg06, nanni13, ventura14}, despite the C-star lifetime is usually shorter than the initial phase, during which the star is oxygen rich.

Understanding the dust formation process is extremely important to predict the variation of the SED of these objects, as they evolve through the AGB. This step is crucial when a detailed comparison with the observations is required. In the bottom-left panel of Fig.~\ref{fdust20} we show the synthetic SED of the $2~M_{\odot}$ star, corresponding to the three evolutionary stages, indicated with arrows in the top-right panel. Note that the luminosity of the star is unchanged within $200~L_{\odot}$ among the three cases, thus the significant difference in the flux distribution is entirely due to the IR emission, which becomes larger and larger as the dust production rate increases.

In this particular case we see that the peak of the SED gradually shifts from
$\sim 3~ \mu$m to $\sim 12~ \mu$m. We note the emission feature related to SiC, at $11.3~ \mu$m, which turns into absorption when the circumstellar envelope becomes extremely thick.

\begin{figure*}
\begin{minipage}{0.48\textwidth}
\resizebox{1.\hsize}{!}{\includegraphics{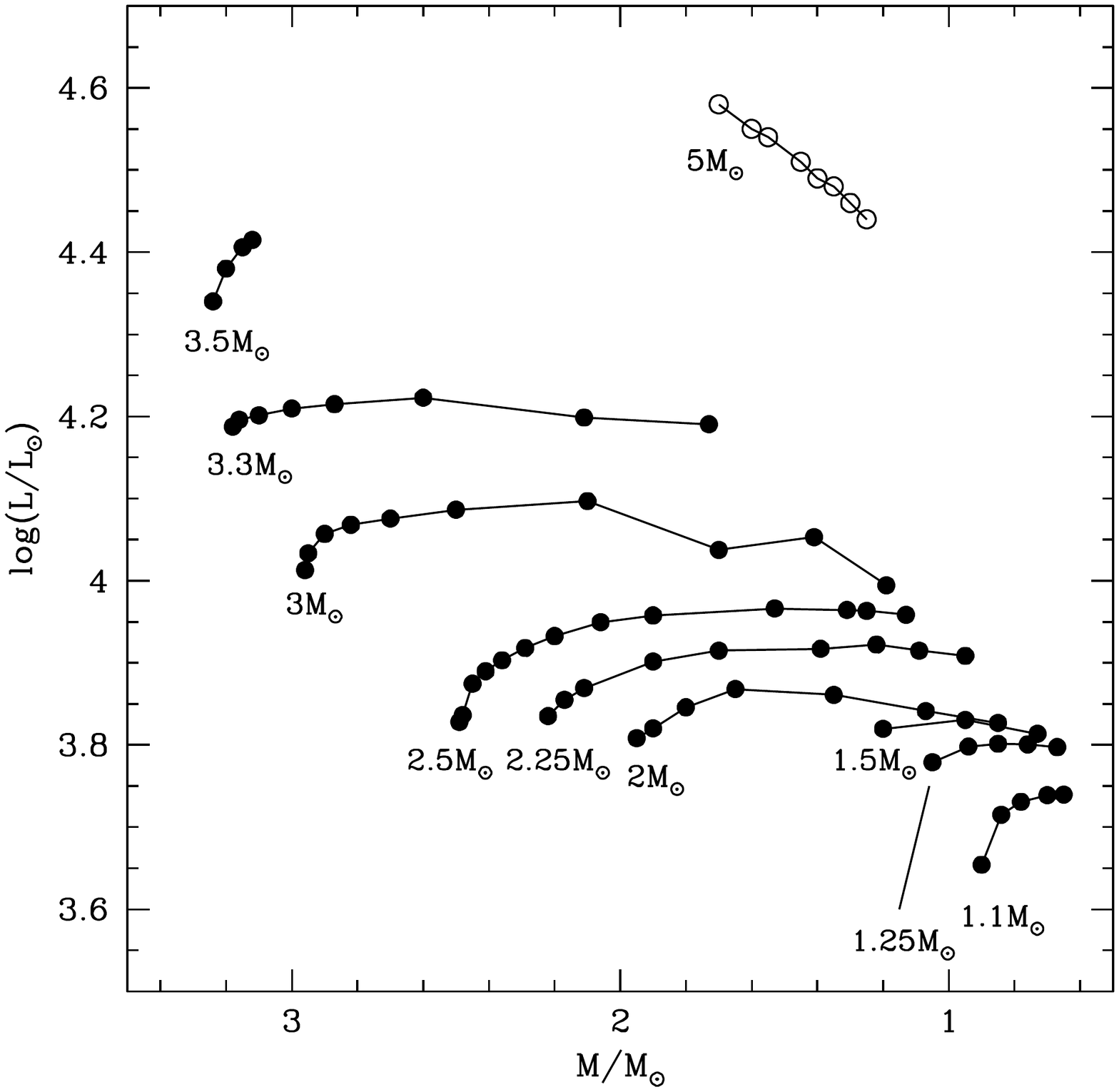}}
\end{minipage}
\begin{minipage}{0.48\textwidth}
\resizebox{1.\hsize}{!}{\includegraphics{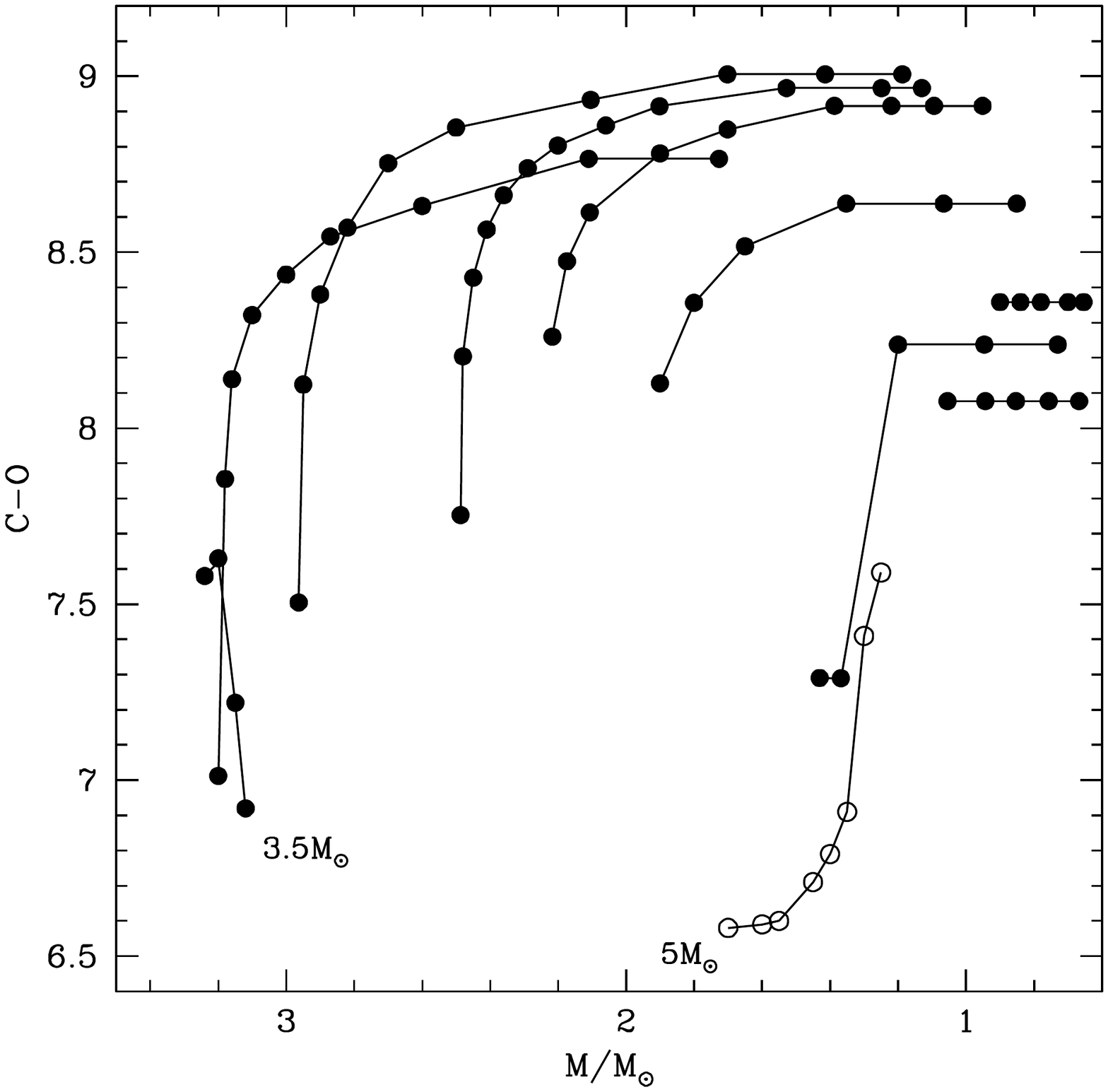}}
\end{minipage}
\vskip-70pt
\begin{minipage}{0.48\textwidth}
\resizebox{1.\hsize}{!}{\includegraphics{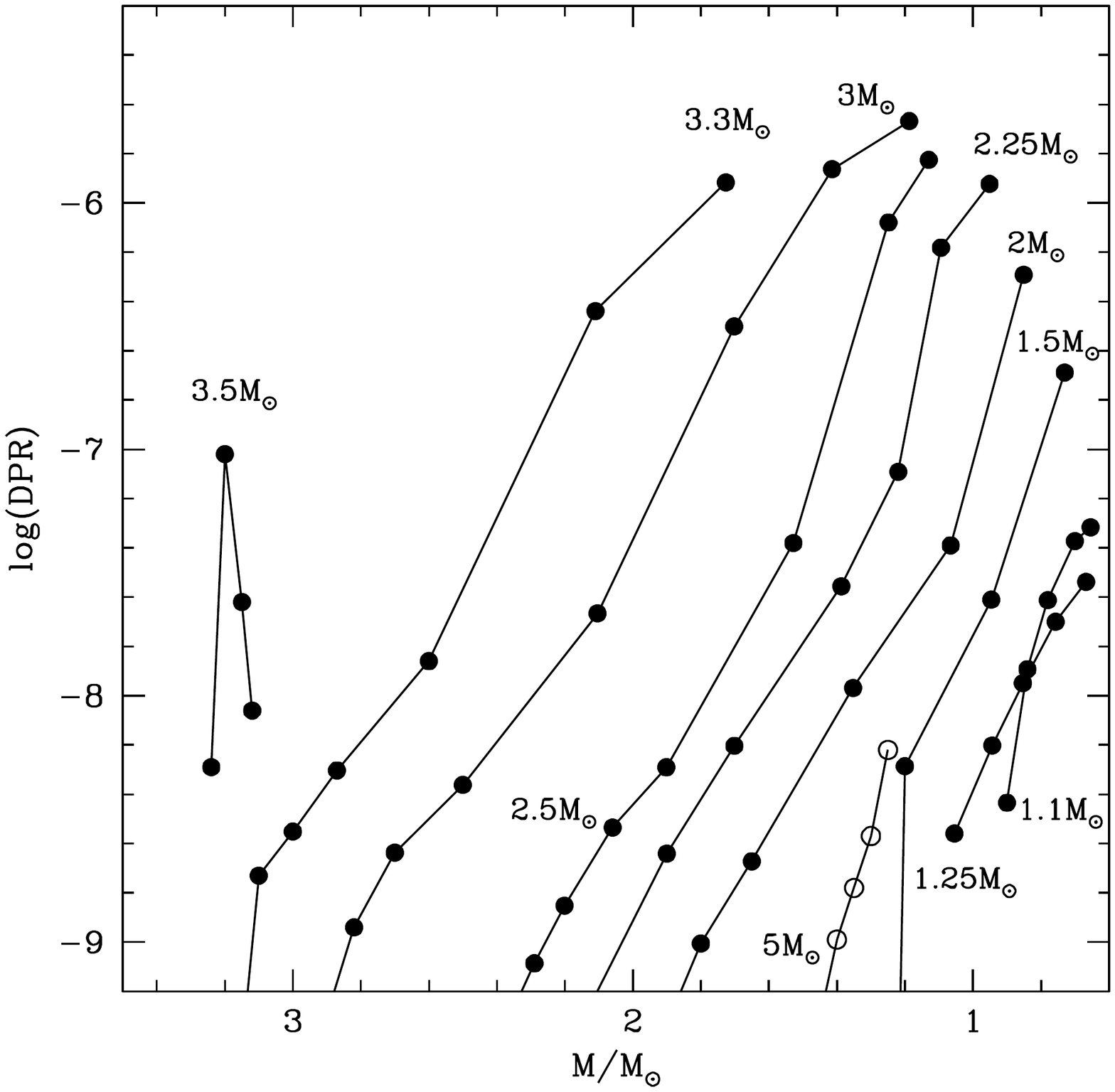}}
\end{minipage}
\begin{minipage}{0.48\textwidth}
\resizebox{1.\hsize}{!}{\includegraphics{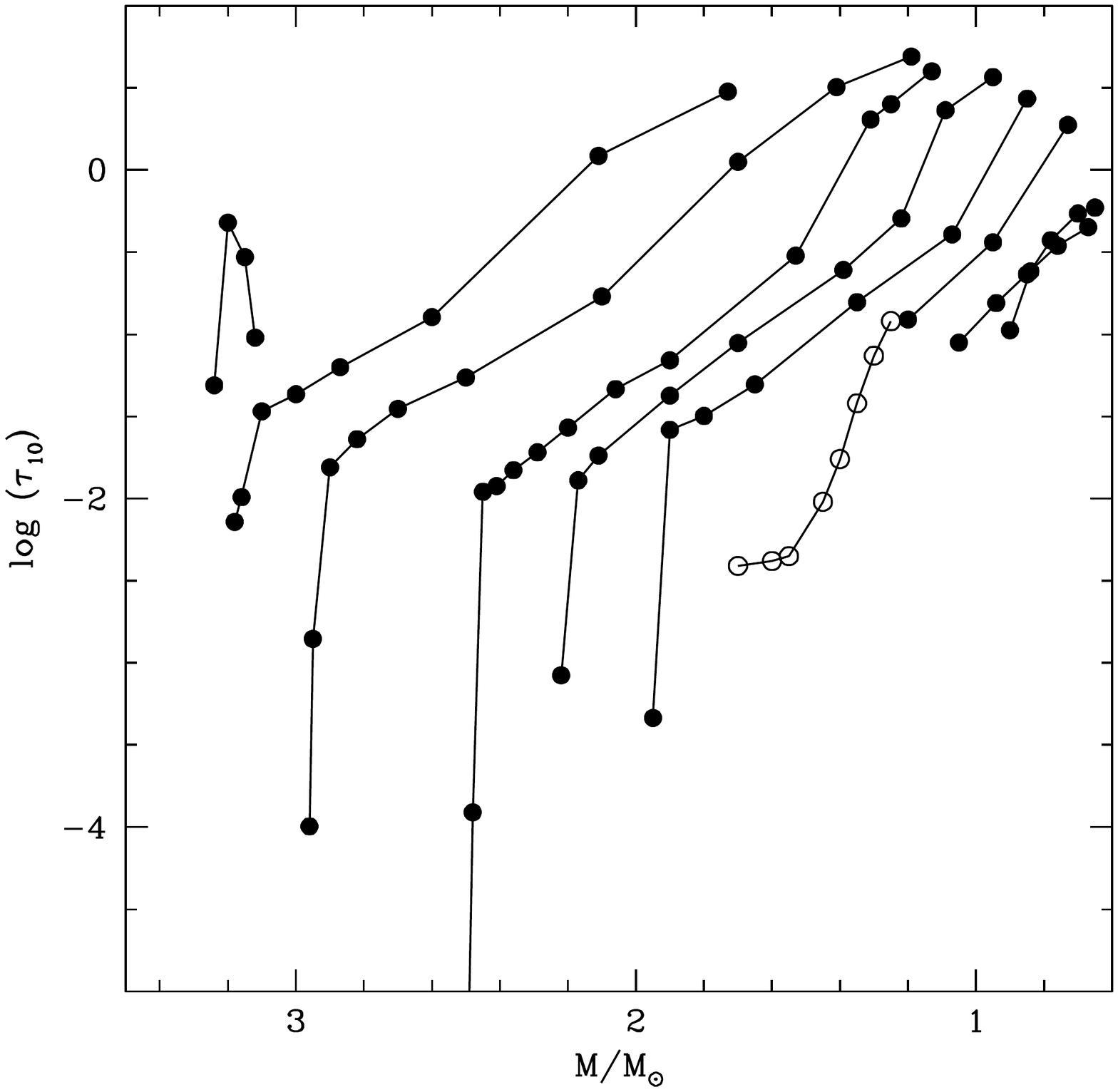}}
\end{minipage}
\vskip-40pt
\caption{Variation of luminosity (top-left panel), carbon excess
C$-$O (top-right, see text for the definition), dust production rate
(bottom-left) and $\tau_{10}$ (bottom-right) of the AGB models used
in the present work. The evolutions are shown with respect to the
current mass of the star on the abscissa. The different lines connect
a few points selected during the evolution trough the C-star phase.
Open points refer to the evolution of a $5~M_{\odot}$ star of metallicity
$Z=0.001$}
\label{fallm}
\end{figure*}

\subsection{Formation and evolution of carbon stars: the role of the
stellar mass}
\label{allm}
The behaviour of C-stars is sensitive to the mass of the progenitor. Higher mass stars are brighter, because their core mass is larger. Furthermore, the higher the
initial mass of the star the higher the mass in the envelope at the beginning of
the TP-AGB phase: this reflects into a higher number of TDU events and,
for a given TDU efficiency, into a more efficient surface carbon enrichment and
a larger dust production rate.

A summary of the main physical and chemical properties of the models used in the
present analysis are given in Table \ref{tabmod}, where for each mass we report 
the evolutionary time, the duration of the TP-AGB and of the carbon-star phase,
the luminosity range during the C-star phase, the final surface carbon-to-oxygen 
ratio and the highest dust production rate and optical depth reached. The latter
two quantities are attained towards the late AGB phases, before the general contraction, which marks the beginning of the post-AGB evolution.

Fig.~\ref{fallm} displays the evolution of stars with mass within the
$1.25-3.3~M_{\odot}$ range, that reach the surface C$/$O$>1$ condition 
and evolve as carbon stars for the remaining part of their evolution.
We also report in the same figure the tracks corresponding to the $3.5~M_{\odot}$ star, which evolves as carbon star during a small fraction of the AGB phase,
before HBB causes the drop in the surface carbon, and of the $5~M_{\odot}$
star of metallicity $Z=0.001$, which becomes C-star in the final part of
the AGB evolution, after HBB is turned off. 

The top-right panel of Fig.~\ref{fallm} shows the carbon excess, C$-$O, defined as the difference between the surface number densities of carbon and oxygen nuclei, normalised to the density of hydrogen: C$-$O$=12+\log[(n($C$)-n($O$))/n($H$)]$. This quantity is relevant to
understand how much carbon dust forms \citep{lars08, bladh19}. In $1.25-1.5~M_{\odot}$ stars we find that C$-$O is slightly below 8; on the other hand, $3~M_{\odot}$ stars reach C$-$O$\sim 9$ during the very final AGB phases.
In the $3.3~M_{\odot}$ case the final carbon excess is smaller than in the
stars of lower mass, because the final TDU episodes are so efficient and deep that some oxygen enrichment of the envelope occurs. The overproduction of
oxygen in carbon stars is consistent with the study of the chemistry of
carbon rich Planetary Nebulae by \citet{garcia16}. In the $3.5~M_{\odot}$ star 
C$-$O$<7.7$, because the ignition of HBB prevents further increase in the
surface carbon and eventually makes the star to become O-rich. The same upper limit approximately holds for the  $5~M_{\odot}$, $Z=0.001$ model, as the accumulation of carbon in the surface regions begins only during the final AGB phases, after HBB is turned off.

The evolution of the DPR, shown in the bottom-left panel of Fig.~\ref{fallm}, reflects the trend with the initial mass followed by the carbon excess. The higher availability of the surface carbon affects dust formation in two
ways: a) stars with a higher carbon mass fraction are more expanded and their effective temperatures are cooler, which favours dust production; b) the higher the percentage of gaseous carbon, the larger the number of molecules available to condense into solid carbon grains. $M \geq 2.5~M_{\odot}$ stars attain DPRs above $10^{-6}~M_{\odot}/$yr during the final phases, whereas in the $M < 2~M_{\odot}$ counterparts DPR is below $10^{-7}~M_{\odot}/$yr. The $2~M_{\odot}$ case discussed in the previous section follows an intermediate behaviour, with a maximum DPR of the order of $4\times 10^{-7}~M_{\odot}/$yr.

The formation of dust is associated with the shift of the SED to IR wavelengths, owing to scattering and absorption of photon by dust grains in the circumstellar envelope, which becomes more and more opaque to the electromagnetic radiation, with the consequent increase in $\tau_{10}$, whose evolution is shown in the bottom, right panel of Fig.~\ref{fallm}. According to the present modelling, we find that the largest optical depths, $\tau_{10} \sim 5$, are attained by $3~M_{\odot}$ stars; on the other hand low mass stars with initial mass $\sim 1.5~M_{\odot}$ reach $\tau_{10} \sim 1$ during the final AGB phases. Based on these arguments \citet{flavia15a} suggested that the most obscured stars in the LMC, identified in the colour-colour plane obtained with the IRAC filters, are the progeny of $2.5-3~M_{\odot}$ stars, formed during the burst in the star formation rate of the LMC, which occurred $\sim 300-600$ Myr ago \citep{harris09}.

\begin{table*}
\caption{Main properties of the $Z=0.008$ models discussed in the text. Only
the mass range of the stars which become carbon stars is considered. The various
columns report the following quantities: 1 - initial mass; 2 - formation epoch;  3 - duration of the TP-AGB phase; 4 - fraction of the AGB phase spent as carbon star;
5 - luminosity range of the inter-pulse phases; 6 - final surface C$/$O; 7/8/9 - largest DPR (7), optical depth (8) and $\lambda$ (9)}             
\label{tabmod}      
\centering          
\begin{tabular}{c c c c c c c c c c}     
\hline       
M$/$M$_{\odot}$ & $\tau_{\rm ev}$ (Myr) & $\tau_{AGB} (10^3$ yr) & $\%($C) &
L$/$L$_{\odot}$ & C$/$O & DPR (M$_{\odot}/$yr)& $\tau_{10}$ & $\lambda $\\ 
\hline                    
1.10 & 6530 & 1350 & 4  & 5000 - 5500 & 1.5 & $4.8\times 10^{-8}$ & 0.60 & 0.28 \\
1.25 & 4320 & 1310 & 8  & 6000 - 6300 & 1.3 & $2.9\times 10^{-8}$ & 0.45 & 0.40 \\
1.50 & 2234 & 1132 & 18 & 6000 - 6700 & 1.4 & $1.7\times 10^{-7}$ & 1.89 & 0.41 \\
2.00 & 1106 & 2400 & 25 & 6500 - 7300 & 2.0 & $5.1\times 10^{-7}$ & 2.7  & 0.41 \\
2.25 & 992  & 2920 & 23 & 8000 - 8200 & 2.8 & $1.2\times 10^{-6}$ & 3.7  & 0.47 \\
2.50 & 609  & 2520 & 37 & 8000 - 9200 & 3.0 & $1.5\times 10^{-6}$ & 4.0  & 0.58 \\
3.00 & 412  & 816  & 68 & 10500-12500 & 2.6 & $2.1\times 10^{-6}$ & 4.9  & 0.80 \\
3.30 & 313  & 450  & 50 & 15000-17000 & 2.3 & $1.1\times 10^{-6}$ & 3.0  & 0.80 \\
\hline
\end{tabular}
\end{table*}

\subsection{Uncertainties in AGB and dust production modelling}
\label{uncert}
The results regarding the characterisation of carbon stars are partly affected by some uncertainties, related to AGB modelling and to the description of dust formation. Before entering the discussion
on the interpretation of the individual sources, we believe important to present
a summary of the impact of these uncertainties on the conlusions reached.

As outlined in section \ref{2msun} and \ref{allm}, the quantity of dust that carbon stars form depends on the amount of $^{12}$C transported by TDU from the internal regions to the convective envelope. This is extremely sensitive to the treatment of the convective borders \citep{herwig00, herwig05, karakas14b}. 
However, this does not represent a serious issue in the present context, because 
the carbon stars population of the LMC has been traditionally used as a living laboratory to test the efficiency of TDU. Among others, we mention the works by \citet{martin93}, \citet{marigo99}, \citet{karakas02}, \citet{devika12}, who provided robust calibrations of the extent of TDU, for stars of different mass.
Thorough investigations aimed at calibrating mass-loss, TDU efficiency
and dust properties of AGB stars in the Magellanic Clouds were recently
presented in \citet{giada19} and \citet{giada20}.

Among the physical quantities relevant to the evolution through the AGB, shown in Fig.~\ref{fevol20}, mass loss is the most poorly known. This is a delicate point, because the description of mass loss determines the rate at which the mass of the envelope is expelled into the interstellar medium, the duration of the AGB 
phase, thus the amount of carbon that is gradually accumulated in the surface regions via TDU \citep{ventura05b}.

Several evolutionary codes adopt the classic period - mass loss rate relation by \citet{vw93}, both for O-rich and carbon stars \citep{cristallo09, karakas14b}.
Some research teams have based the mass loss modelling of carbon stars 
on the theoretical radiation-hydrodynamical models published by the Berlin group 
\citep{wachter02, wachter08}, which consider dust production in C-rich winds, and the effects of radiation pressure on the carbonaceous dust particles formed in the circumstellar envelope \citep{weiss09, ventura18}. The formulae by \citet{wachter02, wachter08} do not include any dependence on the carbon excess with respect to oxygen, which should affect the amount of dust formed in the wind of the stars \citep{lars09, lars10, eriksson14, bladh19}.

The treatment of mass loss affects the description of the dust formation process, because the density of the wind is directly proportional to $\dot M$. In 
optically thin environments we expect that the calculated optical depth shows us some sensitivity to $\dot M$. On the other hand, in the optically thick case some compensation takes place, because the formation of more dust in the regions just above the condensation zone triggers a higher acceleration of the wind, which decreases the density of the gas, thus inhibiting further dust formation: in these cases the description of mass loss affects the tomography of the wind, but the effects on the global amount of dust formed, hence on the optical depth along the observation line, is modest.

The description of the stellar wind and of the dust formation process used in
the present investigation is characterised by uncertainties, the most relevant being the effects of pulsations, the poor knowledge of the sticking coefficients and of the density of seed nuclei, upon which the dust grains grow. 
The primary effect of large amplitude pulsations is the formation of several shock fronts, moving outwards and potentially able to project dense matter to cool regions of the envelope, suitable to condensation of gas molecules into dust \citep{bowen88, bertschinger, fleischer}. The inclusion of the effects of shocks is beyond the current computational capabilities; furthermore, these calculations would be extremely time consuming in the present case, because the system of moments and radiative transfer equations commonly used to model the atmosphere should be applied to all the models in which the AGB evolution is split, usually between $\sim 50000$ and $\sim 100000$, according to the criteria followed to choose the time-steps. On the other hand, what we obtain by use of the stationary wind modelling is the average amount of dust formed in the outflow.

The role of the sticking coefficients of the different dust species is extensively discussed in \citet{fg01, fg02, fg06}. In the present context the key point is
how the choice of the coefficients affects the formation of solid carbon,
which is the species that gives the dominant contribution to the optical depth.
On general grounds, we find that changes in the sticking coefficients can alter the dust density gradient from the carbon dust condensation zone, but the effects on the size reached by carbon grains and on the overall dust formed is on percentage much smaller than the change in $\alpha_C$.

One of the most relevant open points connected with the dust formation in the wind of AGB stars is the nature and the density of seed nuclei, which impact on the density of the dust grains, the fraction $f$ of the gas condensed into dust and the overall scattering coefficient, which is proportional to $f$ \citep{fg01}. In the present computations, following \citet{fg06}, we adopted
$n_d=10^{-13}$ for all the dust species considered. 
However, other choices are possible; for example \citet{nanni13} propose to scale the number density of the seeds of carbon dust with C$-$O. 

The choice of $n_d$ translates into the size of the grains formed: a higher $n_d$ for a given dust species implies higher condensation fractions of the relevant gaseous molecules, which decrease the growth rate $J_i^{\rm gr}$, thus leading to the formation of smaller size particles. If we focus the attention on carbon grains, to determine the degree of obscuration of the star at a given evolutionary stage, we find that the choice of $n_d$ reflects into the size reached by the carbon grains, but has only a small effect on the value of $\tau_{10}$. Therefore, in the interpretation of the IRS spectra, which will be addressed in the following sections, we will safely rely on the highest values of $\tau_{10}$ reported in table \ref{tabmod}. 

On the other hand, we cannot rule out that different dust species have different $n_d$. This would not affect $\tau_{10}$, but would change the relative distribution of the different particles, which would translate into significant alterations in the SED. As an example, we show in the bottom panels of Fig.~\ref{fdust20} how the SED of the star is modified if small percentages of grafite and MgS, are added to SiC and C. 

A description of the wind of carbon stars with such a high predictive power is 
currently unavailable. In the following, we will characterise the individual stars based on our expectations regarding the evolution of the luminosity and the optical depth in stars of various mass and during different evolutionary phases; on the other hand we will be guided by the observations in the attempt of
understanding which dust species form and in what percentages.

\begin{figure*}
\begin{minipage}{0.48\textwidth}
\resizebox{1.\hsize}{!}{\includegraphics{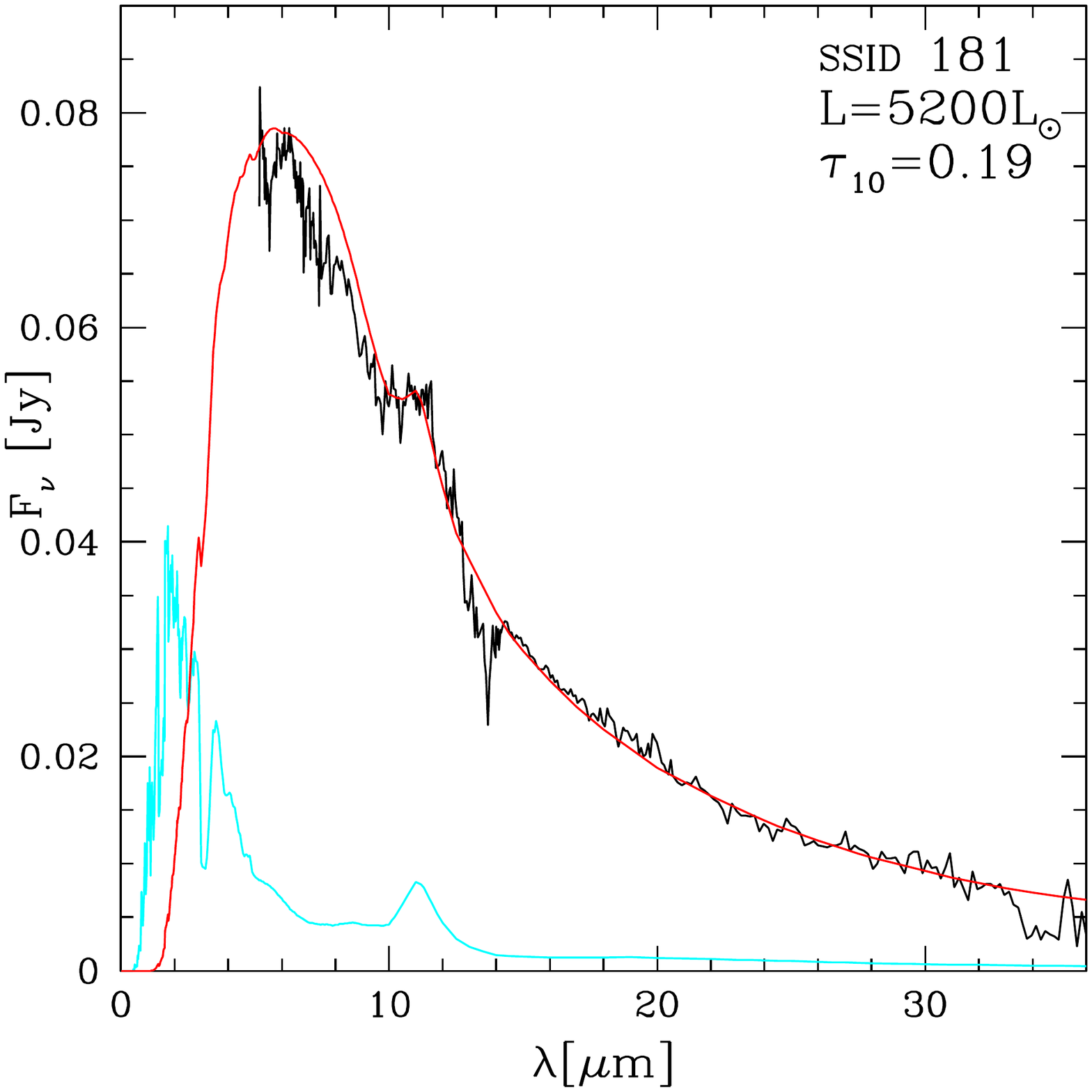}}
\end{minipage}
\begin{minipage}{0.48\textwidth}
\resizebox{1.\hsize}{!}{\includegraphics{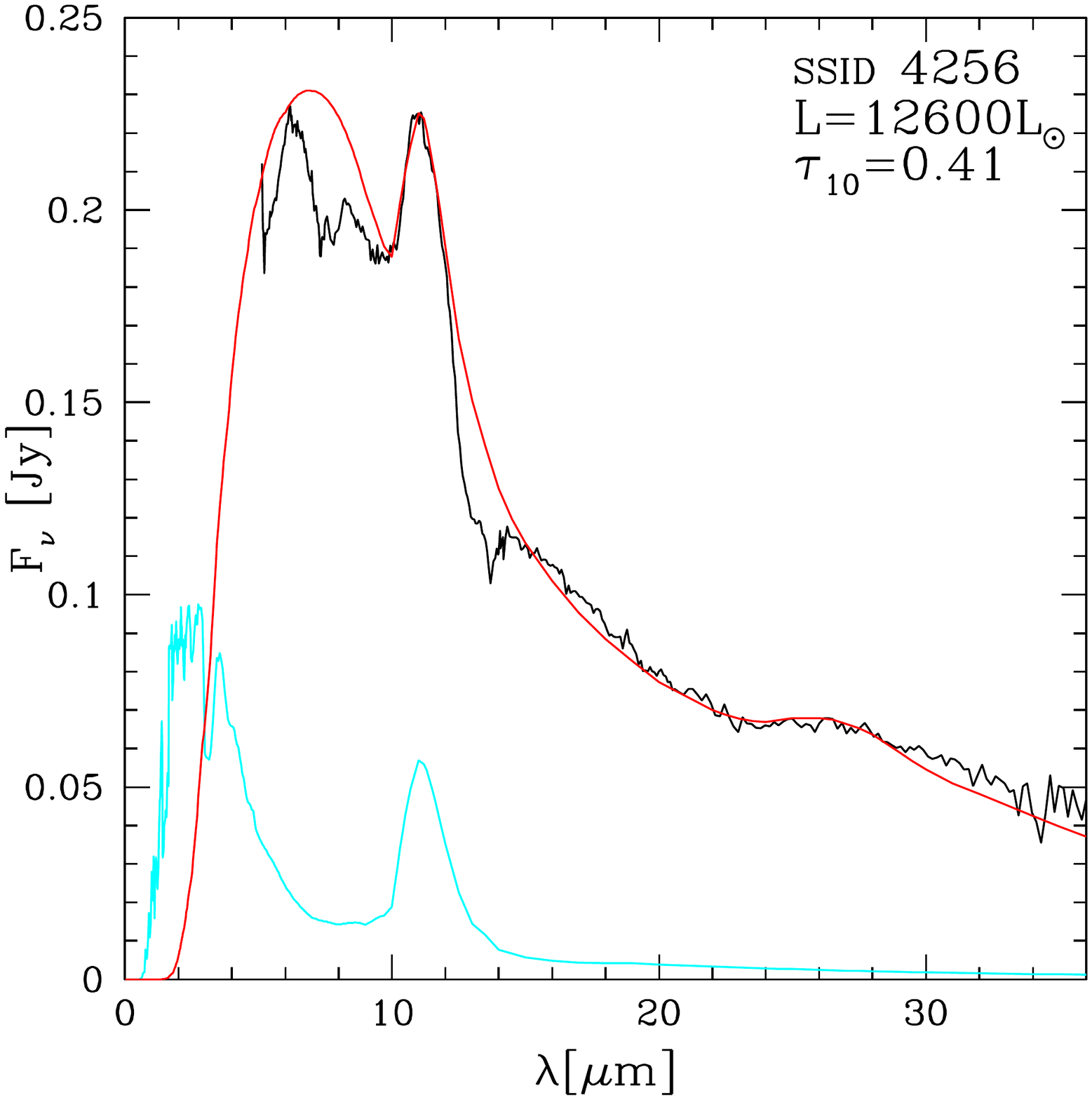}}
\end{minipage}
\vskip-70pt
\begin{minipage}{0.48\textwidth}
\resizebox{1.\hsize}{!}{\includegraphics{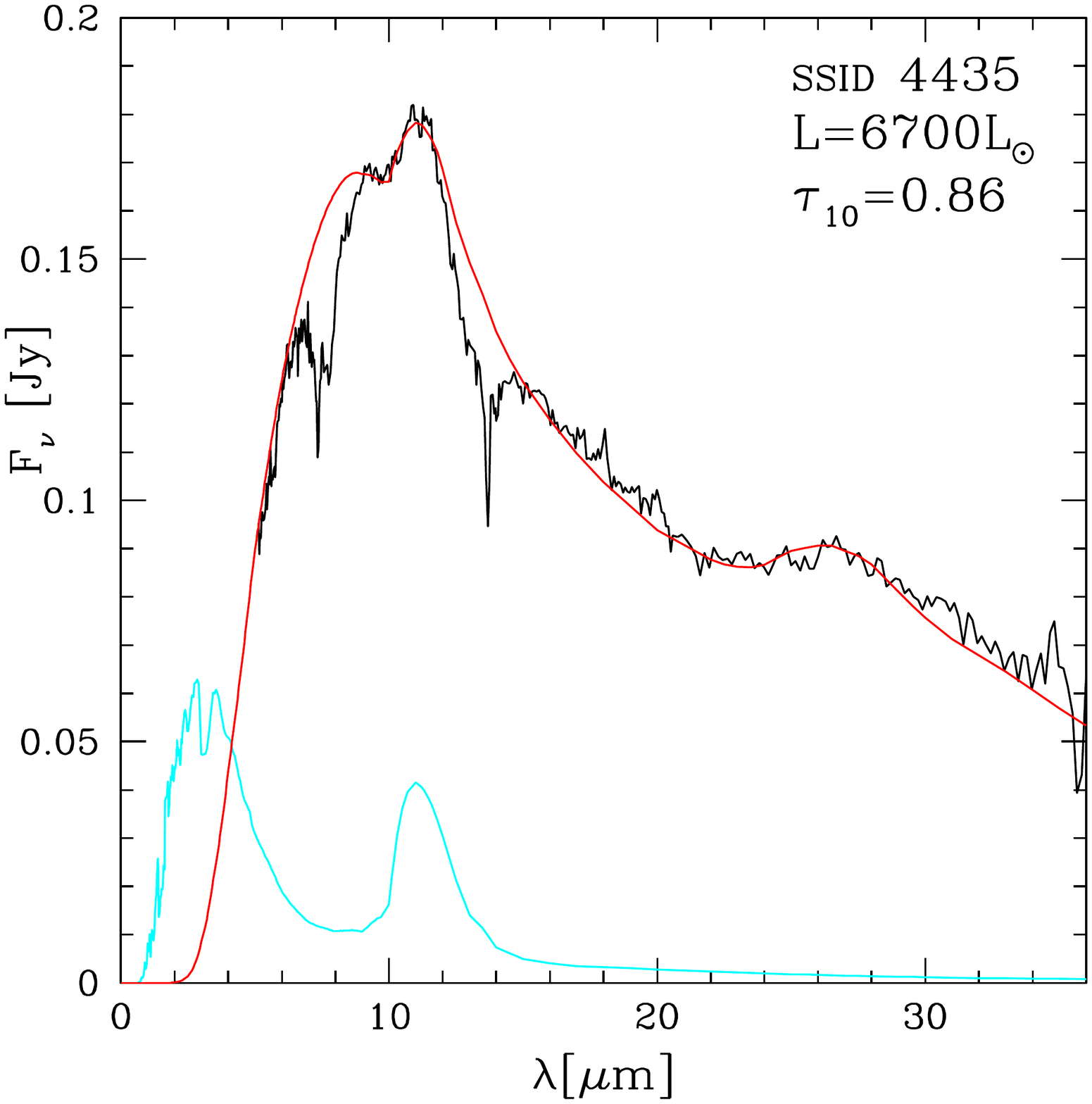}}
\end{minipage}
\begin{minipage}{0.48\textwidth}
\resizebox{1.\hsize}{!}{\includegraphics{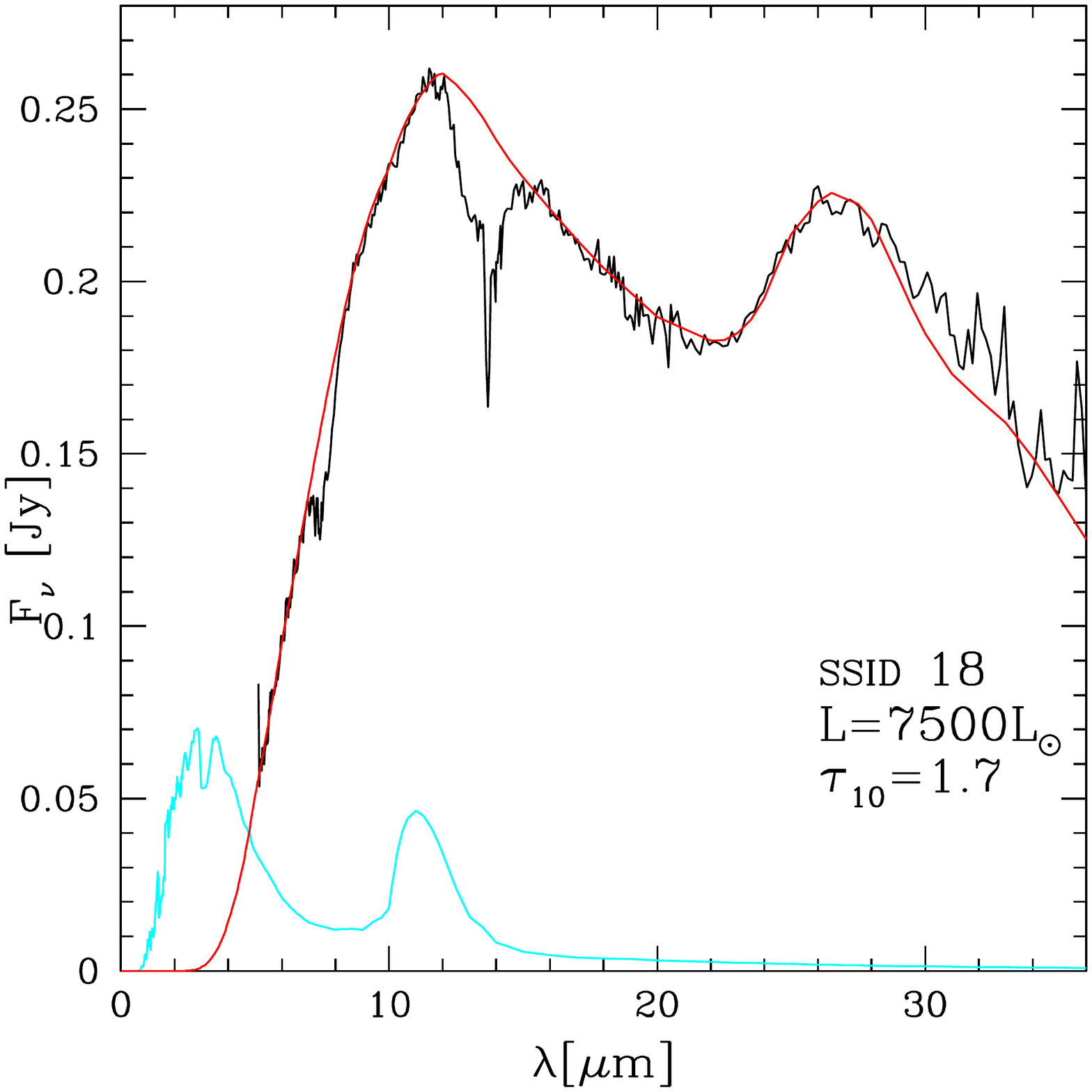}}
\end{minipage}
\vskip-40pt
\caption{Comparison between the IRS (black lines) and the synthetic (red)
spectra of a few sources in the LMC sample of carbon stars, with various luminosities and degrees of obscuration. The values of luminosities and $\tau_{10}$ reported in each panel are those allowing the best fit, corresponding to the red lines. The cyan lines indicate the synthetic SED entering the carbon condensation zone, resulting from absorption of the radiation from the photosphere by SiC particles in the most internal region of the circumstellar envelope.}
\label{fsedex}
\end{figure*}

\longtab{
\centering                                
\begin{longtable}[t]{|c|c|c|c|c|}    
\caption{The interpretation of the sample carbon stars analysed in the present work. The different columns report the estimated luminosity (1), dust mineralogy (2), optical depth $\tau_{10}$ (3). Column 4 reports the luminosity estimated by GS18.}  
\label{tabsample} \\
\hline
 SSID  & L/L$_{\odot}$  &  \%(C,SiC,MgS) &  $\tau_{10}$  &  L/L$_{\odot}$ (GS18)  \\
\hline
  3   & 4100  & (87,13,0)  & 0.1  & 5325 \\
  7   & 7900  & (99,1,0)   & 0.05  &  8170 \\
  9   & 5200  & (81,14,5) & 3.4   & 5376   \\
  18  & 7500  & (83,8,9)  & 1.7   & 8349   \\
  51  & 13000 & (100,0,0) & 0.04  & 12038  \\
  55  & 7700  & (97,2,1)  & 0.1   & 10226  \\
  60  & 3300  & (93,4,3)  & 0.54  & 5268   \\
  65  & 6200  & (78,21,1) & 7.1   & 6169   \\
  66  & 4300  & (100,0,0) & 0.32  & 4740   \\
  80  & 5400  & (90,10,0) & 0.024 & 4918   \\
  98  & 5400  & (97,2,1)  & 0.33 &  7408.  \\
  103 & 4000  & (89,11,0) & 0.075 & 3552 \\
  125 & 8800  & (86,12,2) & 2.3  & 8253 \\
  126 & 12200 & (100,0,0) & 0.06  & 10575  \\
  140 & 8500  & (87,6,7)  & 1.2   & 9439   \\
  141 & 3000  & (100,0,0) & 0.29  & 4213   \\
  145 & 6100  & (100,0,0) & 0.19  & 8311   \\
  156 & 5200  & (90,10,0) & 0.13  & 6474   \\
  167 & 10200 & (74,19,7) & 2.0   & 10863  \\
  181 & 5200  & (95,3,2)  & 0.19  & 8864   \\
 190  & 12000 & (96,2,2)  & 3.4   & 11509 \\
 4000 & 7500  & (90,9,1)  & 0.79  & 12136  \\
 4001 & 7900  & (78,15,7) & 1.25  & 8380   \\
 4002 & 5300  & (89,8,3)  & 0.45  & 6409   \\
 4003 & 5500  & (92,4,4)  & 1.0   & 6902   \\
 4004 & 9900  & (91,7,2)  & 0.67  & 12159  \\
 4012 & 7100 & (99,0,1) & 0.38 & 7719    \\
 4016 & 11500  & (95,4,1)  & 0.7   & 15377 \\
 4021 & 15500  & (100,0,0) & 0.6  & 38738 \\
 4034 & 11500 & (100,0,0) & 0.23 & 15044  \\  
 4037 & 18200 & (99,1,0)  & 0.18  & 19003 \\ 
 4052 & 6900  & (87,7,6)  & 0.65  & 8367  \\ 
 4062 & 11500 & (92,5,3)  & 1.05  & 12738 \\ 
 4067 & 10500 & (94,5,1)  & 0.27  & 13324 \\
 4093 & 5600  & (92,5,3)  & 0.58  & 6884  \\
 4100 & 13200 & (87,7,6)  & 1.1   & 15097 \\ 
 4109 & 27800 & (96,3,1)  & 0.37  & 31412 \\ 
 4150 & 7300  & (86,9,5)  & 0.7   & 8305  \\ 
 4154 & 14700 & (93,4,3)  & 0.54  & 14173 \\ 
 4155 & 10800 & (89,9,2)  & 0.85  & 12405 \\ 
 4171 & 8200  & (78,19,3) & 6.3   & 8030  \\ 
 4185 & 5200  & (74,25,1) & 6.2   & 5968  \\ 
 4197 & 11600 & (85,9,6)  & 1.45  & 11262 \\ 
 4206 & 8900  & (81,16,3) & 1.06 & 9989  \\
 4211 & 12300  & (99,1,0)  & 0.18  & 13782 \\
 4225 & 14900 & (95,3,2) & 0.23  & 14209  \\
 4228 & 9500  & (97,2,1)  & 0.65  & 15443  \\
 4238 & 13800 & (97,0,3)  & 1.15  & 16145  \\
 4240 & 8300  & (97,1,2)  & 1.0   & 9401   \\
 4241 & 10800 & (92,4,4)  & 0.94  & 11913  \\
 4244 & 9700  & (95,5,0)  & 0.51  & 13152  \\
 4246 & 13900 & (85,7,8)  & 1.7   & 15453  \\
 4251 & 10400 & (100,0,0) & 0.62  & 23710  \\
4252 & 8000  & (95,3,2)  & 1.0   & 9756   \\
 4256 & 12600 & (87,11,2) & 0.41 & 17281  \\
 4293 & 7200  & (83,10,7) & 0.95 & 7773  \\
 4299 & 9800 & (100,0,0) & 5.3 & 9310     \\
  4308 & 7000  & (77,23,0) & 5.6   & 6921   \\
 4309 & 11500 & (96,0,4)  & 1.2   & 13882  \\
 4334 & 6500  & (97,2,1) & 0.38 & 9403  \\
 4339 & 5500  & (84,9,7) & 0.9  & 5827 \\
 4385 & 5500  & (89,10,1) & 0.22  & 9593   \\
 4391 & 10800 & (93,4,3)  & 0.56  & 13103  \\
 4401 & 6700  & (87,7,6)  & 1.14  & 7917   \\
 4402 & 7200  & (91,5,4)  & 1.05  & 8354   \\
 4408 & 9300  & (88,7,5)  & 0.6  & 9264 \\
 4411 & 15800 & (97,2,1)  & 0.19  & 15861  \\
 4415 & 4700 & (80,20,0) & 6.6 &  3857   \\
 4421 & 10800 & (94,5,1)  & 0.68  & 11978  \\
 4432 & 5700  & (100,0,0) & 0.005 & 7368   \\
 4435 & 6700  & (94,3,3)  & 0.86  & 7678   \\
 4442 & 7300  & (95,5,0)  & 0.26  & 9348  \\ 
 4447 & 7600  & (97,2,1)  & 0.51  & 10417 \\
 4448 & 5800  & (100,0,0) & 0.03  & 5784   \\
 4451 & 28600 & (89,6,5)  & 0.91  & 24856  \\
 4463 & 6400 & (92,7,1) & 0.21 & 8233      \\
 4469 & 6700 & (97,3,0) & 0.23 & 8772     \\
 4476 & 14000 & (100,0,0) & 0.1   & 14134  \\
 4478 & 7100  & (93,4,3)  & 0.79  & 9851   \\
 4479 & 5500  & (82,16,2) & 0.7   & 7800   \\
 4481 & 6700  & (89,5,6)  & 0.7  & 5932  \\
 4488 & 11600 & (90,10,0) & 0.035 & 11041 \\
 4489 & 5100  & (57,40,3) & 4.3   & 5434   \\
 4491 & 8000  & (97,0,3)  & 0.98  & 8869   \\
 4510 & 10300 & (96,4,0)  & 0.28  & 10608  \\
 4513 & 16100 & (94,5,1)  & 0.62  & 19501  \\
 4519 & 6800  & (88,6,6)  & 0.52  & 8203   \\
 4540 & 32000 & (100,0,0) & 0.06  & 29760  \\
 4556 & 12200 & (82,16,2) & 1.0   & 14349  \\
 4562 & 8400  & (89,6,5)  & 0.97  & 10240  \\
 4565 & 5000  & (96,2,2)  & 0.75  & 7656   \\
 4575 & 12100 & (99,1,0)  & 0.33  & 11879  \\
 4589 & 8800  & (96,0,4)  & 0.3   & 10651  \\
 4593 & 5700  & (83,10,7) & 0.95  & 6149   \\
 4600 & 5800  & (93,4,3)  & 0.85  & 6653   \\
 4604 & 11400 & (94,5,1)  & 0.23  & 15407  \\
 4692 & 5500  & (91,4,5)  & 1.2   & 6132   \\
 4717 & 10300 & (91,5,4)  & 1.0   & 9330   \\
 4722 & 12300 & (90,8,2)  & 0.9   & 16840  \\
 4736 & 8800  & (97,2,1)  & 0.28  & 10851  \\
 4758 & 7500  & (88,11,1) & 0.85  & 10815  \\
 4759 & 7200  & (87,6,7)  & 1.1   & 7043   \\
 4776 & 22200 & (100,0,0) & 0.09  & 23808  \\
 4779 & 16500 & (79,17,4) & 1.2   & 17677  \\
 4780 & 10500 & (93,6,1) & 0.52  & 13949  \\
 4781 & 10200 & (70,27,3) & 5.2   & 10412  \\
 4783 & 10000 & (84,13,3) & 1.05 & 10687 \\
 4794 & 17700 & (80,15,5) & 2.0   & 21262  \\
 4811 & 16200 & (87,6,7)  & 1.13  & 19249  \\
 4812 & 12300 & (96,0,4)  & 1.1   & 14191  \\    
\hline 
\end{longtable}
}

\section{The characterisation of carbon stars in the LMC}
\label{lmcc}
To understand the distribution of carbon stars and the obscuration 
sequences in the observational planes (colour-colour and colour-magnitude diagrams obtained with the {\itshape JWST} filters) we need to characterise the individual sources. To this aim, we first confront the IRS data of each star belonging to  the sample described in section \ref{sample} with the sequence of
synthetic SEDs, which represent the evolution of the spectrum of each model star,
built as described in section \ref{modinput}. The identification of the synthetic
SED that best reproduces the IRS spectrum leads to a robust derivation of 
the luminosity and of the optical depth of each source, which are essential ingredients to deduce the mass of the progenitors (see table \ref{tabmod}) and the current AGB phase they are evolving through. The results are reported in table \ref{tabsample}, which gives the derived values of $L$, $\tau_{10}$ and dust mineralogy for each source investigated.

A few examples of this analysis are shown in Fig.~\ref{fsedex}, reporting the interpretation of the SED of stars of various L and $\tau_{10}$. The main deviations of the synthetic SED from the observed spectrum are in correspondence of the C$_2$H$_2$ molecular bands in the regions $6.6-8.5~\mu$m (centered at $\sim 7.5~\mu$m) and $13.5-13.9~\mu$m \citep{matsuura06} and the CO+C$_3$ band in the region $5.0-6.2~\mu$m \citep{jorgensen00}. The properties of these features have been widely analysed in the past (e.g. \citet{sloan16} and references therein) but, unfortunately, they are poorly or not reproduced by model atmospheres available in the literature, owing to the scarce knowledge of the opacities of such molecules. The $\sim 7.5~\mu$m feature affects the flux in the $6-9 ~\mu$m spectral region, the effect becoming smaller and smaller as $\tau_{10}$ increases, eventually vanishing for $\tau_{10} > 1$. On the other hand the 
C$_2$H$_2$ $\sim 13.7~\mu$m feature, which affects the emission in the $12-16~\mu$m domain, is present in the spectra of carbon stars characterised 
by different $\tau_{10}$, even in those with very large infrared emission. The CO+C$_3$ band is the most tricky to handle in the present
analysis, because it is spread across a spectral region only partially covered
by IRS, which poses some problems in the extrapolation of the flux in the
optical and near-IR domains. For the stars where this ambiguity is most relevant, we will consider the combination of the IRS data with results from photometry, following an approach similar to \citet{martin18}. Fortunately the analysis of the most obscured stars is not affected by the CO+C$_3$ band, as the
SED peaks at longer wavelengths.

\begin{figure}
\resizebox{\hsize}{!}{\includegraphics{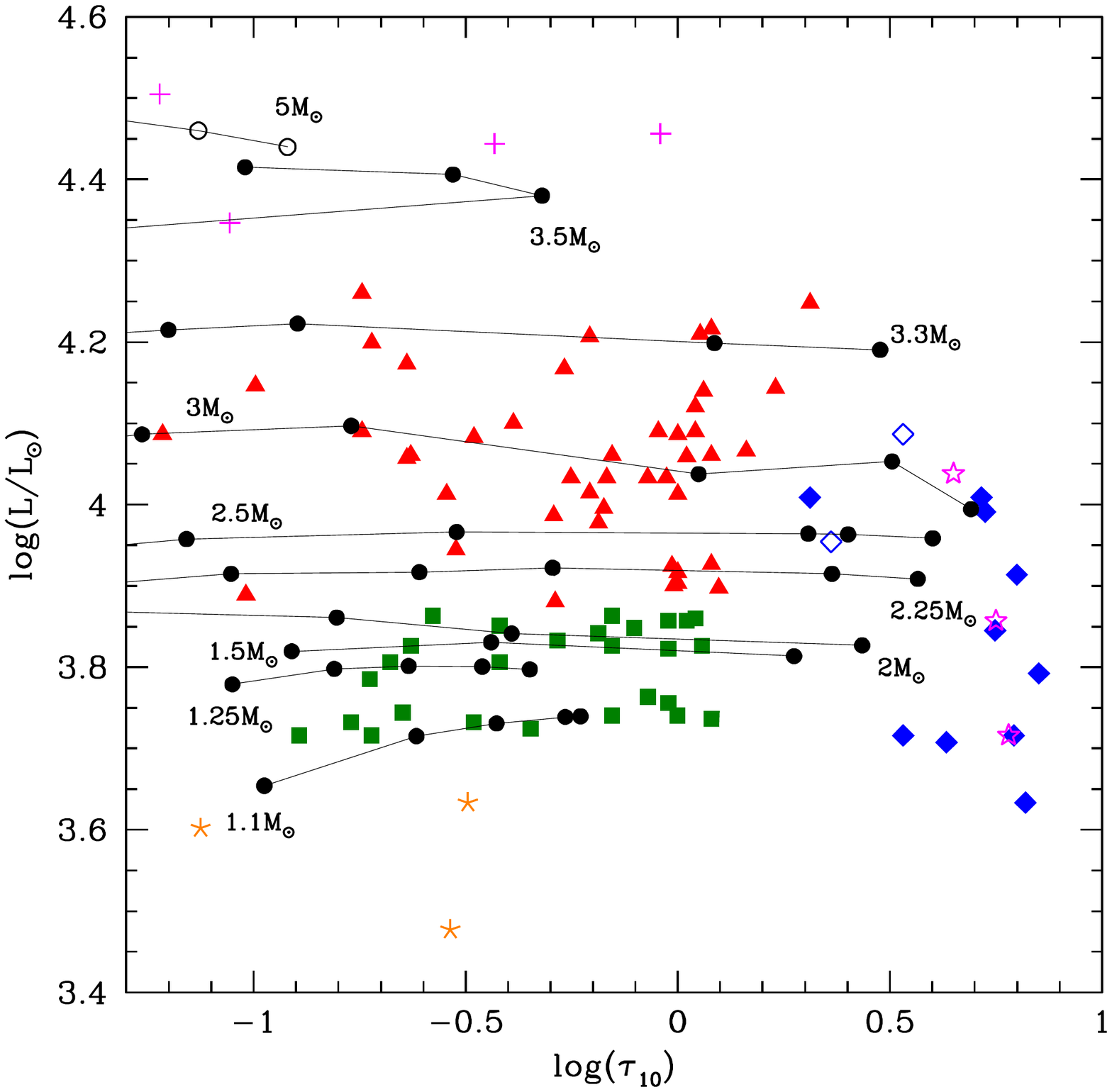}}
\vskip-60pt
\caption{Luminosities and optical depths derived for the stars in the sample
analysed in the present work. Green squares and red triangles indicate the progeny of $M<2~M_{\odot}$ stars and the higher mass counterpart, respectively; orange asterisks refer to stars with luminosities below $5000~L_{\odot}$; magenta crosses indicate stars brighter than $20000~L_{\odot}$; blue diamonds indicate extremely obscured objects, taken from \citet{gruendl08}, plus the source SSID 9; the stars indicated with open diamonds are SSID 125 and SSID 190. The black lines and points in the figure refer to the evolution
of the same models shown in Fig.~\ref{fallm}; the initial masses
are indicated next to the corresponding track. Magenta stars indicate the results obtained by artificially increasing the mass loss rate of the stars of initial mass $1.1~M_{\odot}$, $2.5~M_{\odot}$, $3~M_{\odot}$, after reaching the C-star stage, according to the discussion in section \ref{ero}. 
}
\label{ftaul}
\end{figure}

The result of the interpretation of the observations of the stars in the sample
analysed here is shown in Fig.~\ref{ftaul}, where we report the distribution of
the individual sources on the $\tau_{10}$ versus luminosity plane. Overimposed to the points indicating the stars in the sample discussed here are the tracks, corresponding to the evolution of the AGB stars discussed in section \ref{allm}. 
Green squares in Fig.~\ref{ftaul} indicate stars with luminosities below $7500~L_{\odot}$: these are the oldest objects in the sample, descending from stars of initial mass below $2~M_{\odot}$, formed in epochs earlier than one Gyr ago. Red triangles indicate objects younger than 1 Gyr, descending from stars of initial mass above $2~M_{\odot}$. In particular, those with luminosities above $10^4~L_{\odot}$ are interpreted as the progeny of stars of initial mass $2.5-3.3~M_{\odot}$, formed during the peak in the star formation history of the LMC, between 300 and 600 Myr ago \citep{harris09}. Magenta crosses refer to
bright stars, with luminosities above $20000~L_{\odot}$.

The distribution of the stars in Fig.~\ref{ftaul} is determined by the mass distribution of the progenitors (luminosity) and the amount of carbon accumulated in the surface
regions, which affects the dust formation in the outflow, thus the optical depth. The stars on the right side of the diagram are those producing dust at higher rates. We find that the gas-to-dust ratio, $\Psi$, decreases when moving to higher $\tau_{10}$, ranging from $\Psi \sim 700$, for the stars with little dust, to $\Psi \sim 100$, for the most obscured sources. These results are fully consistent with those obtained by \citet{nanni19b}, who correctly warned against the choice of assuming a constant $\Psi$ when fitting the observed SED. The gas and dust mass loss rates also increase with $\tau_{10}$, until reaching $\dot M \sim 1.5\times 10^{-4}~M_{\odot}/$yr and DPR $\sim 10^{-6}~M_{\odot}/$yr. In the $\tau_{10}<1$ domain the $\dot M$  and DPR values derived here are consistent with \citet{nanni19b}, whereas for the stars with the largest degree of obscuration the present results are similar to \citet{gulli}. 

The tracks reported in Fig.~\ref{ftaul} nicely reproduce the distribution of the sources on the plane, with the exception of a few groups of stars, for which the derivation of the main physical and evolutionary properties is not straightforward. In particular, we refer to: a) the stars with luminosities below $5000~L_{\odot}$, indicated with orange asterisks, which are too faint when compared to the evolutionary tracks; b) the stars on the right  part of the figure, at $\tau_{10}>3$, indicated with blue diamonds, whose degree of obscuration is in excess of the theoretical expectations (see the largest $\tau_{10}$ values expected for stars of different mass and luminosity, reported in table \ref{tabmod}). We discuss the stars outlined in points (a) and (b) and the bright stars indicated with magenta crosses separately.

\subsection{Stars in the post thermal pulse phase}
\begin{figure*}
\begin{minipage}{0.48\textwidth}
\resizebox{1.\hsize}{!}{\includegraphics{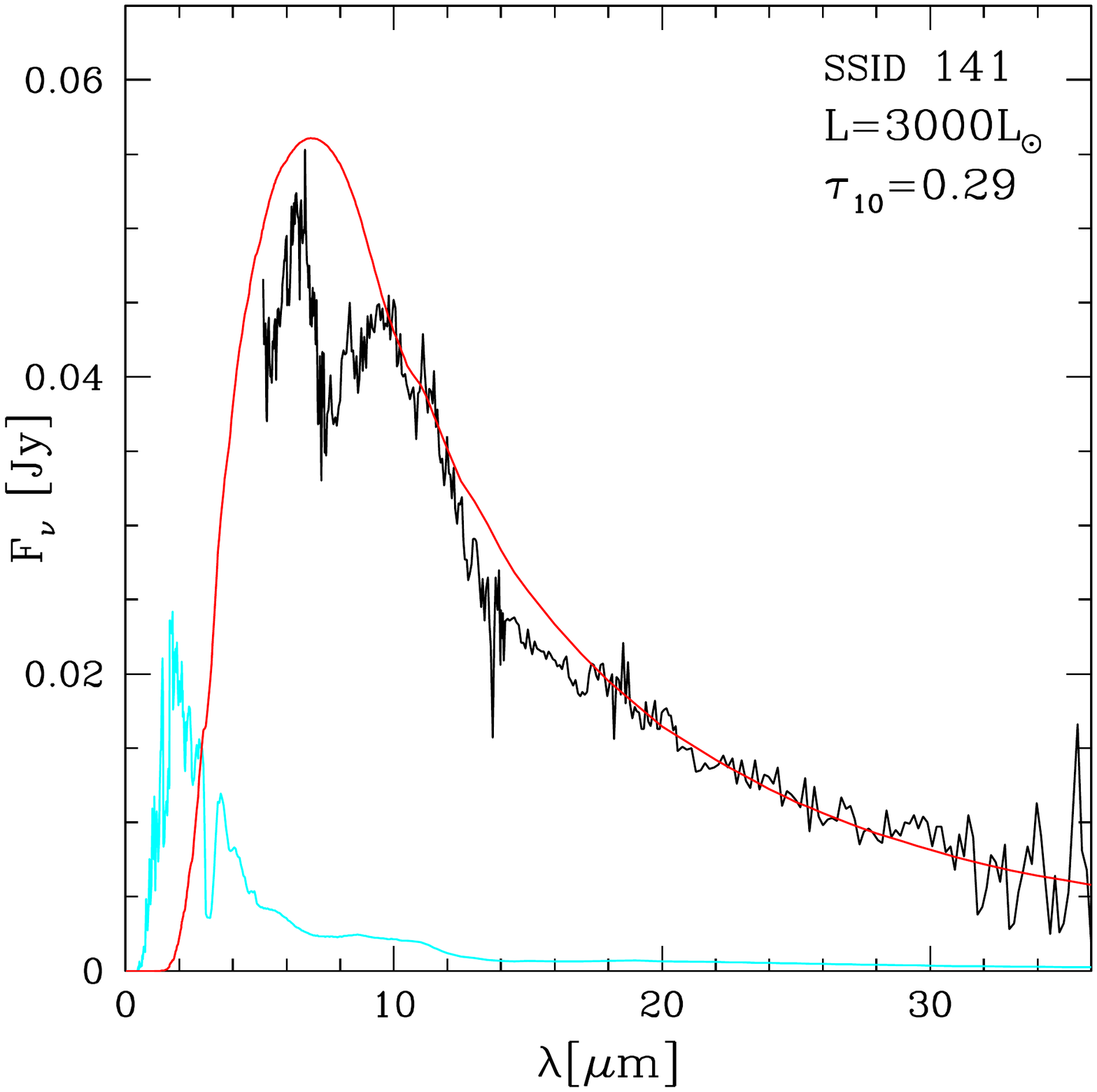}}
\end{minipage}
\begin{minipage}{0.48\textwidth}
\resizebox{1.\hsize}{!}{\includegraphics{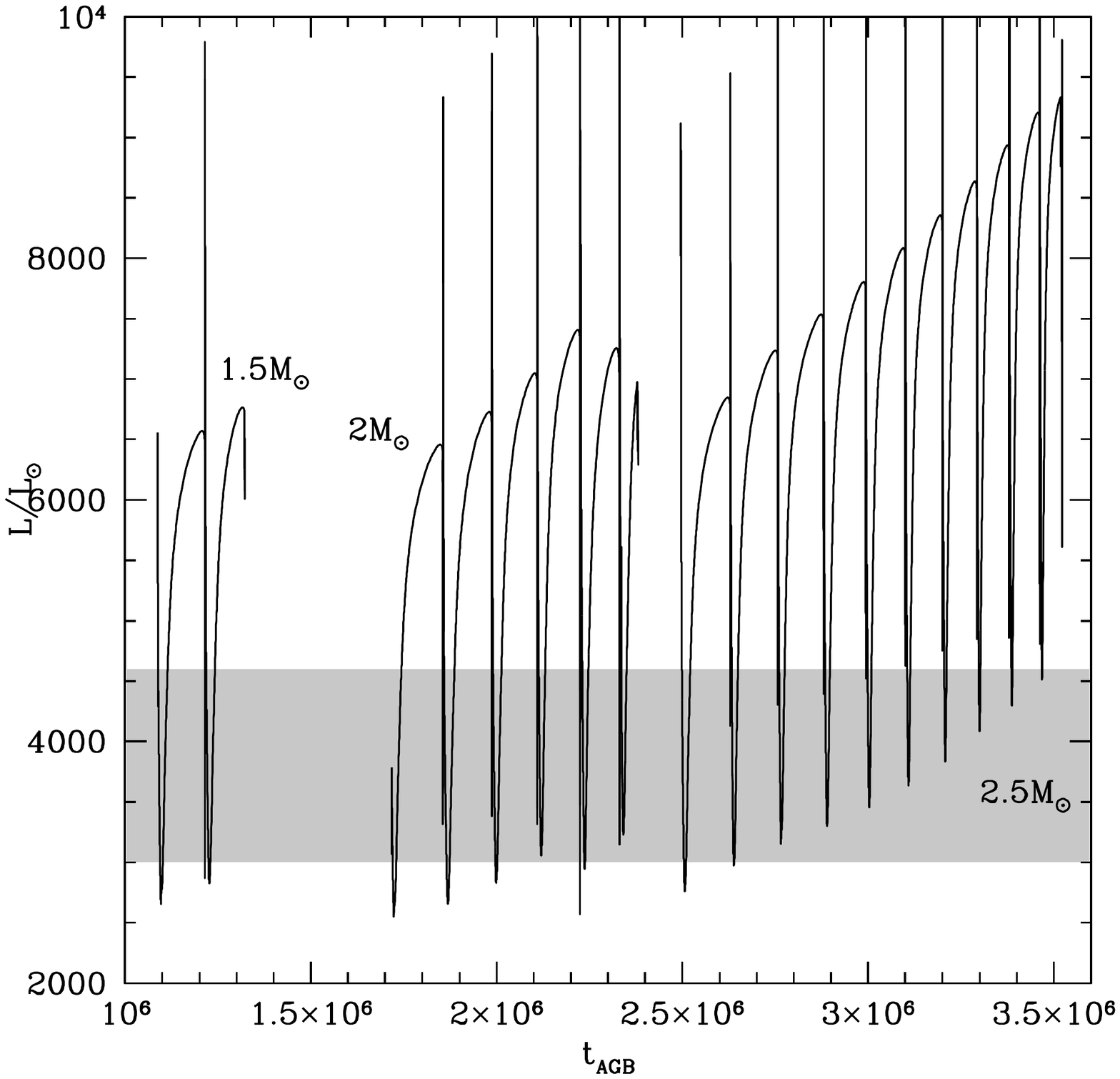}}
\end{minipage}
\vskip-40pt
\caption{Left: IRS SED of the source SSID 141, compared with our interpretation,
based on synthetic modelling. Right: AGB evolution with time of the luminosity of
$1.5, 2, 2.5~M_{\odot}$ stars; the O-rich phase was omitted for readability. The
sequence of the $2.5~M_{\odot}$ was artificially shifted by 1 Myr for readability. The gray shaded region indicates the luminosity range of the
low-luminosity stars SSID 3, SSID 66, SSID 103 and SSID 141.}
\label{flowl}
\end{figure*}

The sources SSID 3, SSID 66, SSID 103 and SSID 141 are characterised by luminosities below $5000~L_{\odot}$. The IRS spectrum of the latter star, with our best-fit interpretation, is shown in the left panel of Fig.~\ref{flowl}. These energy fluxes are at odds with the results shown in the top-left panel of Fig.~\ref{fallm} and in Fig.~\ref{ftaul}, both reporting the luminosities of the stars during the sequence of interpulse phases, when the CNO burning activity is fully recovered, after the temporary extinction owing to the ignition of the thermal pulse. On the other hand the re-ignition of the CNO shell after the TP is extinguished is not immediate, thus the luminosity undergoes a series
of periodic variations, with temporary dips in conjunction with the ignition
of TPs. The right panel of Fig.~\ref{flowl} shows the variation with time of the luminosity of stars of initial mass $1.5, 2, 2.5~M_{\odot}$, during the C-star phase. We see that at the ignition of each TP L drops below $3000~L_{\odot}$, then increases as CNO burning is gradually re-activated. The rise of the luminosity following each TP is sufficiently slow that the possibility of observing stars in such low-luminosity phases is statistically not negligible. This holds particularly for the lowest mass stars considered.

Therefore, we propose that the low-luminosity stars identified in the IRS sample are carbon stars that have recently experienced a TP, and are evolving through a phase when the H-burning shell has not fully recovered its efficiency. This interpretation is consistent with the small degree of obscuration which we derived for these stars, because during these evolutionary phases the stars assume a compact configuration, which prevents the formation of large quantities of dust.

\begin{figure*}
\begin{minipage}{0.32\textwidth}
\resizebox{1.\hsize}{!}{\includegraphics{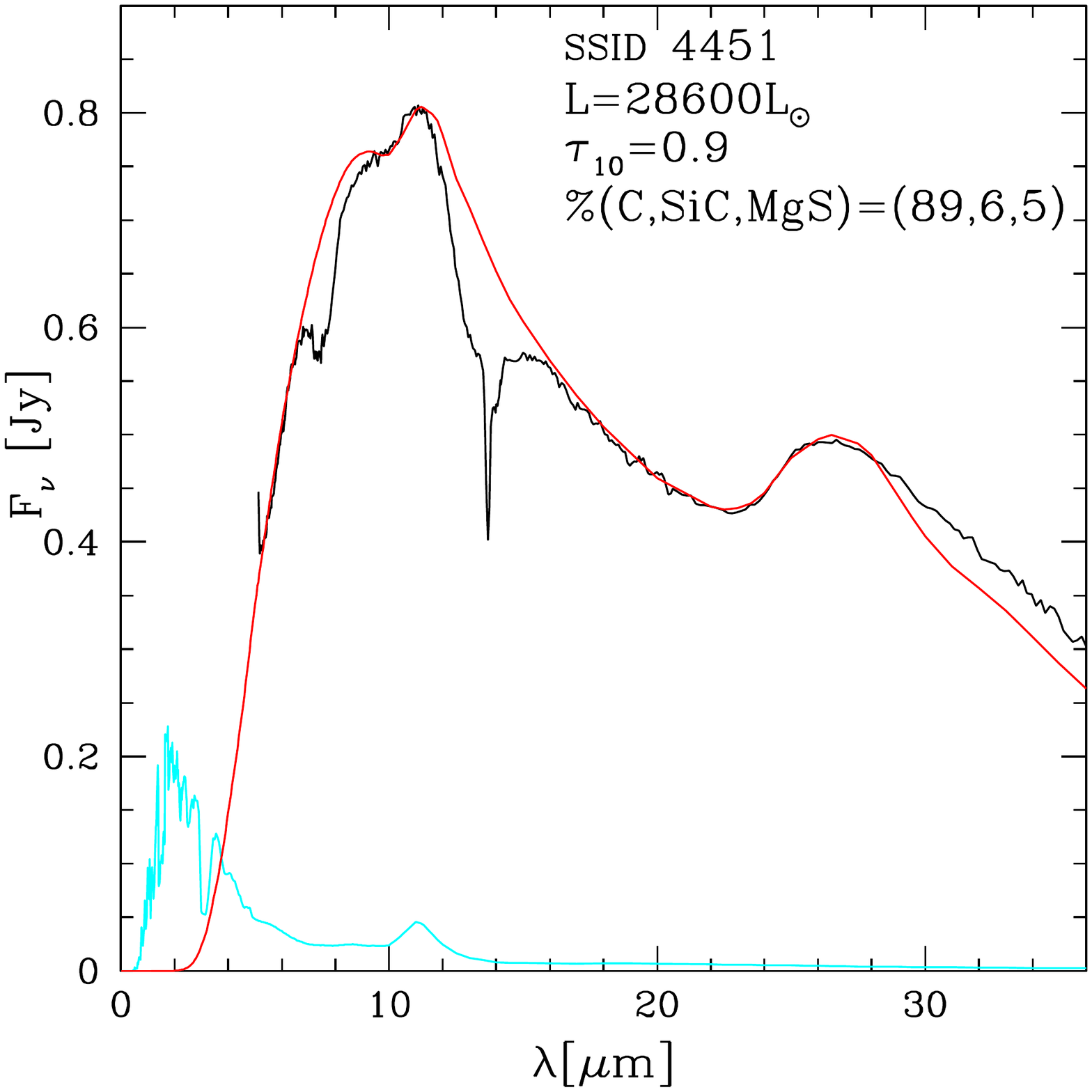}}
\end{minipage}
\begin{minipage}{0.32\textwidth}
\resizebox{1.\hsize}{!}{\includegraphics{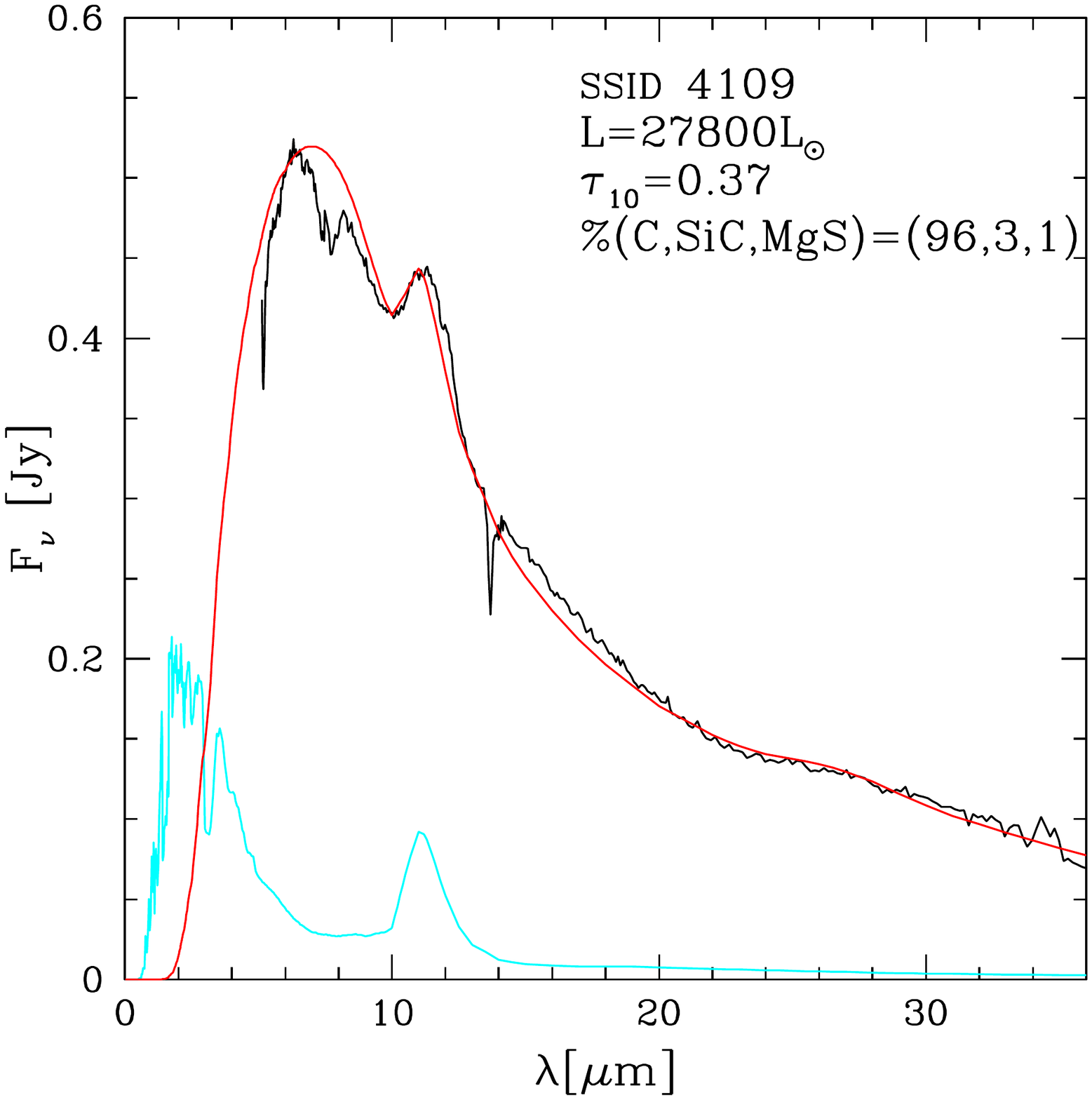}}
\end{minipage}
\begin{minipage}{0.32\textwidth}
\resizebox{1.\hsize}{!}{\includegraphics{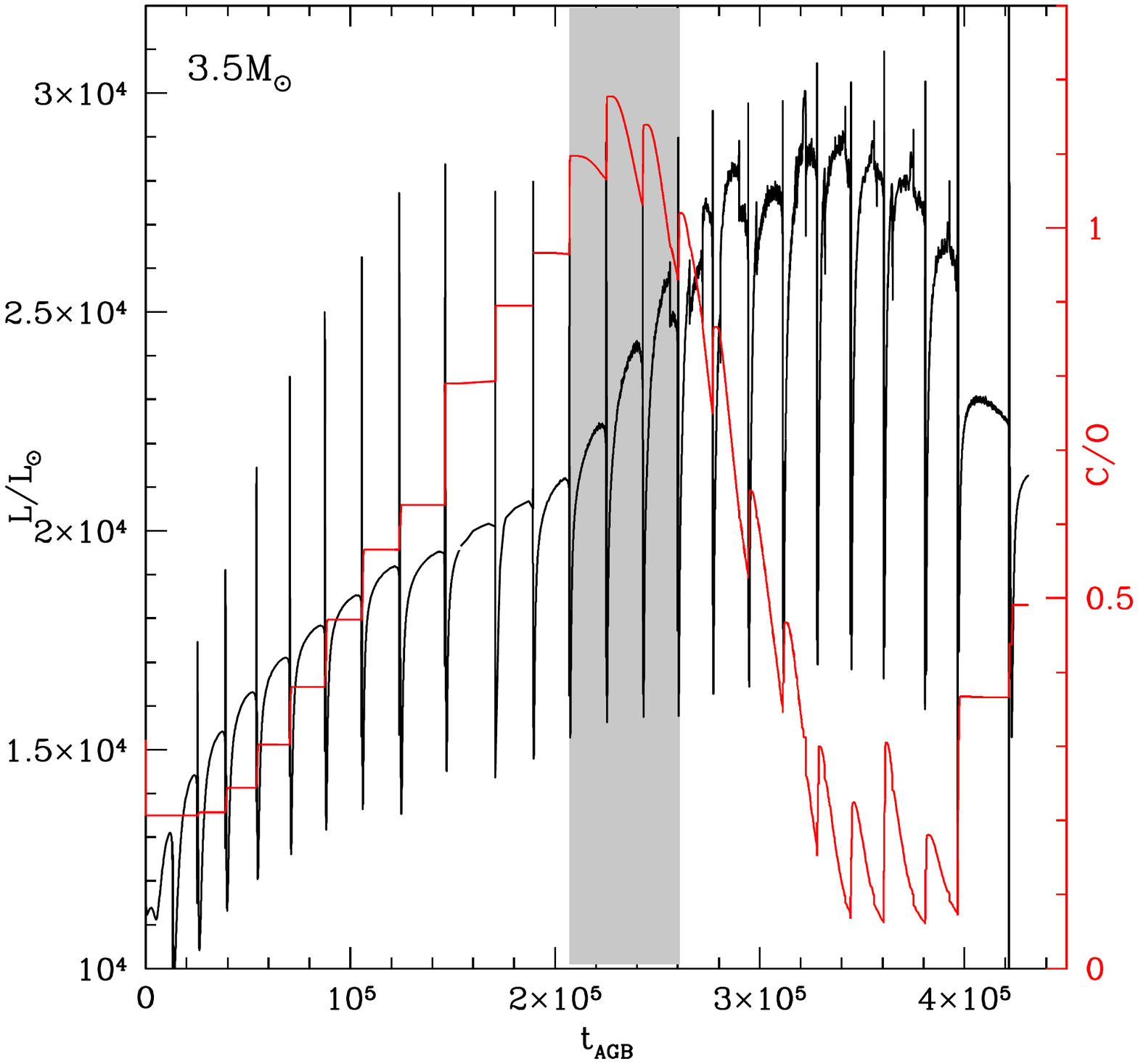}}
\end{minipage}
\vskip-30pt
\caption{IRS and best-fitting SED for the sources SSID 4451 (left panel) and
SSID 4109 (middle). Right: evolution with time (counted since the beginning of the TP-AGB phase) of luminosity (black) and surface C$/$O (red track, scale on the right) of a $3.5~M_{\odot}$ star. The grey-shaded area indicates the phase during which the star is carbon-rich.}
\label{fhbb}
\end{figure*}

\subsection{A signature of hot bottom burning?}
\label{hbb}
The sources indicated with magenta crosses in Fig.~\ref{ftaul} (SSID 4109, 4451, 4540, 4776) are characterised by luminosities above $20000~L_{\odot}$, significantly higher than the other stars in the sample and above the upper limit expected for carbon stars ($\sim 17000~L_{\odot}$), given in section \ref{allm}. These luminosities indicate that the cores of these stars are more massive than their counterparts discussed in section \ref{allm}, which translates into higher mass progenitors. 

The mass threshold given in section \ref{allm} was based on the fact that higher mass stars experience HBB, which destroys the surface carbon, thus preventing the stars from becoming carbon stars. Indeed stars within a narrow mass range, clustering around $3.5~M_{\odot}$, experience a series of TDU events before HBB is activated. These stars reach the C-star stage and evolve as carbon stars for a few TPs, until the beginning of HBB. 

An example of this behaviour is shown in the right panel of Fig.~\ref{fhbb}, reporting the evolution of a $3.5~M_{\odot}$ model, in terms of the surface C$/$O and of the luminosity. The effects of TDU can be seen in the rise of C$/$O after each TP, whereas HBB causes a drop in C$/$O after each TDU event. In this specific case HBB starts after 12 TPs and the stars evolves as C-star during 4 inter-pulse phases, before the surface C$/$O drops below unity.
The ignition of HBB is accompanied by the fast rise in the luminosity of the star, which increases from $\sim 17000~L_{\odot}$ to above $20000~L_{\odot}$. During the C-star phase we find $L \sim 25000~L_{\odot}$, in agreement with the luminosities derived for SSID 4109 and 4451.

\begin{figure*}
\begin{minipage}{0.48\textwidth}
\resizebox{1.\hsize}{!}{\includegraphics{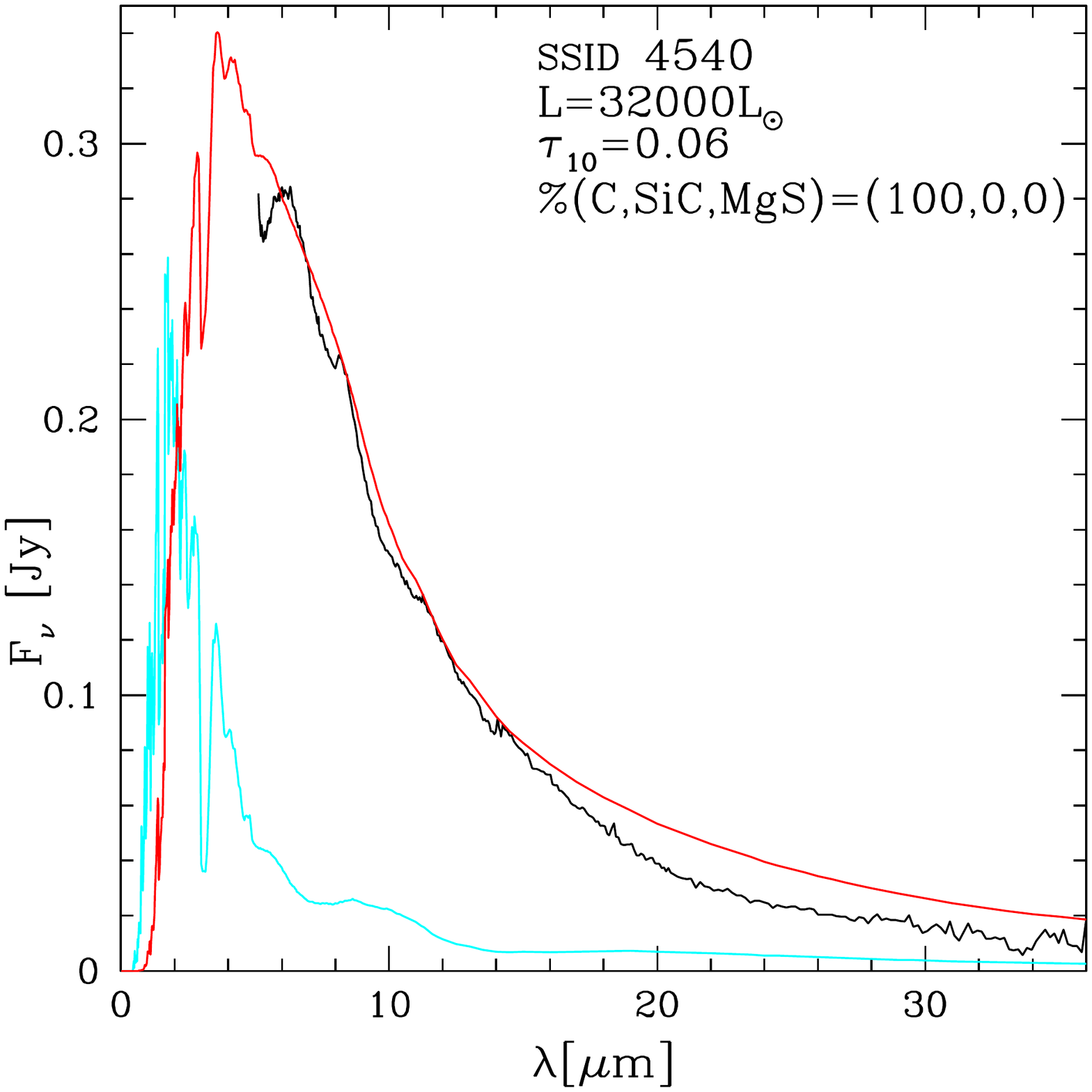}}
\end{minipage}
\begin{minipage}{0.48\textwidth}
\resizebox{1.\hsize}{!}{\includegraphics{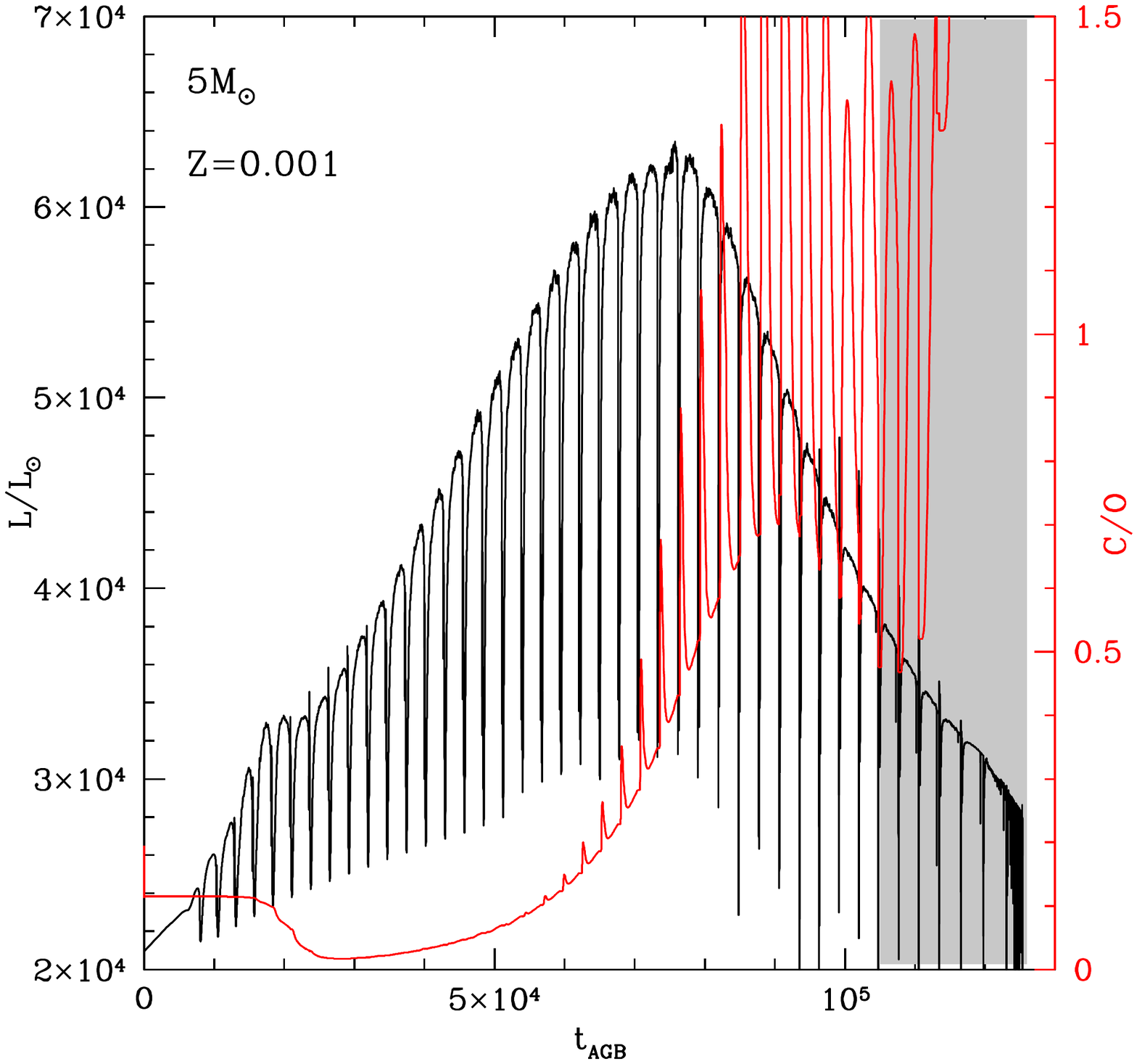}}
\end{minipage}
\vskip-40pt
\caption{Left: IRS SED of the source SSID 4540, compared with our interpretation,
based on synthetic modelling. Right: AGB evolution with time of the luminosity (black line, scale on the left) and of the surface C$/$O ratio (red line, scale on the right) of a $5~M_{\odot}$ model with metallicity $Z=0.001$. The grey-shaded area indicates the evolutionary phase during which
the star evolves as a carbon star.}
\label{fhbb2}
\end{figure*}

We propose that these two sources descend from $3.5-4~M{\odot}$ progenitors, 
formed $\sim 250$ Myr ago, that have reached the C-star phase and that are currently experiencing HBB at the base of the envelope. According to this interpretation, they are evolving through the C-star phase, before becoming again oxygen-rich. If this understanding is correct, these stars are characterised by
a surface C$/$O slightly above unity and they are enriched in lithium, because the time elapsed since the start of HBB is not sufficient to destroy the surface $^3$He, which allows the \citet{cameron} mechanism to operate. 
These sources would be examples of the lithium-rich, carbon stars, which  
\citet{ventura99} proposed as independent distance indicators, based on the robust estimate of the luminosities at which we expect this mechanism to occur. 
We rule out the possibility that the progenitors are higher mass stars formed more recently, because stars of initial mass above $4~M_{\odot}$ experience HBB since the early AGB phases, thus the formation of C-stars is inhibited.

The scenario invoked to explain the IRS spectrum of SSID 4109 and 4451 could potentially work also for SSID 4776. However, the SED of the latter star is peculiar, since we find no evidence of the $11.3~\mu$m SiC feature and of the $30~\mu$m bump: this suggests that the dust is composed exclusively of solid carbon. The situation regarding SSID 4540, whose SED is shown in the left panel of Fig.~\ref{fhbb2}, is even more tricky, because the estimated luminosity,
$32000~L_{\odot}$, is higher than the largest luminosity attained by the
$3.5~M_{\odot}$ model, shown in the right panel of Fig.~\ref{fhbb}.
The large luminosity and the SiC- and MgS-free dust mineralogy lead us to consider the alternative possibility that the progenitors of SSID 4540 and SSID 4776 are metal-poor, massive AGB stars. These stars experience for most of the AGB lifetime vigorous HBB at the base of the external mantle, which causes the depletion of the surface carbon and, unlike their higher metallicity counterparts, a strong reduction of the surface oxygen. During the final AGB phases, when HBB is turned off by the loss of the external envelope, the very low amount of oxygen residual in the external regions allows a few TDU episodes to turn the stars into a C-star. This is shown in the right panel of Fig.~\ref{fhbb2}, reporting the evolution of the luminosity and of the surface C$/$O of a $5~M_{\odot}$, $Z=0.001$ model. The luminosities at which this transition occurs are in the $25000-35000~L_{\odot}$ range, consistent with the estimate done for SSID 4540 and SSID 4776. 

If this is interpretation is correct, these stars share the same origin with the O-rich objects in the LMC, that \citet{ester19} identified as metal-poor stars, in which the significant destruction of the surface oxygen, caused by HBB, makes the dust in their surroundings to be dominated by solid iron grains. According to this hypothesis SSID 4540 and SSID 4776 are currently evolving through the stages immediately following those of the stars studied by \citet{ester19}, after becoming a C-star. They would be the youngest object in the sample studied here, with an age of $\sim 100$ Myr. An important difference with respect to the origin invoked for SSID 4109 and 4451 is the surface chemical composition, which should exhibit an extremely poor oxygen content and no evidence of lithium.

These stars are not expected to produce great quantities of dust, because the carbon excess with respect to oxygen is low, given the effects of HBB experienced during the earlier evolution. This is consistent with the small optical depths, $\tau_{10} = 0.06-0.08$, deduced from the analysis of the SED, and with the shape of the IRS spectrum (see left panel of Fig.~\ref{fhbb2}), which shows no significant IR emission.

\begin{figure*}
\begin{minipage}{0.48\textwidth}
\resizebox{1.\hsize}{!}{\includegraphics{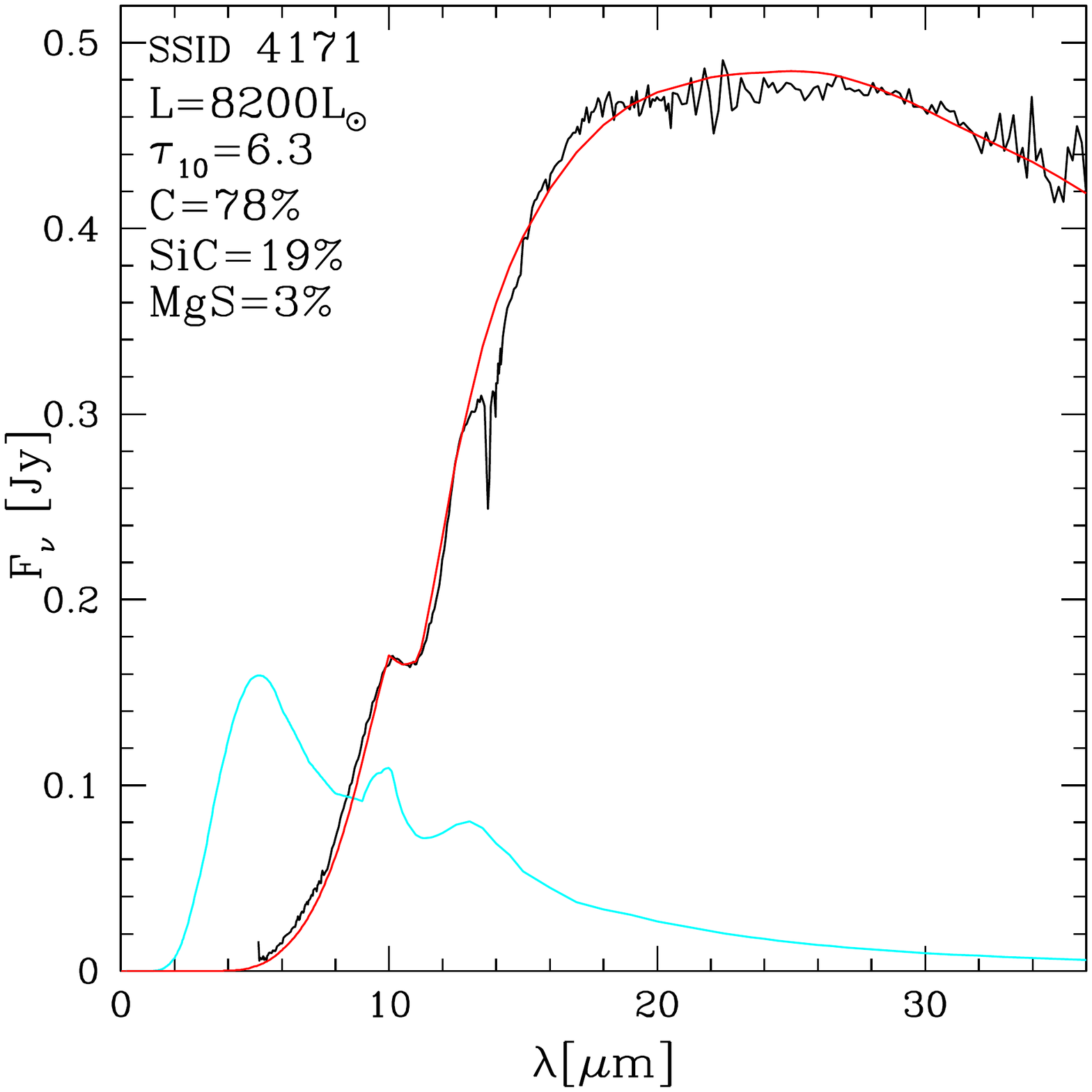}}
\end{minipage}
\begin{minipage}{0.48\textwidth}
\resizebox{1.\hsize}{!}{\includegraphics{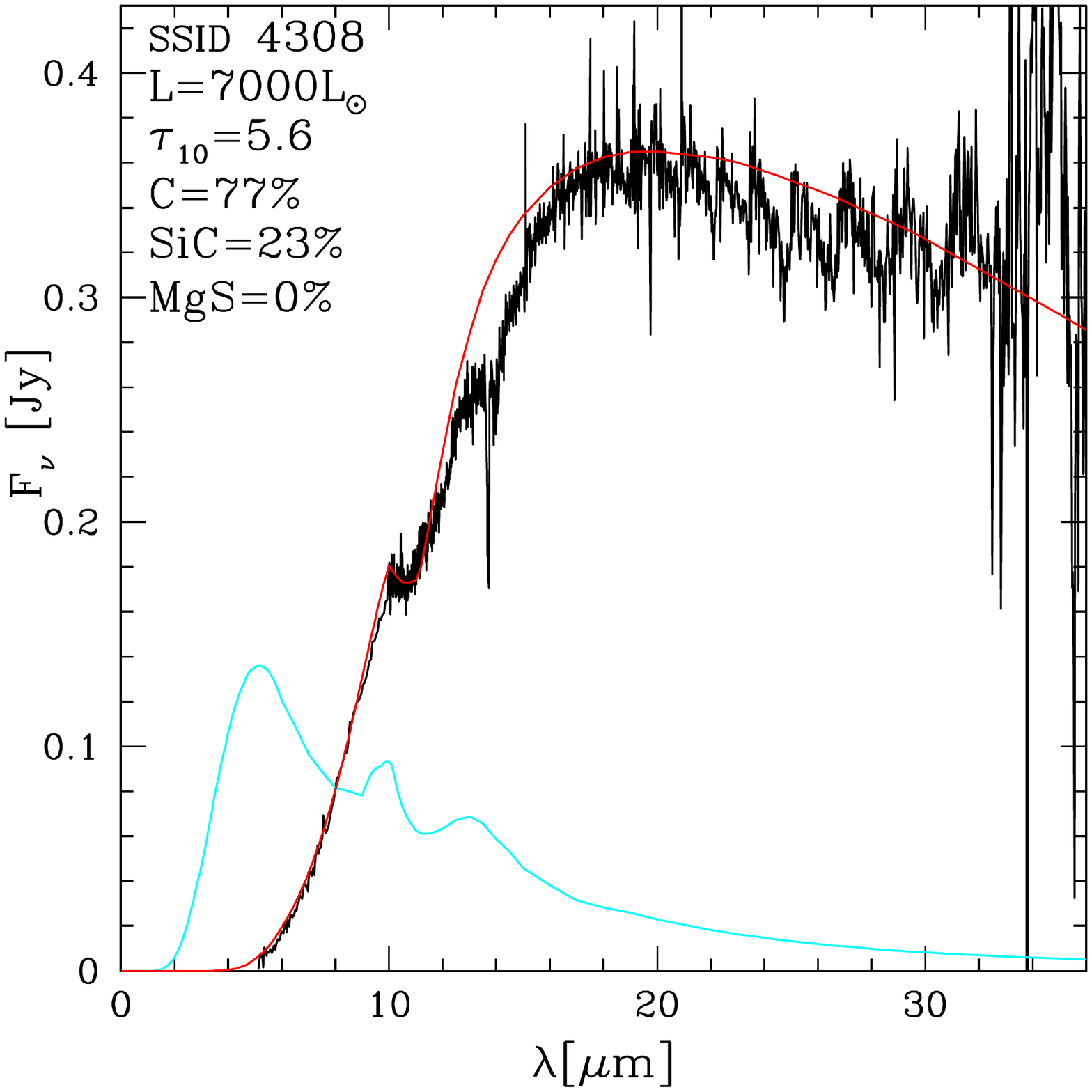}}
\end{minipage}
\vskip-70pt
\begin{minipage}{0.48\textwidth}
\resizebox{1.\hsize}{!}{\includegraphics{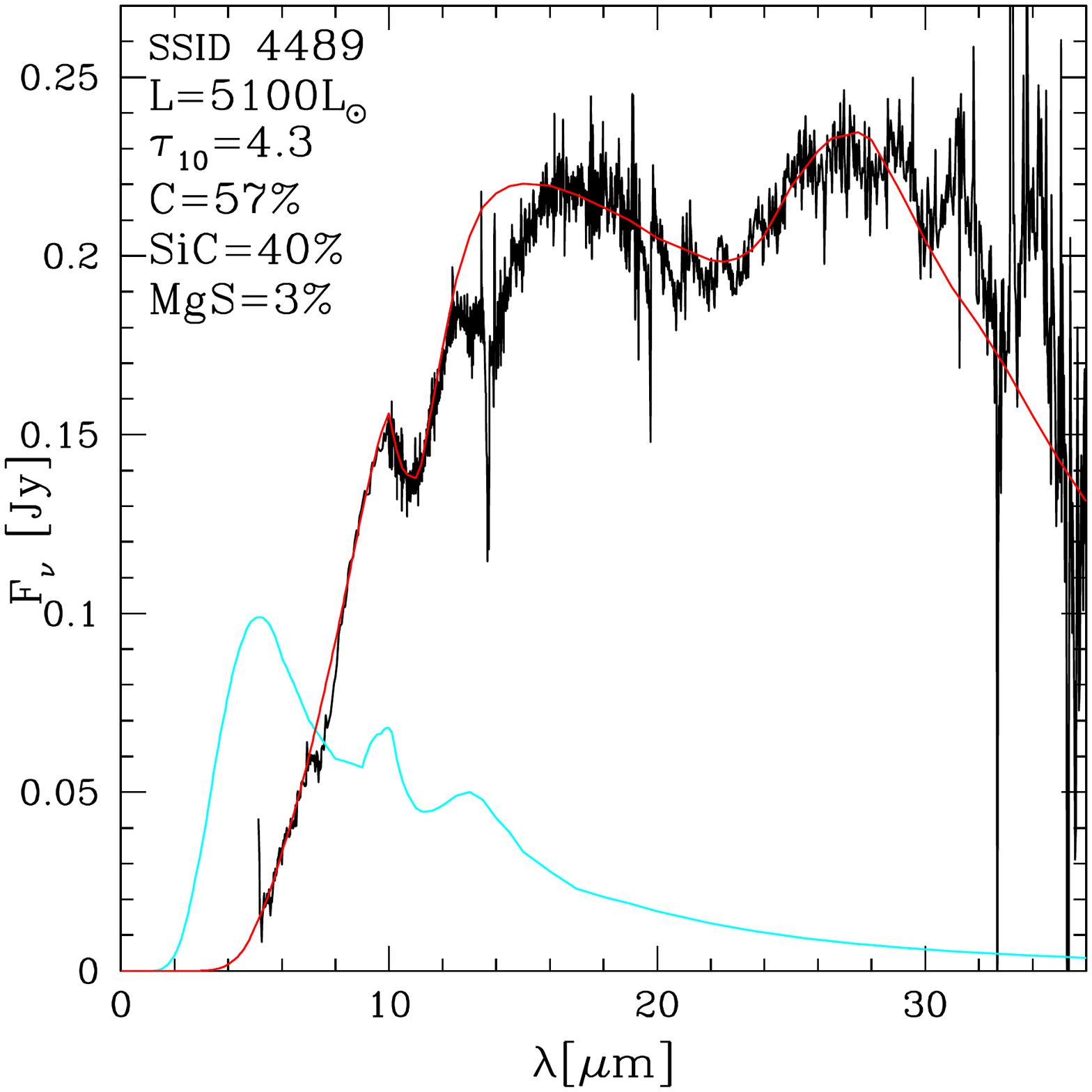}}
\end{minipage}
\begin{minipage}{0.48\textwidth}
\resizebox{1.\hsize}{!}{\includegraphics{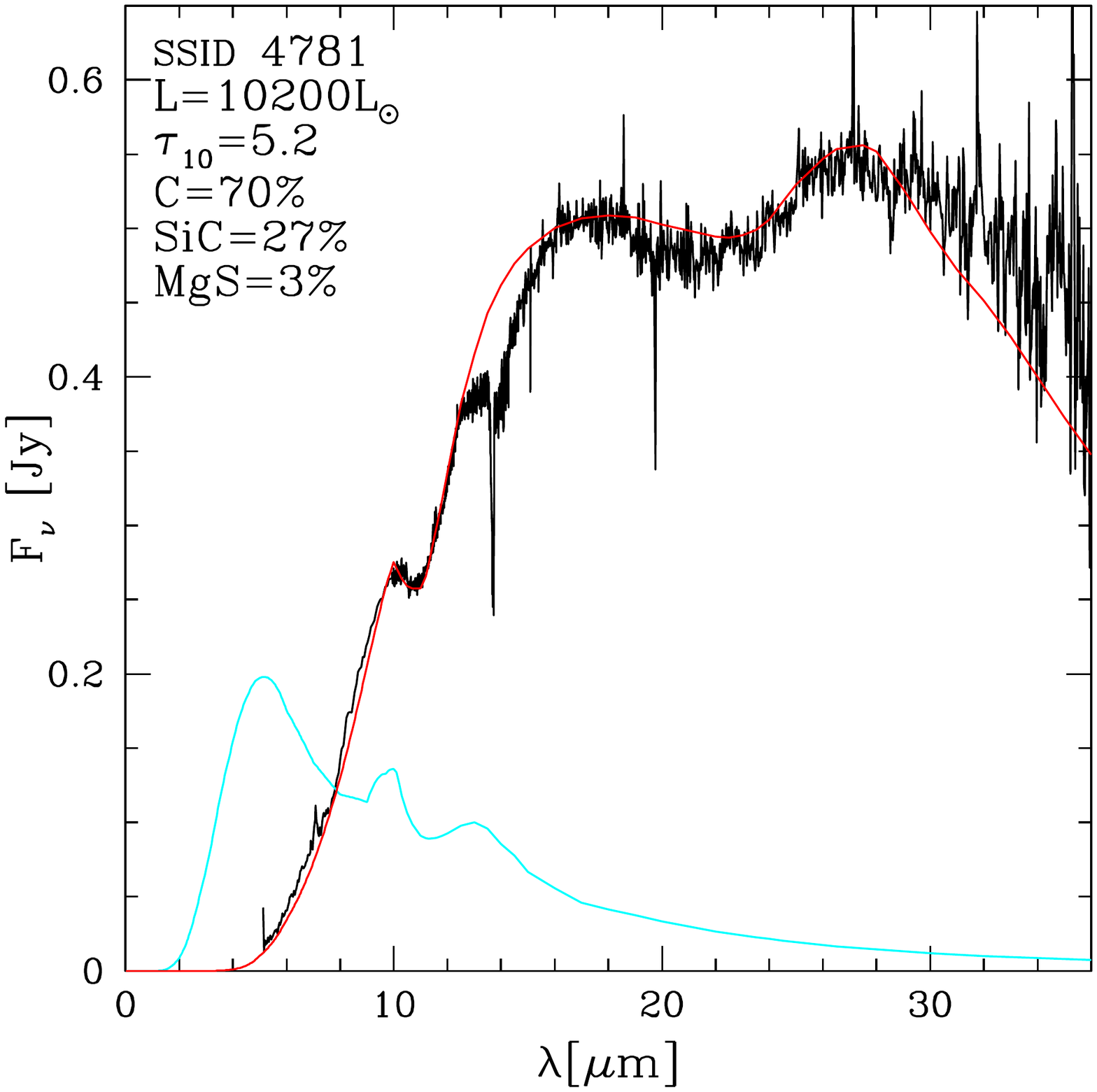}}
\end{minipage}
\vskip-40pt
\caption{SED fitting of 4 out of the extremely red objects studied
by \citet{gruendl08}}
\label{fextr}
\end{figure*}

\subsection{Extreme carbon stars}
\label{ero}
The existence of LMC stars with extremely red mid-IR colours was first
discussed by \citet{gruendl08}, who used IRAC and MIPS photometry and IRS follow-up to identify 7 objects as extreme carbon stars. \citet{gruendl08} underlined the peculiarities of these sources, especially the derived mass loss rates, which the authors claim to span the $(5\times 10^{-5}-2\times 10^{-4})~M_{\odot}/$yr range, significantly higher than those deduced for carbon stars, when radiative transfer modelling is used to reproduce their
photometric properties \citep{jacco99}. 

\begin{figure}
\resizebox{\hsize}{!}{\includegraphics{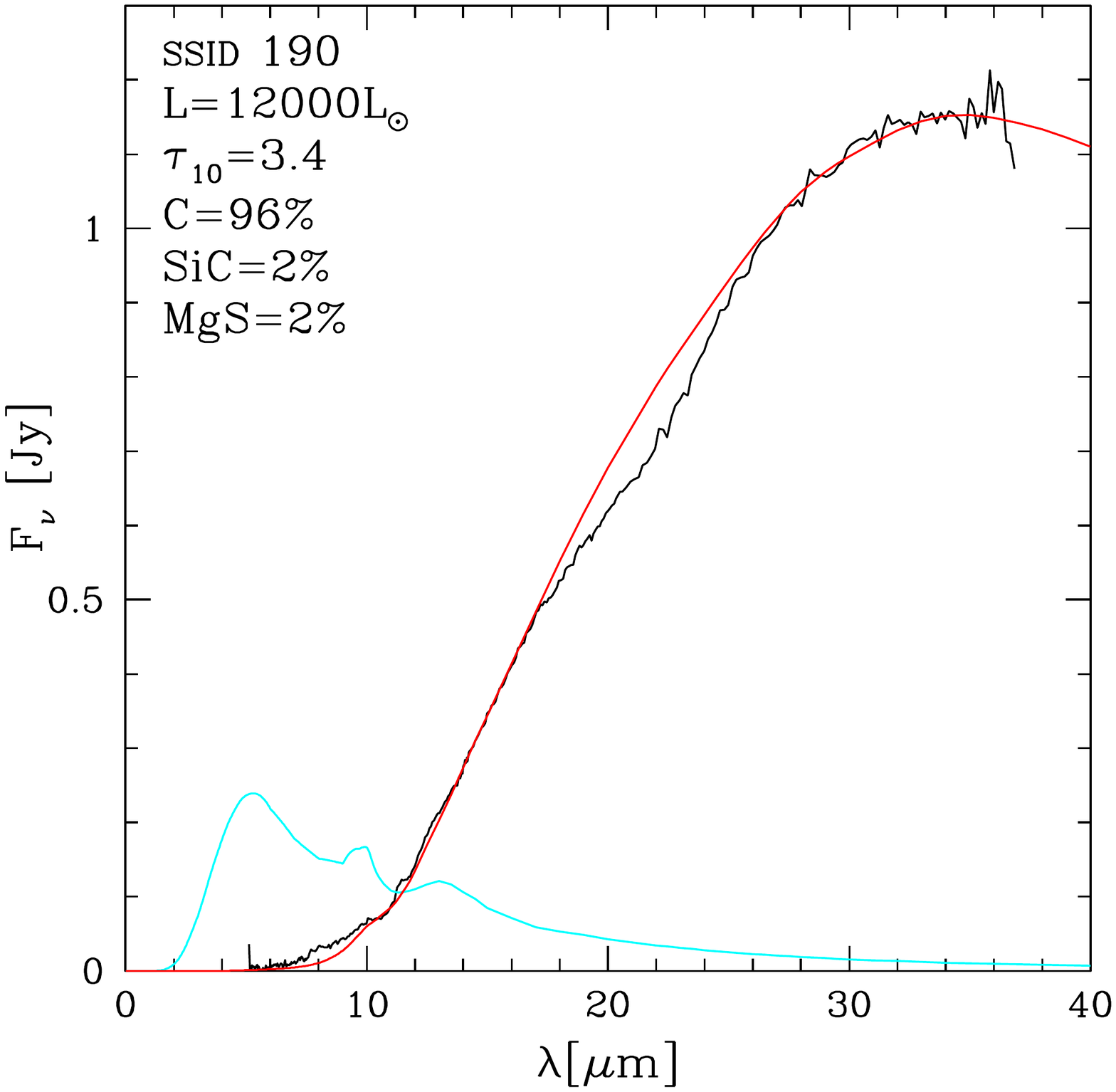}}
\vskip-60pt
\caption{The IRS (black) and synthetic (red) spectra of the source SSID 190.}
\label{fpagb}
\end{figure}

In Fig.~\ref{ftaul} the stars discussed by \citet{gruendl08}, on the right side of the $\tau_{10}-L$ plane, are indicated with blue diamonds. We use the same symbol to identify SSID 9, which we also consider extreme, as it lies on the same region of the diagram populated by the \citet{gruendl08} sources. Fig.~\ref{fextr} shows the IRS spectra of 4 out of these stars, with the parameters required to reproduce the details of the observed SED. In all cases, with the only exception of SSID 4299, a clear SiC absorption feature is visible, indicating an extremely thick dusty region. Note that the interpretation of the spectra of these objects is not affected by the presence of the molecular bands discussed in the first part of this section. 
The sources SSID 125 and SSID 190, belonging to the \citet{gruendl08} sample,
are indicated in Fig.~\ref{ftaul} with open, blue diamonds; these stars are extremely red and show up a peculiar SED, peaking at wavelengths $\lambda > 20~\mu$m. An example is shown in Fig.~\ref{fpagb}.

\begin{figure*}
\begin{minipage}{0.32\textwidth}
\resizebox{1.\hsize}{!}{\includegraphics{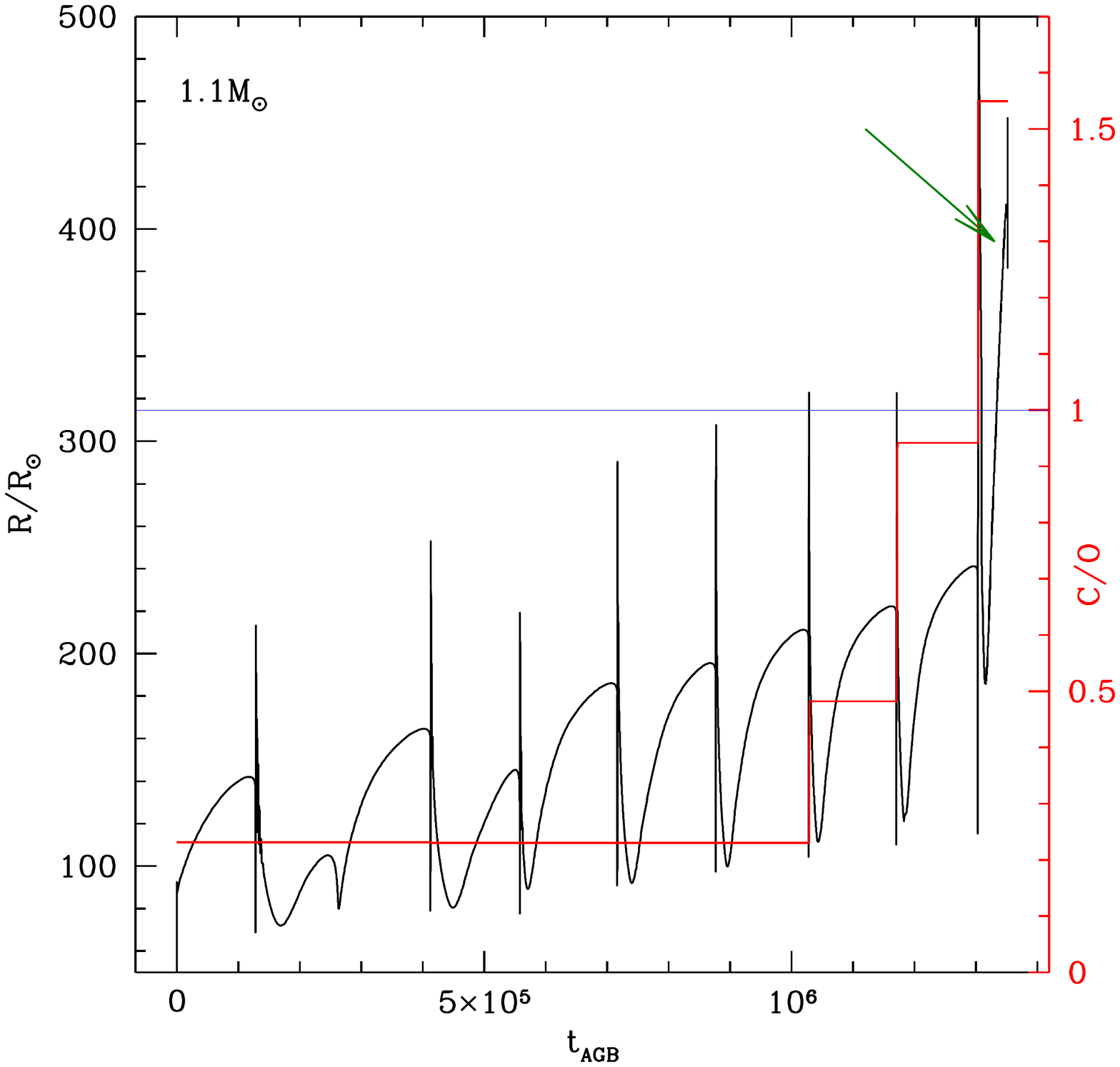}}
\end{minipage}
\begin{minipage}{0.32\textwidth}
\resizebox{1.\hsize}{!}{\includegraphics{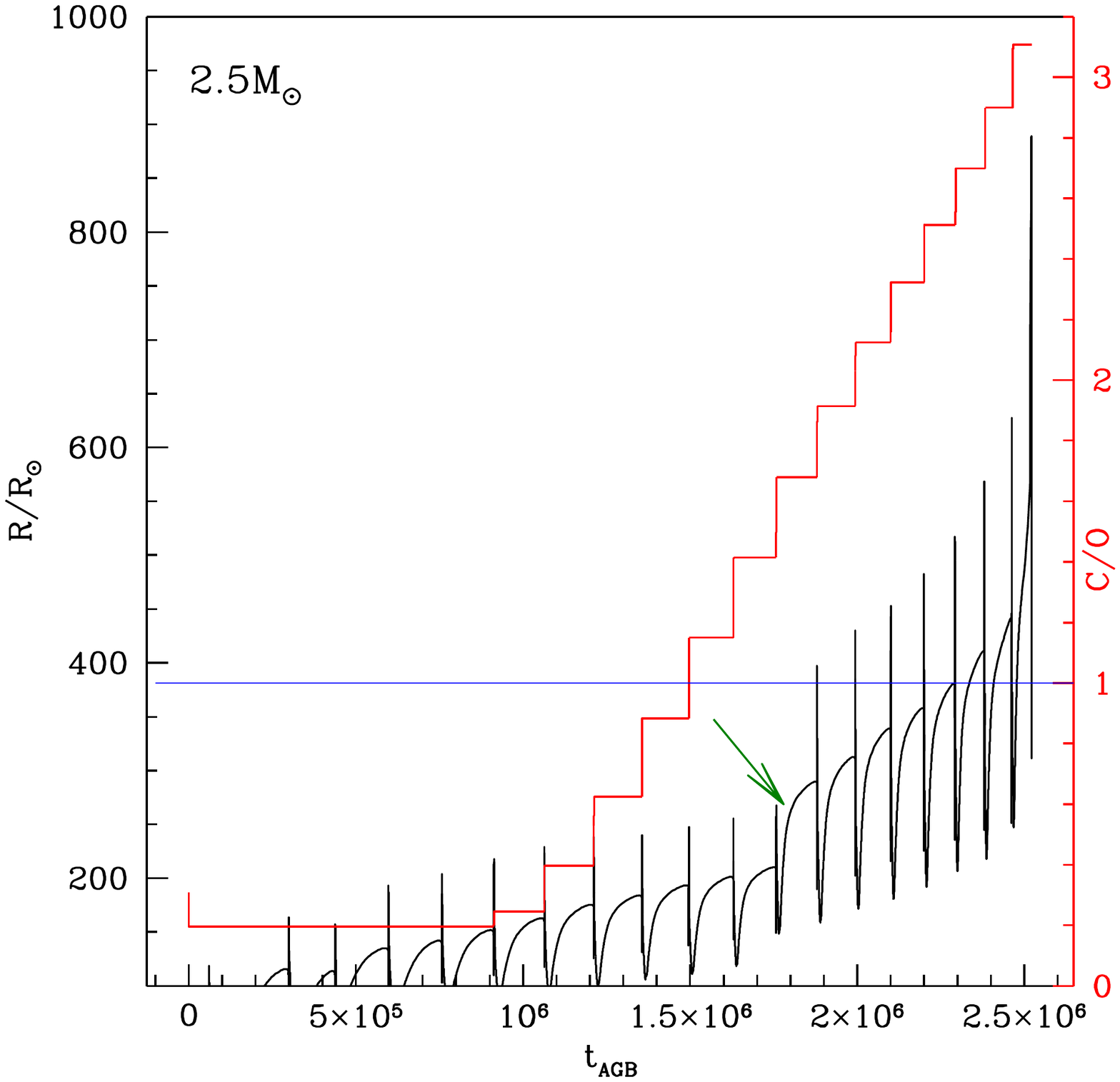}}
\end{minipage}
\begin{minipage}{0.32\textwidth}
\resizebox{1.\hsize}{!}{\includegraphics{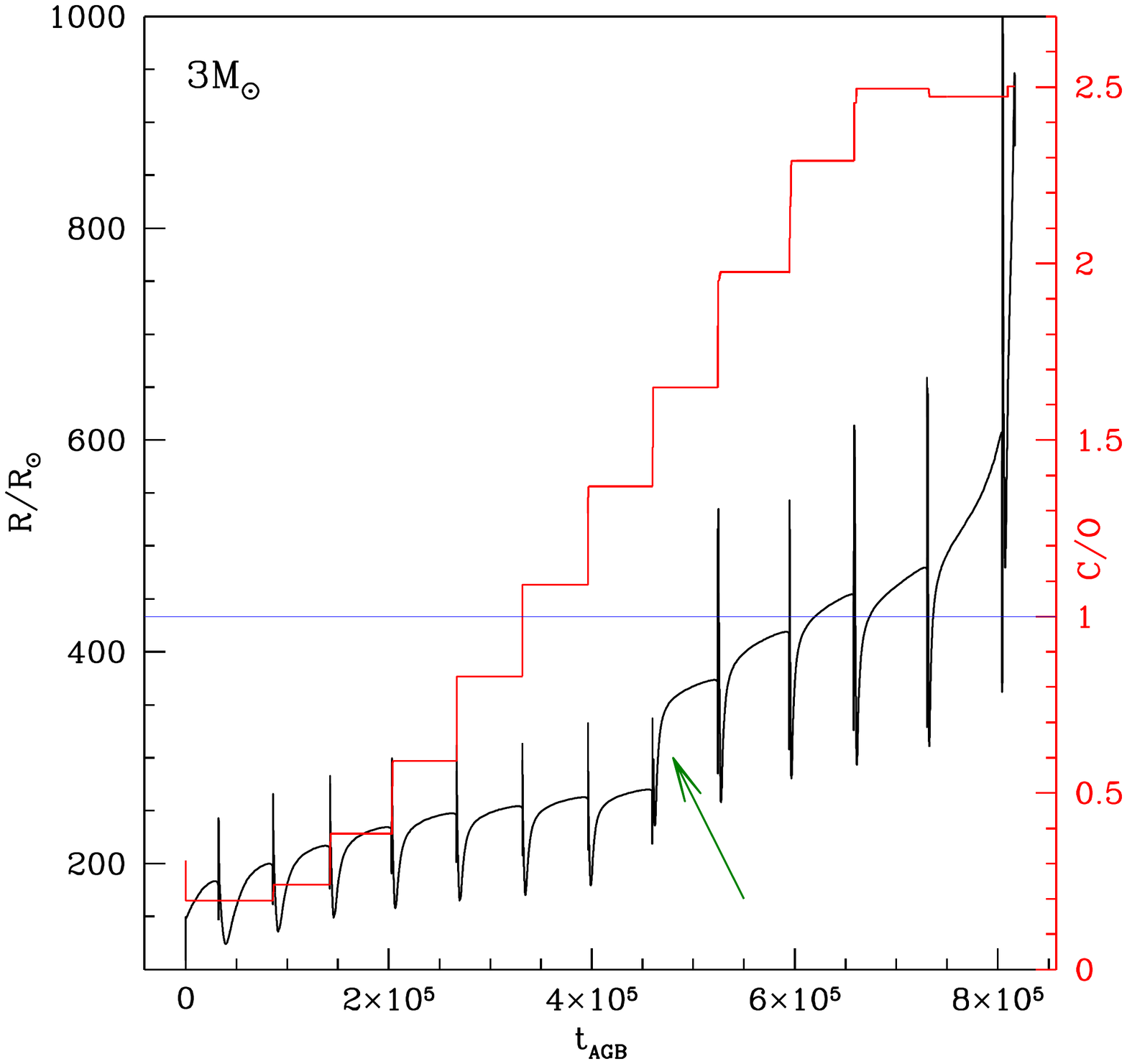}}
\end{minipage}
\vskip-30pt
\caption{Evolution of the radius (black lines, scale on the left) and
surface C$/$O ratio (red, scale on the right) during the AGB phase of stars
of initial mass $1.1~M_{\odot}$ (left panel), $2.5~M_{\odot}$ (middle) and
$3~M_{\odot}$ (right). The thin, blue horizontal lines indicate the C=O
condition, whereas the green arrows indicate the phase of rapid expansion of 
the star, as a consequence of the surface carbon enrichment.
}
\label{fradius}
\end{figure*}

The possible nature of the extreme stars by \citet{gruendl08} was discussed by \citet{sloan16} and \citet{martin18}, who stressed the difficulty in fitting the data of these stars, particularly in the context of modelling dust formation in radial outflows, departing from the surface of single carbon stars. An interesting overview of the critical issues to face when modelling stars with a very large infrared excess is found in \citet{nanni19b}, who explored the role of the mass loss rate, carbon excess, optical constants of solid carbon dust.

We confirm here that the interpretation of these objects, based on the standard
modelling so far used to describe the other stars analysed, is extremely hard.
The results presented in Fig.~\ref{ftaul} and \ref{fextr} outline that the optical depths required to reproduce the observations, which span the 
$3 < \tau_{10} < 7$ range, are significantly higher than the majority of
the largest $\tau_{10}$ expected on the basis of theoretical modelling, reported in table \ref{tabmod}. Indeed the synthetic SEDs shown in the 4 panels of Fig.~\ref{fextr} and the parameters corresponding to these stars reported in table \ref{tabsample} were obtained by artificially increasing the optical depth of the most obscured model, taken among the points of the evolutionary sequences used to characterise all the other stars in the sample.

The top-right panel of Fig.~\ref{fallm} shows that optical depths $\tau_{10} \sim 5$ are reached during the latest AGB phases of $2-3~M_{\odot}$ stars. These largely obscured models might partly explain, as far as the degree of obscuration is concerned, the SSID 4299 and SSID 4781 data (see bottom-right panel of Fig.~\ref{fextr}), the brightest sources in this group, for which we estimate a luminosity around $10000~L_{\odot}$. However, there is no way to account for the observations of the fainter objects, particularly those with estimated luminosities in the $L<6000~L_{\odot}$ range: these energy fluxes indicate $M \sim 1.1-1.5~M_{\odot}$ progenitors, which are not expected to evolve to such extreme $\tau_{10}$, owing to the relatively low carbon excess reached (see Fig.~\ref{fallm}).

We cannot rule out that the stationary wind model used in the present investigation to model dust formation might underestimate the amount of dust formed in the wind of AGB stars. As discussed in section \ref{uncert}, the present model neglects the effects of pulsations and the pulsation-induced shocks, which might drive dense gas clouds into external regions of the outflow, where dust condensation would be favoured by the cool temperatures. Furthermore, here we do not account for gas-to-dust drift, which might affect the structure of the wind and cause higher dust yields. \citet{sandin20} have recently performed unique state-of-the-art simulations of C-star wind formation, confirming that drift is significant at low mass loss rates and becomes less and less important as the rate goes up. This means we cannot firmly conclude that the dust yields should be significantly higher unless we redo the stellar evolution modeling including a correction for drift. The effect on the integrated DPR depends on the evolution of the mass loss rate, which in turn depends on how the stellar parameters and carbon excess evolve. A further critical point related
to the analysis presented here, discussed in section \ref{uncert}, is that the mass-loss rate, described according to \citet{wachter02}, is independent of the carbon excess.

Despite the above uncertainties, we believe that there is no room to obtain $\tau_{10}$ values significantly higher than those reported in table \ref{tabmod}. This is definitively impossible in case of $M < 2~M_{\odot}$ stars, for the reasons given above. For what attains the stars of mass above $2~M_{\odot}$, the large values of $\tau_{10}$ reached are mainly caused by the high mass loss rates experienced during the latest AGB phases, which are slightly below $\sim 2\times 10^{-4}~M_{\odot}/$yr. We consider these rates as upper limits, because their calculation is based on the description by \citet{wachter02, wachter08}, which might overestimate the true mass loss rates \citep{bladh19}. 

A possible reason for the discrepancy between the $ \tau_{10}$ obtained by
modelling and the degree of obscuration derived from the analysis of the
spectra is that the winds are significantly denser than predicted by the theoretical modelling, suggesting that dust is formed at higher rates than predicted by stellar evolution calculations. We reconsidered the C-star phase of the models presented in section \ref{cstars}, and artificially increased the mass loss rate to be used in the modelling of dust formation, until obtaining the values of $\tau_{10}$ required to fit the IRS spectra. We found consistency with mass loss rates of the order of $\dot M \sim 5\times 10^{-4}~M_{\odot}/$yr. It goes without saying that these calculations are based on the assumption that the outflow is radially symmetric, which might be not the case for these peculiar objects; yet we consider this exercise as a tool to infer a reliable estimate of the order of magnitude of the mass loss rate of the star, required to produce wind densities compatible with the observations. 

\citet{sloan16} and \citet{martin18} hinted at the possibility that the extreme
carbon stars are part of binary systems. \citet{flavia20} proposed that
these carbon stars, with a low-mass companion, have recently filled the Roche lobe. Fig.~\ref{fradius} shows the AGB variation with time of the radius and the surface C$/$O of stars of different initial mass. We see that in all cases the stars undergo a phase of rapid expansion, soon after they become carbon stars. The fast increase in the stellar radius enhances the probability that the Roche lobe is overfilled, giving the start to a phase of strong mass loss, which we artificially impose starting from the corresponding evolutionary points, indicated with green arrows in Fig.~\ref{fradius}. As previously discussed, use of $\dot M = 5\times 10^{-4}~M_{\odot}/$yr leads to efficient formation of dust, in quantities sufficiently large to reconcile the optical depth of the envelope with the degree of obscuration indicated by the observations. As shown in Fig.~\ref{ftaul}, the results obtained, indicated with magenta stars, are in substantial agreement with the $\tau_{10}$ and luminosity values of the extreme stars derived from the analysis of the IRS spectra. For SSID 125 and SSID 190, Dell'Agli et al. (2020) proposed a slightly different
interpretation, based on the cool dust temperature required to fit the
IRS spectra, of the order of $\sim 350$ K, which might indicate that the
dust layer is moving away from the system; this is also consistent with the
derived optical depths, $\tau_{10} \sim 2-3$, among the lowest in the extreme stars sub-sample, which might be due to the decrease in the gas densities, caused by the expansion. These findings possibly suggest that these systems have recently moved away from the AGB and have started the post-AGB phase; this understanding would be consistent with the interpretation proposed by \citet{martin18}.

In all cases, the interpretation based on the fact that these stars are evolving through, or have recently crossed a common envelope phase is consistent with the relative small fraction of solid carbon dust, in most cases below $80\%$, because the fast loss of the external envelope prevents the accumulation of large quantities of carbon in the surface regions.

If this understanding is correct, these stars would produce dust at rates
in excess of $10^{-6}~M_{\odot}/$yr, which would make them extraordinarily efficient dust manufacturers. The entire DPR of galaxies would be dominated by a paucity of these systems.

\section{The winds of carbon stars: what we learn from IRS spectra?}
\label{spectra}
We discussed in section \ref{lmcc} that the luminosity and the optical depth of the individual sources can be safely derived by comparing the observed spectrum with the synthetic SED, calculated on the basis of a basic wind model, which assumes that the dust is composed only of carbon and SiC.

On the other hand we know that other dust species form in the winds of carbon stars; furthermore, it is possible that not all the solid carbon forms in the amorphous state, but that a fraction of it might be in other aggregates, such as graphite. The comprehension of the details of the dust formation process demands the knowledge of the mineralogy of the dust formed, which consists into the determination of the abundances of the different species considered, or, equivalently, the individual contributions to the overall cross-section. This ambitious task requires the interpretation of the whole SED, including the details of the features and the slopes of the stellar spectra, in the whole wavelength interval covered by IRS. 

\begin{figure}
 \resizebox{\hsize}{!}{\includegraphics{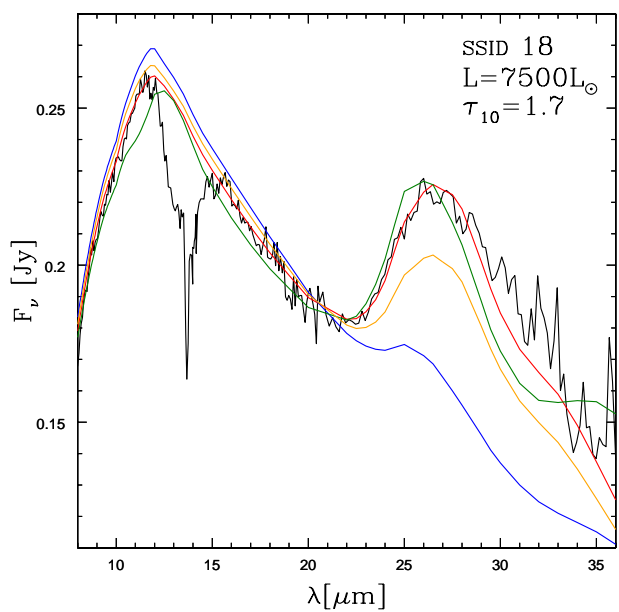}}
\vskip-60pt
\caption{IRS spectrum of SSID 18 (black line) compared with synthetic SEDs, corresponding to the same luminosity and $\tau_{10}$, and different percentages of SiC+MgS and of the relative width of the MgS mantle ($a_{man}$) on top of SiC core. The various lines correspond to the following: red - our best-fit model, with SiC+MgS=9\% and $a_{man}=0.05~\mu m$; green - SiC+MgS=35\% and $a_{man}=0.02~\mu m$; blue - SiC+MgS=9\% and $a_{man}=0.02~\mu m$; orange - SiC+MgS=6\% and $a_{man}=0.05~\mu m$.
}
\label{fmgs18}
\end{figure}

This step is important for two reasons: a) the derivation of the details of the
composition of dust will provide precious information on the structure of the winds of carbon stars and on the efficiency of the formation process of the various dust species; b) the interpretation of the observations from {\itshape JWST} requires the knowledge of the dust species in the circumstellar envelope, since most of the mid-IR {\itshape JWST} filters cover spectral regions partly or fully overlapped with the afore mentioned features (see bottom panels of Fig.~\ref{fdust20}).

The search for the parameters allowing the best fit of the SED of the individual sources was done by eye for each star. We did not use any automatic procedure, despite being less time consuming, because in the present context the priority is the correct reproduction of the whole SED. We now discuss in detail the dust species other than solid carbon, and how their presence reflects onto the SED.

\subsection{Silicon carbide}
SiC is by far the second most abundant species beyond carbon. The presence of SiC dust was used by several authors \citep{martin07, martin09, martin18, srinivasan09, srinivasan10} to reproduce the spectra of carbon stars, based on the observation of the well distinguished SiC feature, centered at $11.3 ~ \mu$m. This can be seen in the synthetic SED reported in the bottom panels of Fig.~\ref{fdust20} and in the IRS spectra shown in Fig.~\ref{fsedex}. As discussed in section \ref{2msun}, SiC is the most stable species: therefore, to build the synthetic SED we assumed that the radiation from the photosphere of the star is first reprocessed by an internal SiC layer, then by a more external dusty shell, where SiC grains cohabit with carbon particles and other species. While we may safely assume that SiC grains are the only solid particles formed in the internal part of the outflow, in the more external regions of the circumstellar envelope the presence of SiC could be associated with the formation of MgS \citep{svetlana08}. Indeed, as it will
be discussed in the following section, SiC is suited as a substrate for MgS precipitation, due to the obvious similarities of structure and bonding properties of SiC and MgS.

\begin{figure*}
\begin{minipage}{0.48\textwidth}
\resizebox{1.\hsize}{!}{\includegraphics{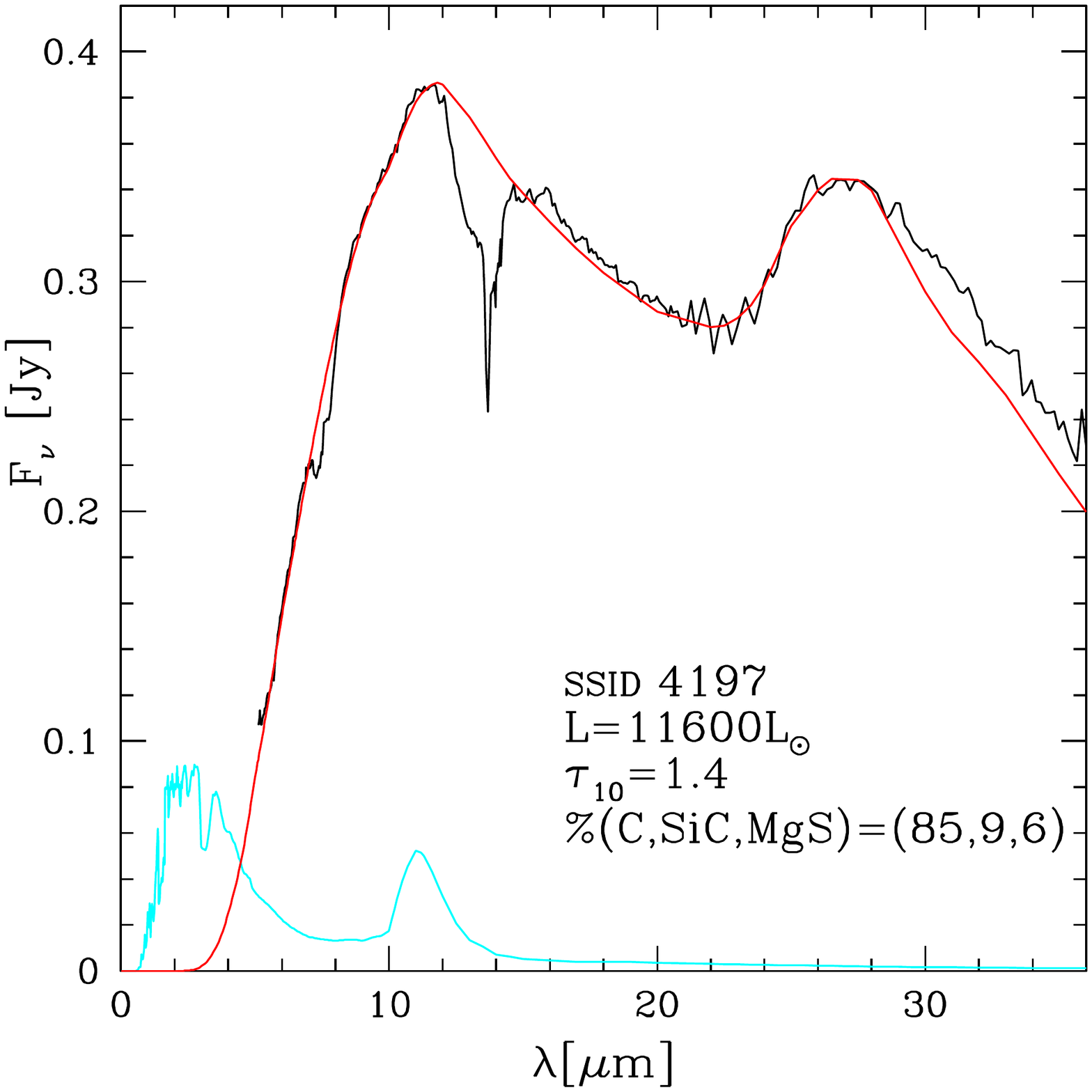}}
\end{minipage}
\begin{minipage}{0.48\textwidth}
\resizebox{1.\hsize}{!}{\includegraphics{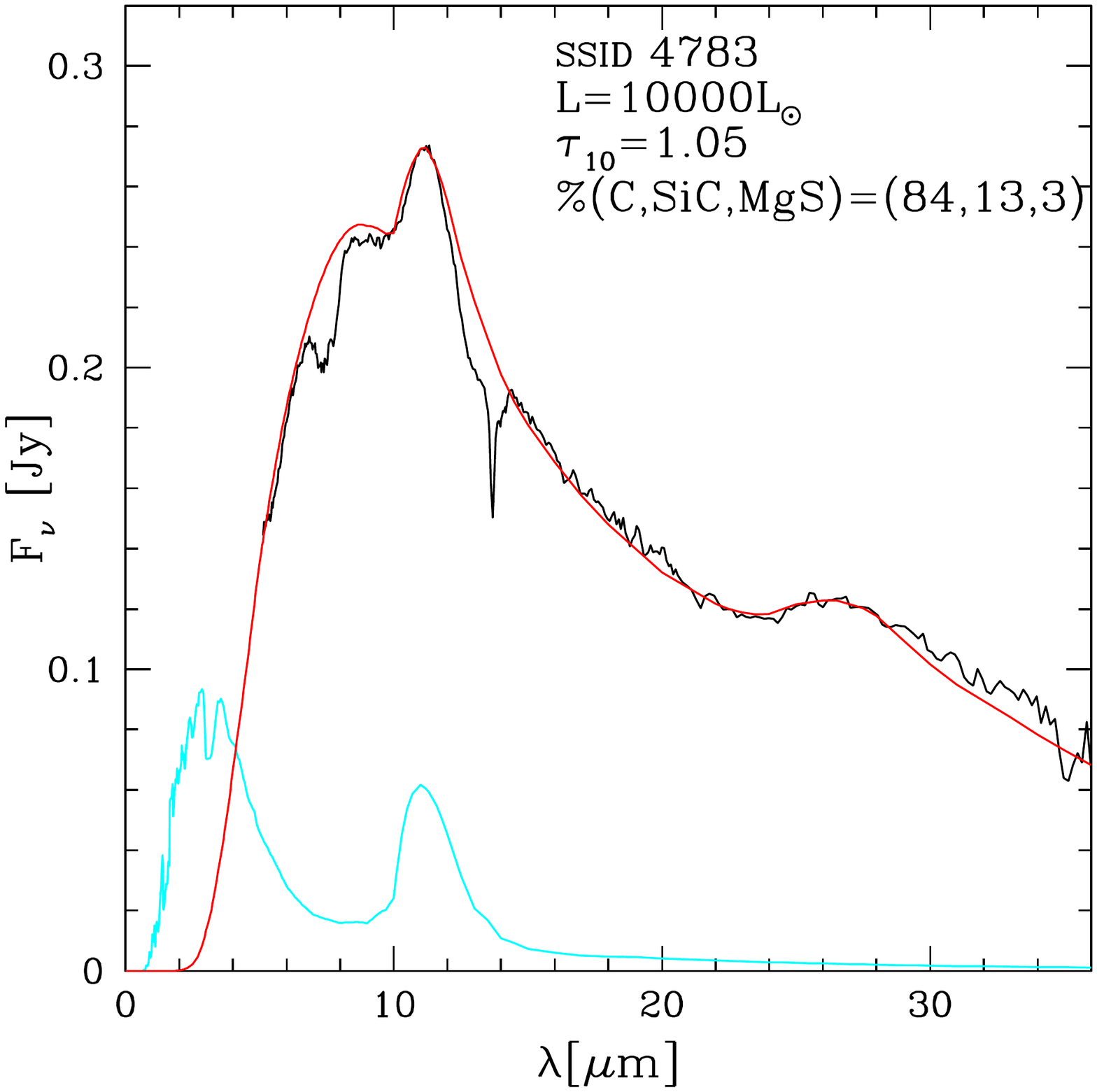}}
\end{minipage}
\vskip-70pt
\begin{minipage}{0.48\textwidth}
\resizebox{1.\hsize}{!}{\includegraphics{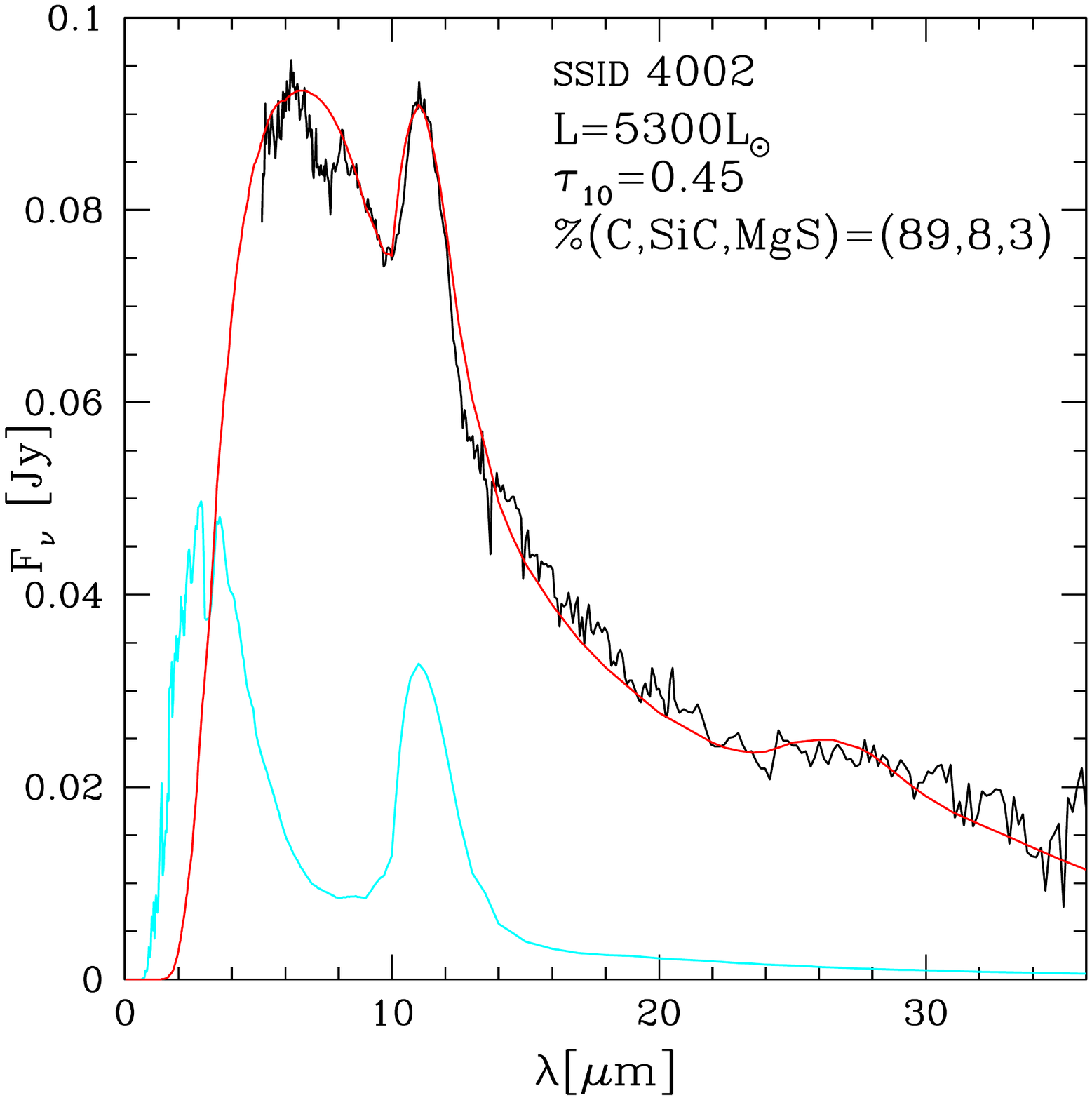}}
\end{minipage}
\begin{minipage}{0.48\textwidth}
\resizebox{1.\hsize}{!}{\includegraphics{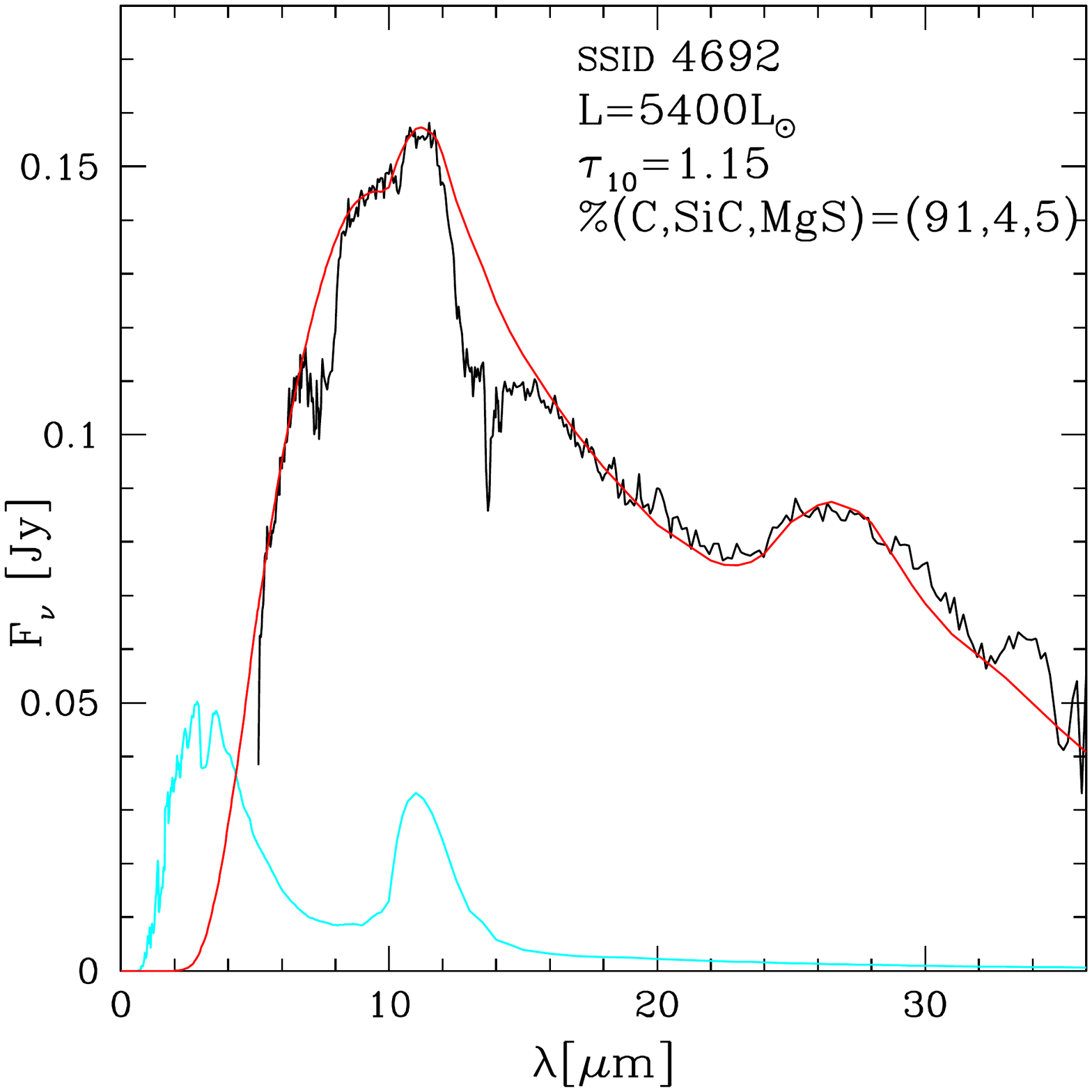}}
\end{minipage}
\vskip-40pt
\caption{IRS spectra of the sources SSID 4197 (top-left panel), SSID 4783 (top-right), SSID 4002 (bottom-left) and SSID 4692 (bottom-right), with the best fit models, obtained with the luminosities, optical depths and dust mineralogy indicated in the individual panels}
\label{fmgs}
\end{figure*}

\subsection{MgS}
The spectrum of several carbon stars is characterised by a wide bump in the spectral region around the $25-30 ~ \mu$m region, which can be seen in the bottom panels of Fig.~\ref{fsedex}. The presence of such a band was detected in the observations of carbon stars in the Milky Way \citep{hony02, volk02} and in the Magellanic Clouds \citep{albert06}. \citet{goebel85} first proposed that this feature can be associated with MgS, because the latter compound has a feature in the same spectral region. The various possibilities proposed so far to explain the $30 ~ \mu$m bump were discussed in the recent review by \citet{volk20}. Here we base on the analysis done by \citet{sloan14}, who concluded that MgS is the best candidate to account for this spectroscopic evidence.

A significant step forward in the modelling of MgS production in the circumstellar envelope of carbon stars was done by \citet{svetlana08}, who included the growth of MgS grains in the description of the stellar wind. In the paper by \citet{svetlana08} it is shown that the formation of pure MgS dust cannot account for the observed feature, because the formation of MgS
is expected to take place at temperatures $\sim 900$ K, after the wind has been
accelerated by the formation of carbon dust, and the density has dropped to values too small to allow a significant growth of MgS particles. We did some tests after including the formation of MgS in the set of equations used here and confirmed the results by \citet{svetlana08}; the largest size reached by MgS grains is below $0.01~\mu$m, which corresponds to cross-sections far too small to account for the observed feature.

To model the MgS feature we considered the idea proposed by \citet{svetlana08}, that the growth of MgS occurs via precipitation on SiC grains. This choice opens the possibility of forming bigger size grains, because the SiC particles formed in the internal regions enter the MgS condensation zone with size in the $0.03-0.07 ~ \mu$m range \citep{nanni13, ventura14}. We find that precipitation of MgS on these already formed SiC grains would further increase the dimension of these particles by $\sim 0.02 ~ \mu$m, if we assume that the deposition begins from the point where the pure MgS dust would start to form; this result is consistent with the analysis by \citet{svetlana08}.

\begin{figure}
 \resizebox{\hsize}{!}{\includegraphics{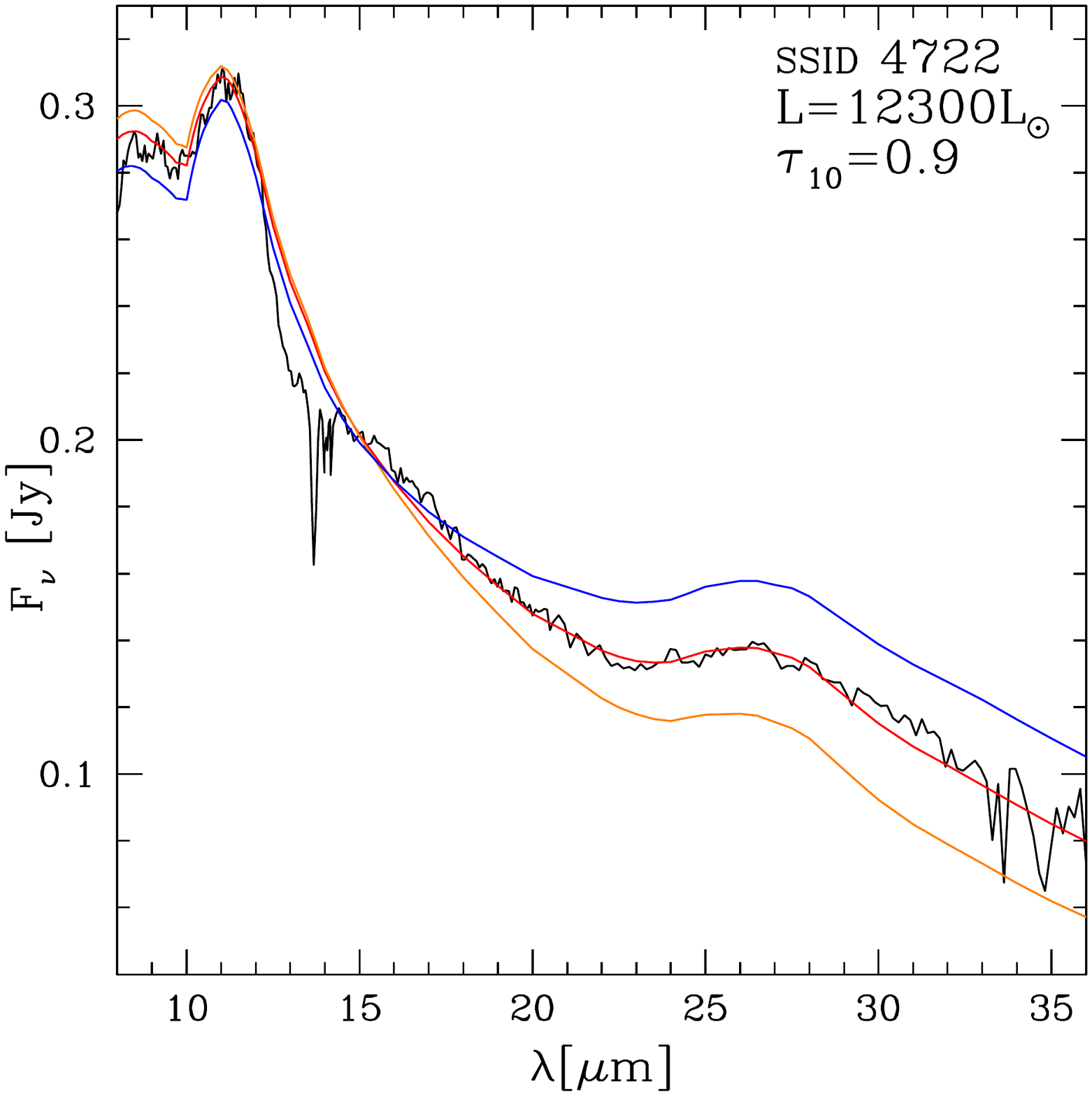}}
\vskip-60pt
\caption{IRS spectrum of SSID 4722 compared with synthetic SED, differing for the percentage of graphite. The different lines refer to percentages of $0 \%$
(orange), $15 \%$ (red), $30 \%$ (blue). 
}
\label{fgraf4722}
\end{figure}

To better understand how the presence of SiC+MgS dust affects the SED of carbon stars, we calculated a series of synthetic SEDs, where the condensation point and the percentage contribution of this compound were allowed to vary within reasonable ranges. The detailed fit of the MgS feature in the SED allows the determination of the percentage of the SiC+MgS particles with respect to the total dust formed, and of the fractional width of the MgS mantle with respect to the total dimension of the SiC+MgS grains. The latter quantity is related to the condensation point, as a larger mantle is formed when the condensation zone is more internal. An example of this kind of analysis is shown in Fig.~\ref{fmgs18}, where the effects of changing the percentage of SiC+MgS dust and of the size of the MgS mantle are indicated.

We find that in the majority of the sources analysed the fraction of MgS+SiC particles is around $5 \%$, of the same order of pure SiC grains. The detailed fit of the morphology of the $25-30 ~ \mu$m feature demands that the width of the MgS mantle accounts for $30-40 \%$ of the overall size of the SiC+MgS grains, which is consistent with assuming that precipitation of MgS onto SiC cores begins in a more internal region of the outflow than found for pure MgS dust. A few examples of stars with SED characterised by SiC and MgS features are shown in Fig.~\ref{fmgs}.

The conclusions we draw for the present analysis are the following:
\begin{enumerate} 
\item{
A fraction around $50\%$ of the SiC particles formed in the internal regions of the circumstellar envelope act as seeds for the deposition of MgS. This is consistent with the study of the $30~\mu$m feature in carbon stars published
in \citet{messenger13}, who found a tight correlation between the $11.3 ~ \mu$m and the $30~\mu$m features in the spectra of Galactic carbon stars, concluding that the carriers of these features are strongly related each other. SiC and SiC+MgS dust make up around $10\%$ of the total dust, this result being substantially independent of the degree of obscuration of the star}.
\item{Formation of SiC+MgS particles takes place in the same region of the circumstellar envelope where solid carbon condensation occurs. 
}
\end{enumerate}

\begin{figure*}
\begin{minipage}{0.48\textwidth}
\resizebox{1.\hsize}{!}{\includegraphics{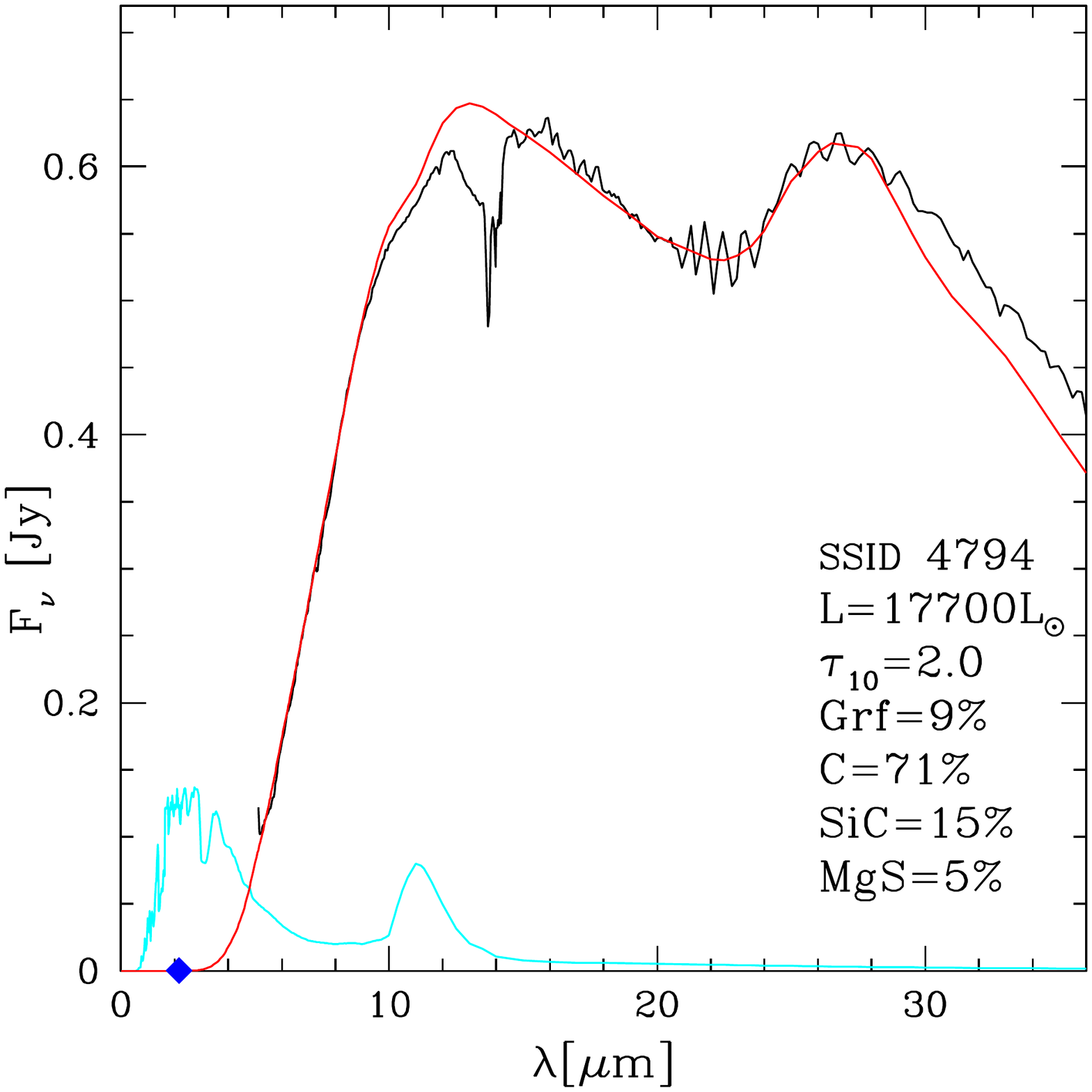}}
\end{minipage}
\begin{minipage}{0.48\textwidth}
\resizebox{1.\hsize}{!}{\includegraphics{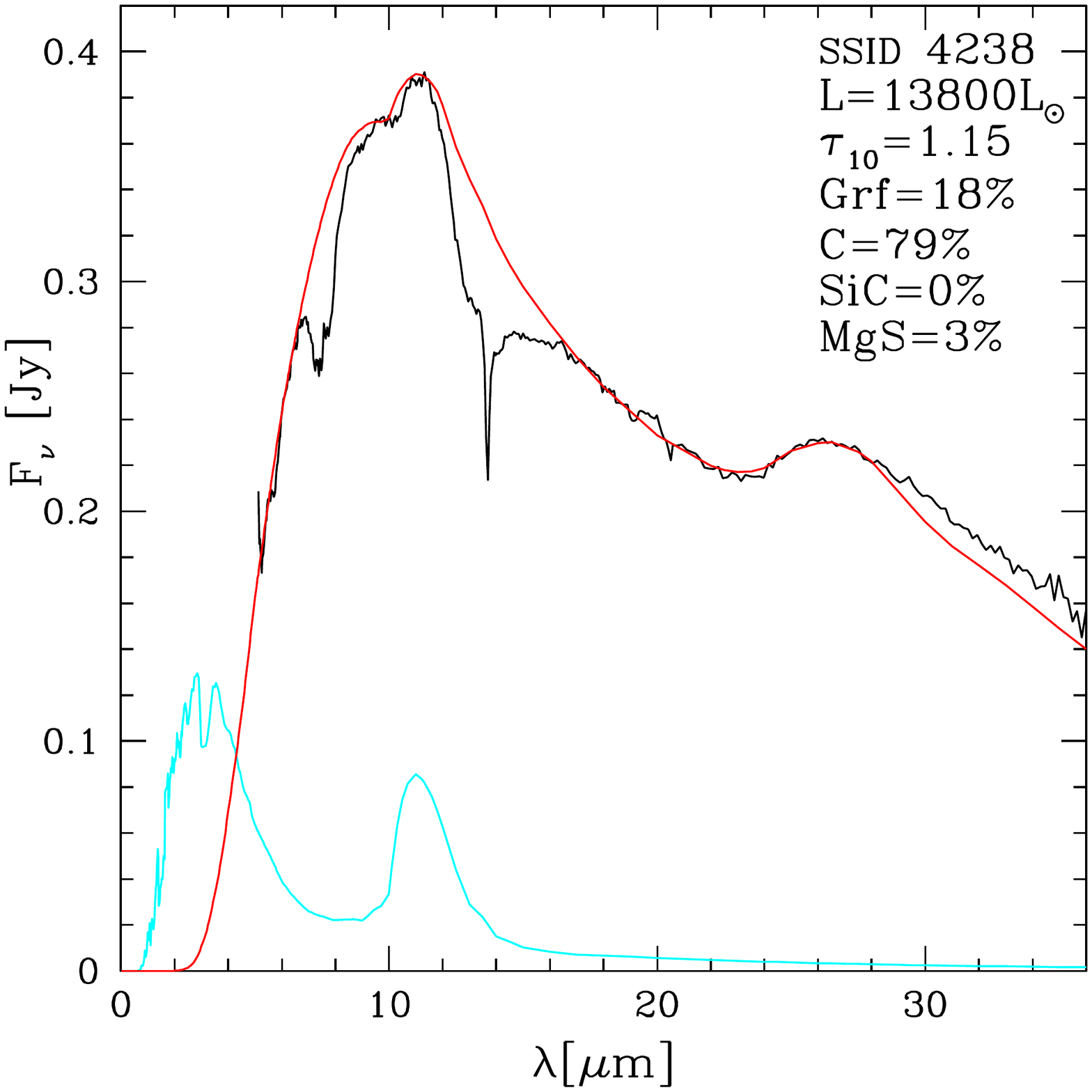}}
\end{minipage}
\vskip-70pt
\begin{minipage}{0.48\textwidth}
\resizebox{1.\hsize}{!}{\includegraphics{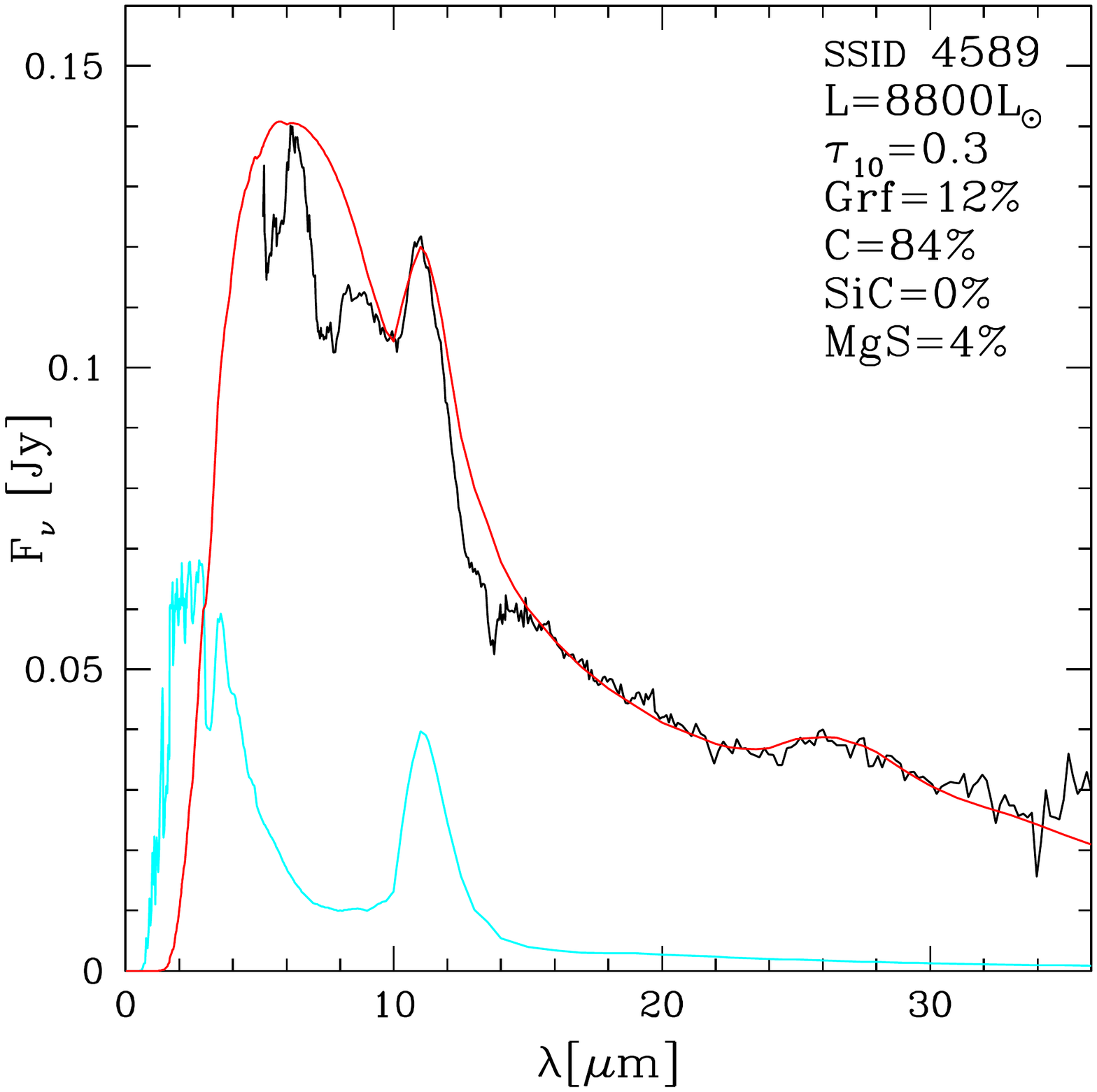}}
\end{minipage}
\begin{minipage}{0.48\textwidth}
\resizebox{1.\hsize}{!}{\includegraphics{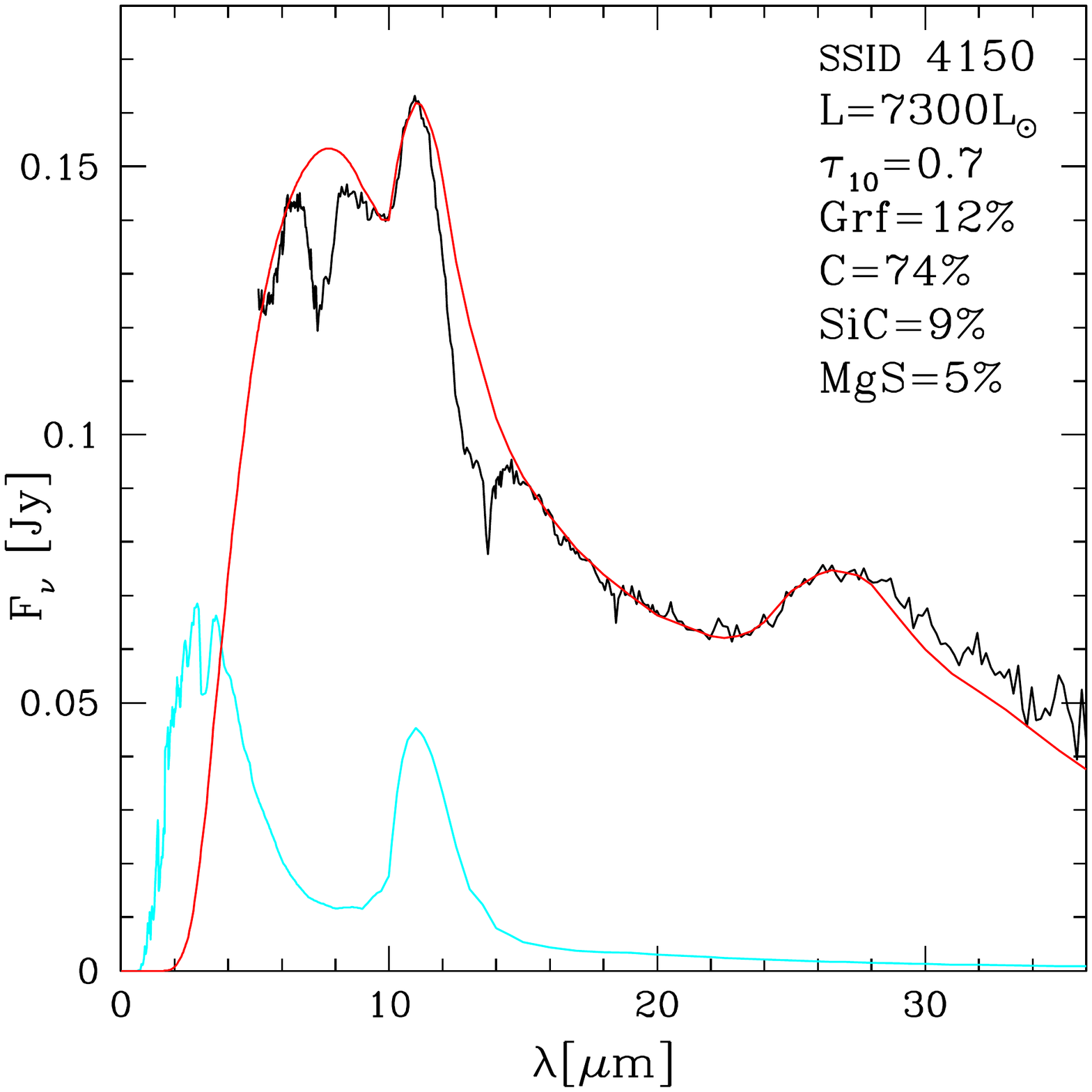}}
\end{minipage}
\vskip-40pt
\caption{IRS spectra of the sources SSID 4794 (top-left panel), SSID 4238 (top-right),
SSID 4589 (bottom-left) and SSID 4150 (bottom-right), with the best fit models,
obtained with the luminosities, optical depths and dust mineralogy indicated in the
individual panels}
\label{fgraf}
\end{figure*}

\subsection{Graphite}
While it is generally recognized that carbon dust is the primary source of extinction in the envelope of carbon stars, the debate is still open regarding the possibility that the dominant species is under the form of amorphous carbon or graphite. The discovery of pre-solar graphite grains from meteorites \citep{zinner08, xu16} indicates that the graphite contribution must be not negligible. \citet{speck09}, based on results from radiative transfer modelling, suggested that graphite is the dominant dust species instead of amorphous carbon.
On the other hand, \citet{andersen03} argued that the absence of the narrow band at $11.52~\mu$m in most observed spectra and the fact that the slope of infrared graphite spectra follows $\lambda^{-2}$ imply that graphite formation in AGB winds is unlikely. On general grounds, the physical conditions in AGB atmospheres seem to favour amorphous carbon dust, since graphite formation requires temperatures that significantly limit the growth interval in a typical C-star.

The stationary wind model adopted here allows the determination of the growth rate of carbon dust and the calculation of the extinction related to carbon grains; however, we cannot derive any indication on whether the carbon particles formed are under the form of amorphous carbon or graphite. Our choice is to derive the relative contribution from amorphous carbon and graphite by looking for the best fit of the SED of carbon stars in the sample examined here. The approach we followed is the following: we first derived the luminosity, optical depth and percentages of SiC and MgS according to the method so far discussed, then assumed a variable percentage of  graphite with respect to carbon, until reaching full consistency with the observed SED.

An example of the analysis done is shown in Fig.~\ref{fgraf4722}, where we show the interpretation of the SED of the source SSID 4722, whose spectrum exhibits
a fairly large level of infrared emission, with $\tau_{10}$ slightly below unity. Changing the fraction of graphite with respect to the total of carbon dust (hence amorphous carbon + graphite) does not sensitively alter the morphology of the SED in the wavelength region close to the emission peak (in the example shown in Fig.~\ref{fgraf4722}), but has a strong effect on the $\lambda > 20~\mu$m domain, which is lifted by the presence of graphite.

Other results, where a not negligible fraction of graphite is required to reproduce the observed SED of the stars in the $\lambda > 20~\mu$m spectral region, can be seen in Fig.~\ref{fgraf}, which shows the comparison between the IRS and the synthetic spectra, for stars of various luminosities and $\tau_{10}$.

From the analysis of the sources examined, we deduce that amorphous carbon is the dominant dust component. Graphite is formed in the circumstellar envelope of
$\tau_{10} > 0.1$ stars, in percentages growing with $\tau_{10}$, ranging from
a few percents to $\sim 20\%$.

\section{The {\itshape JWST} observational planes}
\label{JWST}
It is extremely important to understand how carbon stars are expected to
distribute on the observational planes that will be built with the
mid-IR filters of the MIRI camera. This will be crucial to interpret the data collected by {\itshape JWST} and to select the combination of filters allowing the best determination of the degree of obscuration and of the mineralogy of the stars, and the identification of the progenitors, in terms of formation epoch and chemical composition. 

We discussed in section \ref{spectra} the spectral features in the SED of carbon stars, the most important being the $11.3~\mu$m feature and the bump at $\sim 30~\mu$m. These spectral features can also be seen in the synthetic spectra shown in Fig.~\ref{fdust20} (the $30~\mu$m feature is not present in the left panel, because in these simulations only amorphous carbon and SiC were considered).

We outlined that the dust species responsible for the presence of these features 
do not provide any significant contribution to the overall degree of obscuration of the star, which is mainly given by absorption by solid carbon grains. On the other hand, reprocessing of the radiation by SiC and MgS particles affects the morphology of the SED in regions of the spectrum which fall inside the transmission curves of the majority of the MIRI filters. As clear in Fig.~\ref{fdust20} and \ref{fsedex}, the MIRI filters covered by the IRS spectral range and falling in a wavelength interval substantially clear of features associated with dust are F770W, F1000W and F1800W. 

F770W is affected by the molecular band centered at $7.5 ~\mu$m, which becomes less and less deep as the degree of obscuration increases, and has no relevance in the spectra of the stars with the largest IR emission. This behaviour can be seen, e.g., in the sequence of spectra in Fig.\ref{fsedex}, where we note the little incidence of this feature in the interpretation of the spectrum of the most obscured source SSID 18, shown in the bottom-right panel. According to the present analysis, we find that in the $\tau_{10} > 1$ domain the acetylene $7.5~\mu$m feature is not relevant in the determination of the main properties of the stars, whereas the impact on the analysis of the poorly obscured objects can be safely managed by introducing a correction factor to account for the depression of the flux in that spectral region. We will return to this point shortly.

The wavelength interval covered by F1000W is only partially overlapped on the SiC feature, provided that the latter is very prominent. We will see that this does not affect the general colour-$\tau_{10}$ trend while, on the other hand, use of this filter allows drawing information on the metallicity of the stars.

The spectral region covered by the transmission curve of F1800W (see bottom panels Fig.~\ref{fdust20}) is not affected by molecular or dust features. This
suggests to use the F1800W flux in the interpretation of the SED of carbon stars.
The only caution to take into account is that the reprocessing of the radiation by MgS and graphite particles makes the SED flatter in the $12-20 ~\mu$m region, which reflects into a higher F1800W flux. This is clear from the comparison between the synthetic SED shown in the bottom-left and bottom-right panels of Fig.~\ref{fdust20}. The F1800W fluxes based on synthetic models where the presence of MgS and graphite dust is neglected are slightly underestimated.

Based on these arguments, we consider the colour-magnitude ([F770W]-[F1800W], [F1800W]) and ([F1000W]-[F1800W], [F1800W]) planes as the most suitable to study the obscuration sequences of carbon stars, where the characterisation of the individual objects is fairly independent of the details of the dust mineralogy and scarcely affected by the details of the morphology of the most relevant spectral features.

\begin{figure*}
\begin{minipage}{0.48\textwidth}
\resizebox{1.\hsize}{!}{\includegraphics{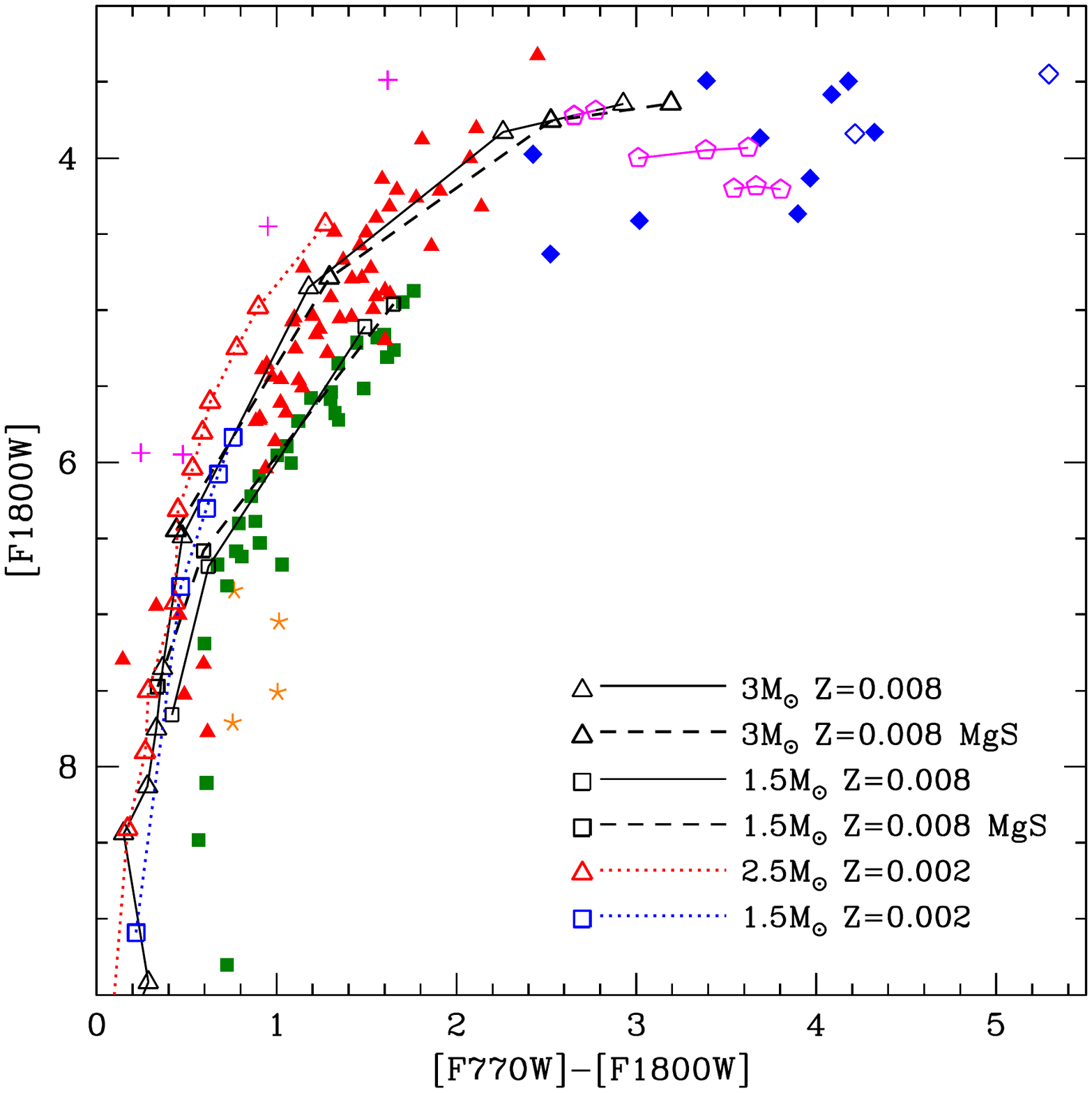}}
\end{minipage}
\begin{minipage}{0.48\textwidth}
\resizebox{1.\hsize}{!}{\includegraphics{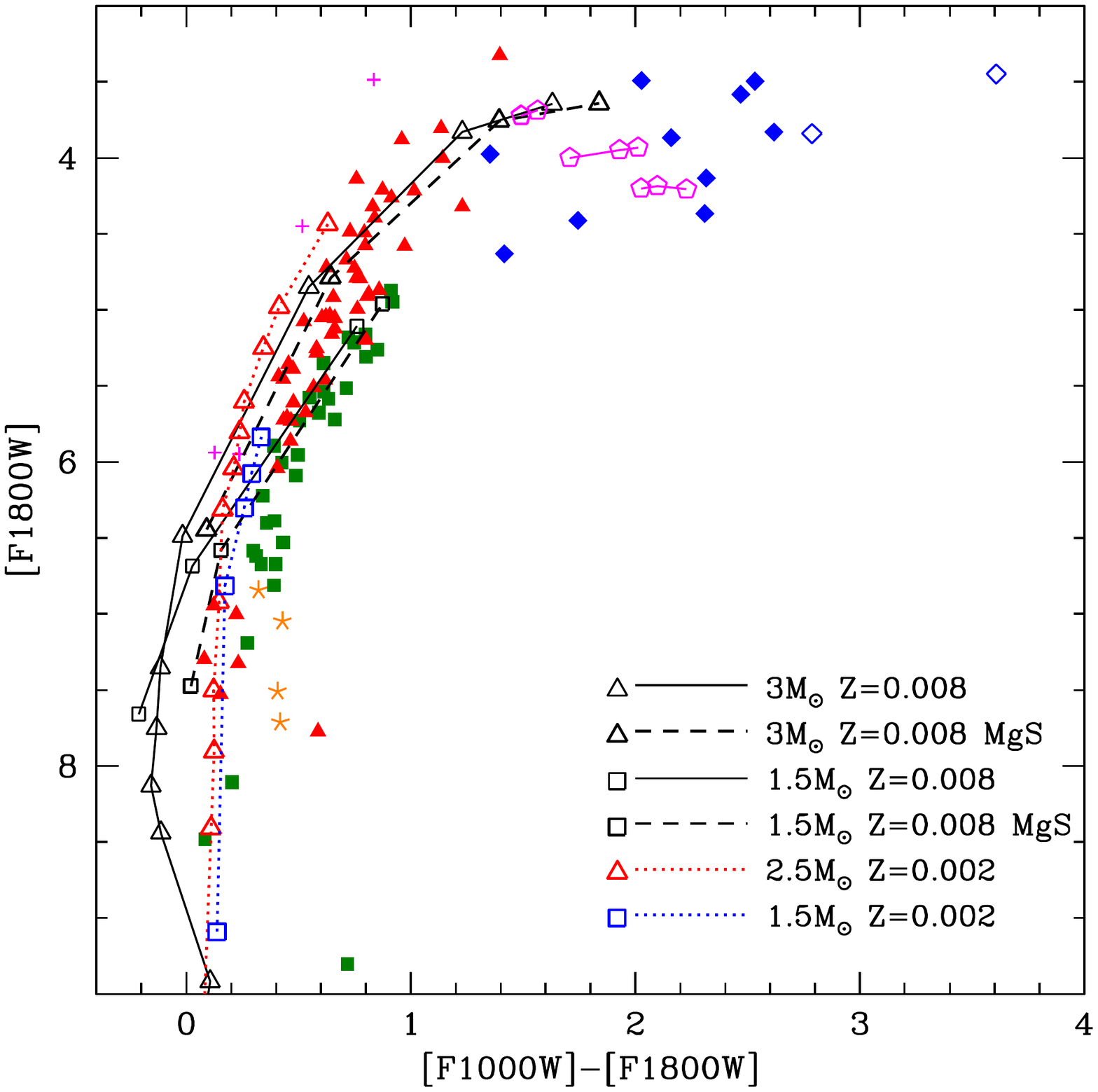}}
\end{minipage}
\vskip-40pt
\caption{Left: Stars in the sample described in section \ref{sample} are shown in the colour-magnitude ([F770W]-[F1800W], [F1800W]) plane. Same symbols of Fig.~\ref{ftaul} were adopted. 
Black squares and triangles indicate the expected evolution of stars of metallicity $Z=0.008$ and initial mass $1.5~M_{\odot}$ and $3~M_{\odot}$, respectively. Dashed lines refer to the expected evolution of $1.5~M_{\odot}$ and $3~M_{\odot}$ stars, when the presence of Sic+MgS and graphite grains are considered in the computation of the synthetic SED. Blue squares and red triangles, connected with dotted tracks, refer to the evolution of $Z=0.002$ stars of initial mass $1.5~M_{\odot}$ and $2.5~M_{\odot}$, respectively. 
Magenta pentagons indicate the tracks of $1.1~M_{\odot}$, $2.5~M_{\odot}$ and $3~M_{\odot}$ (from fainter to brighter) models, obtained by artificially increasing the mass loss rate after the jump in the stellar radius caused by the increase in the surface carbon, according to the discussion in section \ref{ero}. Right: distribution of stars in the colour-magnitude ([F1000W]-[F1800W], [F1800W]) plane. 
}
\label{fcmd1}
\end{figure*}

The left panel of Fig.~\ref{fcmd1} shows the distribution of the sample of carbon stars described in section \ref{sample} on the ([F770W]-[F1800W], [F1800W]) plane. The sources within the subsample that we analysed in detail in the present investigation are shown with different symbols. In the figure we show as solid lines the tracks corresponding to the sequences of synthetic SED of $Z=0.008$ stars of initial mass $1.5~M_{\odot}$ (black squares) and $3~M_{\odot}$ (black triangles), which were calculated by assuming a dust composition made up only of solid carbon and SiC. The dashed lines indicate the tracks of the same stars, recalculated to account for the presence of SiC+MgS and graphite grains, in quantities consistent with the discussion in section \ref{spectra}. The dashed lines connecting the blue squares and red triangles refer to the evolution of $1.5~M_{\odot}$ and $2.5~M_{\odot}$ stars of metallicity $Z=0.002$, taken as a representative of a metal-poor population. In this case we do not show the
corresponding tracks which consider SiC and MgS, because little formation of
these species is expected in the wind of metal-poor stars.

The evolutionary tracks provide a nice fit of the observations, with the exceptions of the stars discussed in section \ref{ero}. For the latter
objects we followed the indications given in section \ref{ero}, i.e.
we modelled dust production by assuming a higher mass loss rate since the
phase during which the stars experience a fast expansion, owing to the
carbon enrichment of the surface regions. The results obtained, shown
with magenta stars on the plane, are in fair agreement with the colour and
magnitudes of the extreme stars.

The $\sim 0.2$ mag colour shift between the synthetic and the observed colours of the bluest stars in the sample, characterised by little dust in the circumstellar envelope, is connected with the absorption C$_2$H$_2$ feature at $7.5~\mu$m, discussed earlier in this section. This is the correction to be applied to the synthetic ([F770W]-[F1800W]) colour of carbon stars with little dust in the circumstellar envelope.

The stars on this colour-magnitude plane define a well distinguished obscuration sequence, spanning the colour range $0<$ [F770W]-[F1800W] $<4.5$ and the magnitude interval $3.5<$ [F1800W] $<8.5$. ([F770W]-[F1800W]) is a reliable indicator of the degree of obscuration of the stars: we find a tight correlation between this colour and the optical depth, which extends from [F770W]-[F1800W] $ \sim 0.5$, $\tau_{10} \sim 0.1$, to the objects with the largest infrared emission, with [F770W]-[F1800W] $\sim 4.5$ and $\tau_{10} \sim 7$.

Roughly, we have

\begin{multline}
\log \tau_{10} \sim -0.1\times ([F770W]-[F1800W])^2 + \\ 
+ 0.9\times ([F770W]-[F1800W]) -1.2
\end{multline}

For a given ([F770W]-[F1800W]) the observed distribution exhibits a $\sim 1$ mag spread in [F1800W], which is due to differences in the luminosity of the stars, which cover the $5000-17000~L_{\odot}$ range (see Fig.~\ref{ftaul}). The position on the plane is connected with the formation epoch of the sources, the stars formed more recently being in the upper part of the diagram.

The bright sources discussed in section \ref{hbb}, which we associate either to
stars currently experiencing HBB or to metal-poor, massive AGB stars during the
late AGB phases, make a sort of upper envelope of the distribution of the stars
on this plane.

\begin{figure*}
\begin{minipage}{0.48\textwidth}
\resizebox{1.\hsize}{!}{\includegraphics{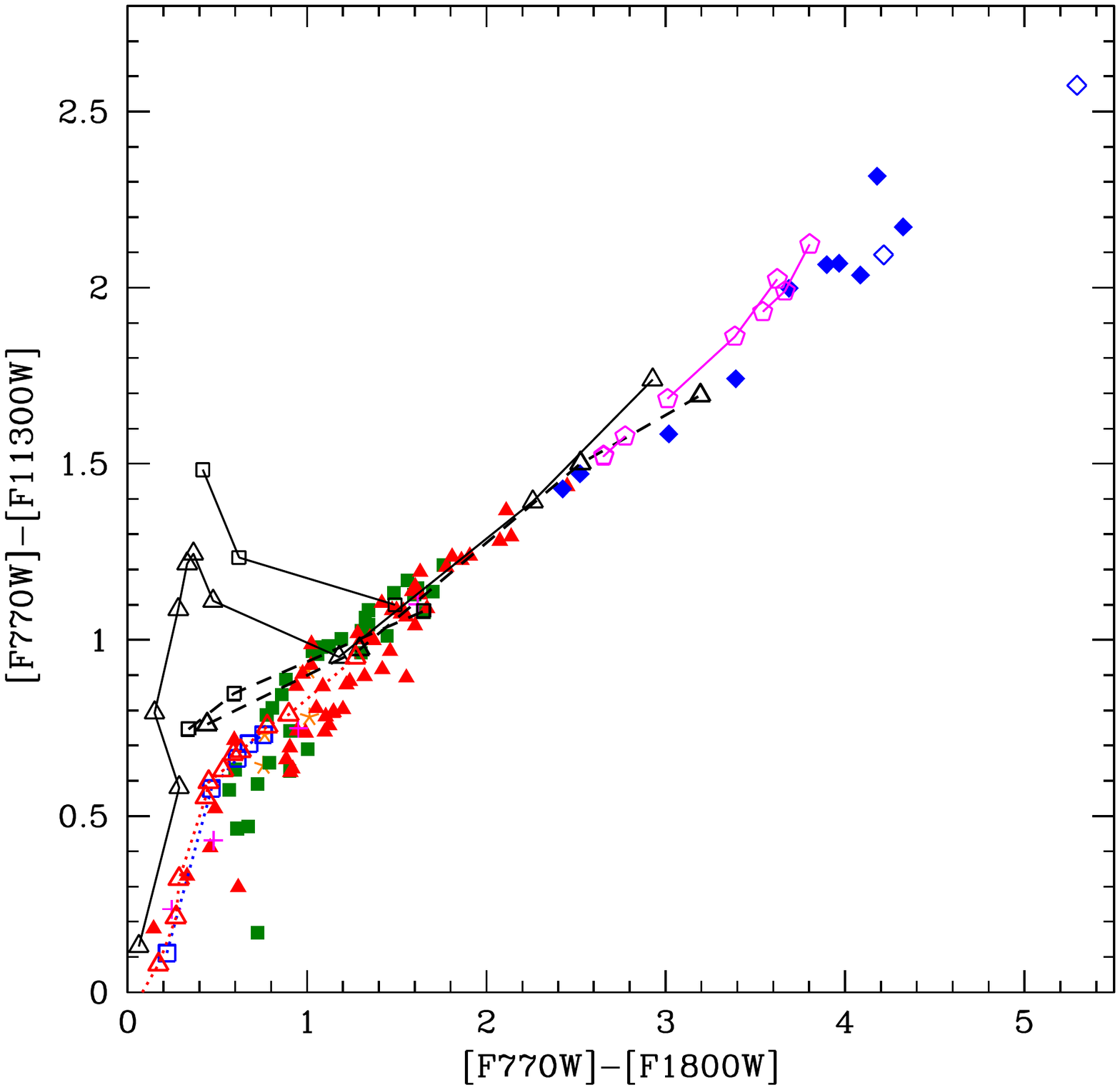}}
\end{minipage}
\begin{minipage}{0.48\textwidth}
\resizebox{1.\hsize}{!}{\includegraphics{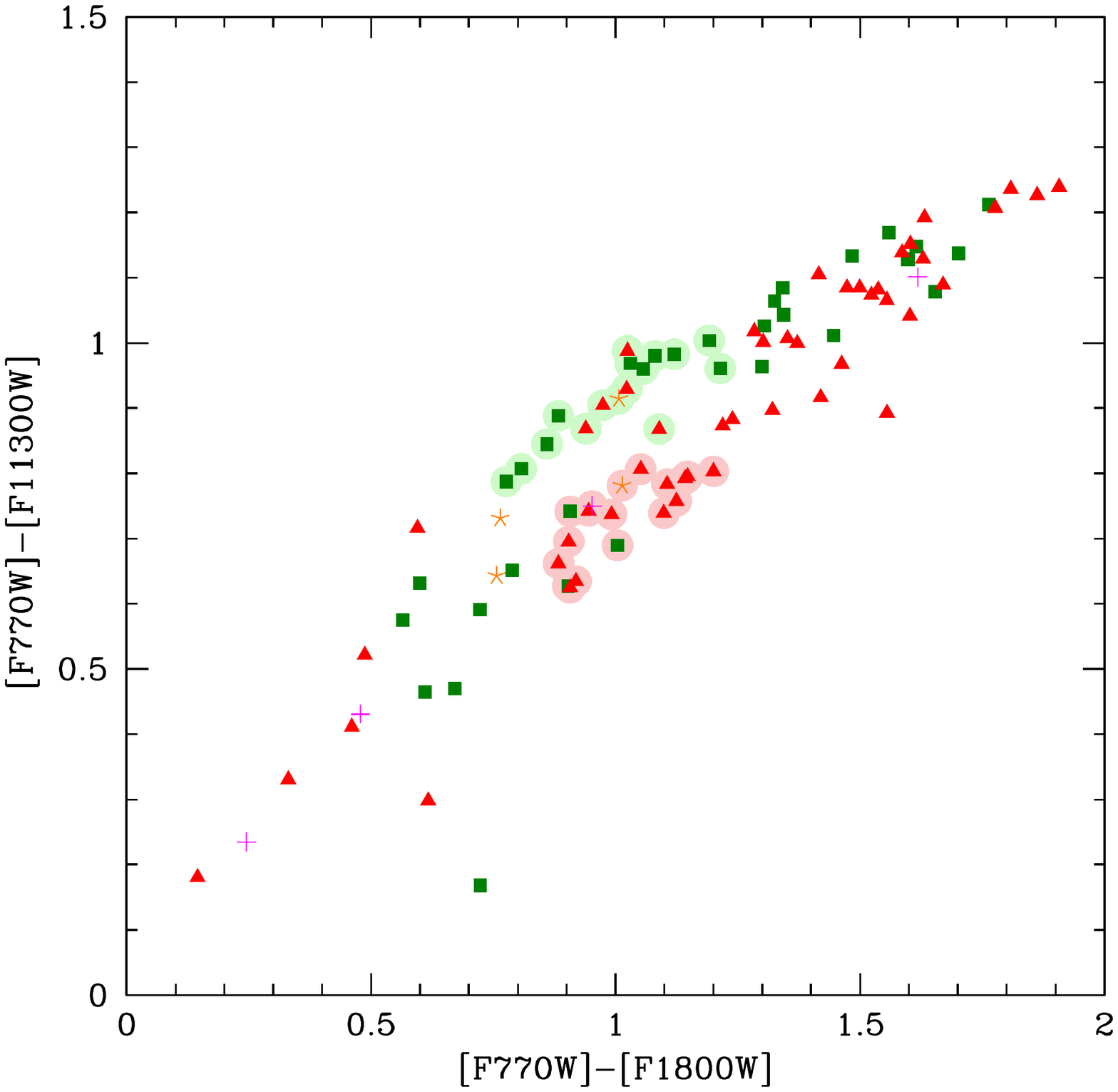}}
\end{minipage}
\vskip-40pt
\caption{Left: position of the stars in the sample presented in section \ref{sample} in the colour-colour ([F770W]-[F1800W], [F770W]-[F1130W]) plane. The symbols for the stars and the evolutionary tracks are the same as in Fig.~\ref{fcmd1}. Right: zoom of the same plane shown in the right panel, to outline the bifurcation between stars of different metallicity.}
\label{fccd}
\end{figure*}
 
The tracks based on models that consider SiC+MgS and graphite are slightly redder than those which take into account only solid carbon and SiC. This difference, which is visible only for the most obscured stars, is related to the increase in the F1800W flux, consistently with the discussion done at the beginning of the present section.

The tracks corresponding to metal-poor stars are bluer than the $Z=0.008$
tracks. This difference is visible only for the stars characterised by optical depths $\tau_{10} > 0.1$, whereas they are substantially overlapped in the low $\tau_{10}$ domain. This is explained by the lower rate of dust formation in metal-poor stars, related to the absence of SiC and to the higher effective temperatures, which inhibits the formation of great quantities of dust \citep{ventura14}. The scarcity of stars in the region of the plane covered by the track of the metal-poor, $2.5~M_{\odot}$ indicates the poor star formation of metal-poor stars occurred in the LMC in the last 1 Gyr, in agreement with the studies on the star formation in the LMC \citep{harris09}.

The distribution of the sample stars in the colour-magnitude ([F1000W]-[F1800W], [F1800W]) plane is shown in the right panel of Fig.~\ref{fcmd1}, with the same evolutionary tracks reported in the left panel. The trend traced by the observations is similar to the one in the ([F770W]-[F1800W], [F1800W]) diagram, with the difference that the colour range is in this case $\sim 2.5$ mag wide: as expected, ([F770W]-[F1800W]) is more sensitive to $\tau_{10}$ in comparison to ([F1000W]-[F1800W]). The relationship between ([F1000W]-[F1800W]) and $\tau_{10}$
can be approximated by

\begin{multline}
\log \tau_{10} \sim -0.2\times ([F1000W]-[F1800W])^2 + \\
+ 1.13\times ([F1000W]-[F1800W]) -0.75
\end{multline}

In the lower-left side of the plane, populated by stars with little dust,
the low-metallicity tracks are bluer than their $Z=0.008$ counterparts: this is connected with the SiC feature, present in the SED of $Z=0.008$ objects and absent in that of metal-poor stars, which causes a slight lift of the 
F1000W flux. We note that the metal-poor tracks are overlapped on the observations, suggesting that a significant fraction of stars with little or no dust in the circumstellar envelope belong to a low-metallicity population. We will return to this point later in this section.

Fig.~\ref{fccd} shows the distribution of the stars in the colour-colour
([F770W]-[F1800W], [F770W]-[F1130W]) plane. The same tracks as in
Fig.~\ref{fcmd1} are shown.

The obscuration pattern traced by the sample stars is clear in this plane,
and can be roughly approximated by a straight line, with slope 0.5. 
Generally speaking, ([F770W]-[F1130W]) gets redder and redder as $\tau_{10}$
increases. For $\tau_{10} < 1$ this trend is mostly due to the presence of the SiC feature in the SED, which increases the flux in the $11.3 ~\mu$m region
(see the blue and red lines in the bottom panels of Fig.~\ref{fdust20}).
For higher $\tau_{10}$ the increase in ([F770W]-[F1130W]) is related to
the shift of the SED towards the mid-IR part of the spectral range, the peak
being at wavelengths $\lambda > 10~\mu$m (see green line in the bottom panels 
of Fig.~\ref{fdust20}).

The tracks of metal-poor models are below those of the more metal-rich stars,
because the F1130W flux is smaller, owing to the little (or no) quantity of SiC
in the circumstellar envelope. The difference between the $Z=0.002$ and $Z=0.008$ tracks are negligible at large $\tau_{10}$, because under those conditions the SiC feature becomes less prominent and the SED shifts to longer wavelengths.

The evolutionary tracks in this plane reproduce the extension of the trend traced
by the stars in the sample. The extreme stars are once more an exception on this side, their colours being reproduced only by invoking higher mass loss rates, according to the discussion in section \ref{ero}. In
the region of the plane at [F770W]-[F1130W] $<0.8$ there is an offset between the $Z=0.008$ tracks and the observations, even when the correction for the
presence of MgS+SiC and graphite is considered. The $Z=0.002$ tracks are in much better agreement with the observational evidence in this region of the colour-colour plane, which is populated by stars with little dust in their surroundings, with $\tau_{10} < 0.5$.
Most of these stars descend from low-mass progenitors, with initial mass below
$\sim 1.5~M_{\odot}$, formed in epochs older than 2 Gyr.
In the spectrum of the majority of these stars there is no evidence for the SiC feature, which is the reason why their position is better reproduced by $Z=0.002$ models. While we cannot rule out that some unknown mechanisms inhibits the formation of SiC particles under particular circumstances, the most plausible explanation for this result is that the low-mass stars with little degree of obscuration in the sample discussed in section \ref{sample} are mostly metal-poor stars; this is consistent with the age-metallicity relationship of the LMC, studied in \citet{harris09}, and with the study by \citet{flavia15a}, who claimed the presence of a significant fraction of low-metallicity objects in the LMC stars descending from $M < 2~M_\odot$ progenitors. The right panel of Fig.~\ref{fccd} shows a zoom of the colour-colour plane in the $0<$ [F770W]-[F1800W] $<2$ region. Here we note a clear dichotomy in the distribution of the stars, which we interpret as a metallicity spread: the stars painted in pink, with no SiC, correspond to the metal-poor population, whereas the green counterparts represent the more metal-rich stellar component. These spread vanishes for $0<$ [F770W]-[F1800W] $<1.5$, because metal-poor stars are not expected to evolve to such red colours.

\begin{figure}
 \resizebox{\hsize}{!}{\includegraphics{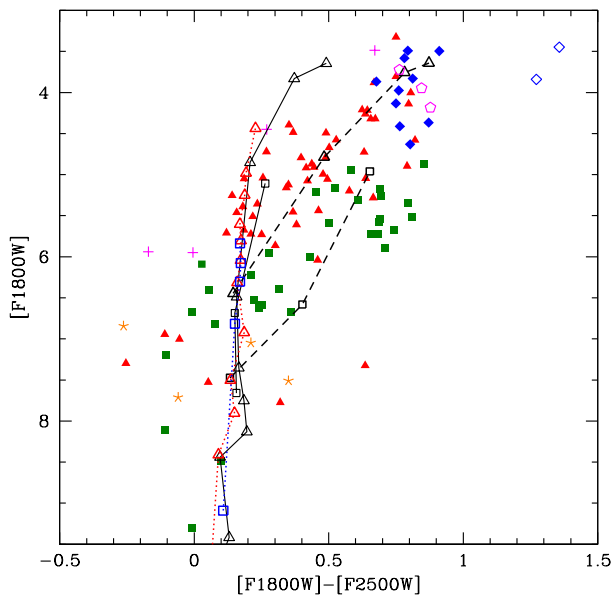}}
\vskip-60pt
\caption{Position of the stars in the sample presented in section \ref{sample}
in the colour-magnitude ([F1800W]-[F2550W], [F1800W]) plane. The symbols for the
stars and the evolutionary tracks are the same as in Fig.~\ref{fcmd1}.
 }
\label{fcmd2}
\end{figure}

We now turn our attention to the colour-magnitude ([F1800W]-[F2550W], [F1800W]) diagram, shown in Fig.~\ref{fcmd2}, with the same symbols used in Fig.~\ref{fcmd1} and Fig.~\ref{fccd}.

In this plane, as expected, the use of the tracks that consider the formation of SiC+MgS is mandatory, because the transmission curve of the F2550W filter is entirely overlapped with the $30~\mu$m bump. The significant discrepancy between the tracks calculated with or without SiC+MgS can be understood by confronting the synthetic SEDs shown in the bottom panels of Fig.~\ref{fdust20}: especially in the higher $\tau_{10}$ cases the SEDs are noticeably different, particularly in the F2550W flux. This despite the assumed MgS percentage is only $5\%$ of the total dust.

We see in Fig.~\ref{fcmd2} that the tracks which consider MgS are in much better agreement with the observations and are overlapped with most of the stars with $F1800W<6$. 

The colour range of the sample stars in the ([F1800W]-[F2550W], [F1800W]) plane
is $\sim 1$ mag. This is significantly smaller than in the ([F770W]-[F1800W], [F1800W]) plane (see Fig.~\ref{fcmd1}), where the colour difference between the stars with no dust and the most obscured ones is $\sim 4$ mag. This difference is explained by the saturation of ([F1800W]-[F2550W]) for $\tau_{10} > 3$, because the SED becomes approximately horizontal in the $\lambda > 20 ~\mu$m domain. 

The low-metallicity tracks shown in Fig.~\ref{fcmd2} define an approximately vertical trend, which is because there is no $30~\mu$m bump in the spectra of these stars and the slope in the $15-30~\mu$m spectral region is practically unchanged. The latter point is linked to the fact that low-metallicity stars are not expected to evolve at $\tau_{10} > 1$. We note that the $Z=0.002$ tracks nicely reproduce the position of the stars with [F1800W]-[F2550W] $\sim 0.1-0.2$ and $4 <$ [F1800W] $< 6$. These sources are the same located in the [F770W]-[F1130W] $< 0.8$ region in the colour-colour plane shown in Fig.~\ref{fccd}, which we identified as low-metallicity stars. This conclusion is further reinforced by the present analysis, suggesting a criterion to disentangle stars of different chemical composition in the ([F1800W]-[F2550W], [F1800W]) diagram.

Recent studies focused on the distributions of evolved stars in the observational planes built with the MIRI filters were published
by \citet{jones17} and \citet{kraemer17}; a similar discussion is also given
in \citet{martin18}. \citet{kraemer17} investigated various colour-colour 
planes, to study how the sequences of carbon-rich and oxygen-rich AGBs, young stellar objects and red super giants can be clearly separated. Unfortunately
a straight comparison with the present analysis is not straightforward, because
no colour-magnitude plane is considered and in all the planes proposed by \citet{kraemer17} the F560W flux is used, whereas we ruled out this filter,
as the IRS spectra used here do not allow a reliable estimate of [F560W].

\citet{jones17} presented an exhaustive analysis on how different classes
of objects will populate the observational planes, outlining that the 
planes allowing the best disentangling of the various types of objects 
are based on the combination of F1000W and F2100W MIRI filters. Indeed this
is a valid criterion to separate evolved stars from other objects, but significant overlapping of O-rich and C-rich AGB and of RSG stars prevents
from using this filter combination to study the C-star population, which is
the primary goal of the present work. We find that use of F1800W allows
a better analysis of the C-star population than F2100W, as the latter
is influenced by the $25-30~\mu$m bump. \citet{jones17} showed that the
plane where carbon stars are most easily identified is the colour-magnitude
([F1500W]-[F2100W], [F1500W]) diagram. Use of this plane definitively allows
to disentangle carbon stars with low and moderate infrared excess from the
other classes of objects, whereas those with the largest degree of obscuration are overlapped with O-rich AGBs and red supergiants. Here we decided to
avoid use of F1500W, because, as discussed in section \ref{lmcc}, it is affected by the C$_2$H$_2$ band centered in the $13.5-13.9~\mu$m spectral region. 

A wider exploration of the {\itshape JWST} potentialities was presented in \citet{martin18}.
The advantage of the latter work compared to \citet{kraemer17}, \citet{jones17}
and to the present investigation is the wider spectral region considered,
which is not limited to the wavelengths covered by MIRI. In \citet{martin18}
it is shown that the C-star sequence is expected to spread over a large
magnitude interval in the ([F150W]-[F360W], [F360W]) plane, with the C-star
band being substantially separated from other objects. Regarding the mid-IR
domain, \citet{martin18} proposed the colour-colour ([F770W]-[F1800W], [F560W]-[F770W]) plane to distinguish carbon stars from oxygen-rich AGBs. Interestingly, \citet{martin18} suggest use of the F560W, F770W and F1800W filters to study obscured AGBs, which are among those recommended here, with the exception of F560W, for the reasons given above.

\section{Conclusions}
We study a sample of carbon stars in the LMC, for which the mid-IR spectral energy distribution was obtained by means of the IRS spectrograph onboard of {\itshape Spitzer}.
To characterise the sources in this sample we calculated stellar models evolved through the AGB phase, until the almost complete ejection of the external convective envelope. For each stellar mass, we first modelled dust formation in the outflow to determine the amount and the mineralogy of the dust formed during various evolutionary phases. These results, combined with the main stellar parameters derived from stellar evolution modelling, were the ingredients used to determine a sequence of synthetic SEDs, which allowed us to describe how the SED of these stars changes as they evolve through the AGB.

The comparison between the IRS spectra and the synthetic SEDs leads to a reliable determination of the optical depth and the luminosity of the individual sources, which, in turn, can be used to deduce the mass and formation epoch of the progenitors and to provide an estimate of the amount of carbon accumulated in the surface regions via repeated TDU episodes. The majority of the sources 
analysed has luminosities in the $5000-17000~L_{\odot}$ range and
optical depths $\tau_{10}<3$, in agreement with the results from carbon star
modelling. Approximately half of the stars examined descend from $M<2~M_{\odot}$ progenitors, older than
1 Gyr, the remaining half being composed by younger objects, formed not later
than 1 Gyr ago, the progeny of $M>2~M_{\odot}$ stars. 

The sample discussed in the present work includes a paucity of objects with 
luminosities below $5000~L_{\odot}$, which we interpret as stars that have not
yet fully resumed CNO burning after having experienced a TP, and 4 very bright stars with $L>20000~L_{\odot}$, which we suggest to descend either from $\sim 3.5~M_{\odot}$ stars currently experiencing HBB, or from metal-poor, $\sim 5~M_{\odot}$ stars, evolving through the latest AGB phases. We also find a group of extremely red stars, which can be hardly explained within the context of the obscuration path followed by single carbon stars; we believe that they belong to binary systems, in which a common envelope phase caused a significant increase in the mass loss rate, which enhanced dust formation in the outflow; two out of these objects have likely left the AGB and are moving to the post-AGB phase.

The detailed comparison between the IRS data and the synthetic spectra allows
to fix important properties of the dynamics and the dust composition of the
stellar wind. Most of the observed SEDs show up a prominent bump, in the
$25-30~\mu$m spectral region, whose origin is still debated. Our results confirm previous claims, that MgS alone cannot account for the observational evidence, because the grains of this dust species form in an external region of the outflow, where the growth rate is so small that the amount of MgS formed is negligible. Conversely, when assuming that MgS forms by precipitation on pre-existing SiC grains, we find a nice agreement with the observations, provided that the width of the MgS mantle is within $20-40\%$ of the SiC core: this suggests that precipitation of MgS begins in a regions of the outflow more internal than the zone where pure MgS forms, approximately in the same region where condensation of gaseous carbon into solid particles takes place.

In the {\itshape JWST} perspective, we propose the colour-magnitude ([F770W]-[F1800W], [F1800W]) and ([F1000W]-[F1800W], [F1800W]) diagrams as the planes where the
obscuration sequences of carbon stars of different mass can be more clearly distinguished. The former plane allows a better determination of the optical depth, given the strong sensitivity of the [F770W]-[F1800W] colour to $\tau_{10}$; on the other hand, in the latter plane the evolutionary tracks are more sensitive to the metallicity, which allows disentangling stars of
different chemical composition. Interesting information can be deduced also
by the analysis of the distribution of the stars in the ([F1800W]-[F2550W], [F1800W]) plane. [F1800W]-[F2550W] is not particularly sensitive to the optical depth; on the other hand the position of the stars in this diagram critically depends on the presence of the $30~\mu$m bump, which, if we assume due to MgS, renders this colour extremely sensitive to the metallicity. Furthermore, this is the plane where the stars on leave from the AGB can be more clearly separated from the remaining carbon star population.

\label{end}

\begin{acknowledgements}

DAGH acknowledges support from the State Research Agency (AEI) of the Spanish Ministry of Science, Innovation and Universities (MCIU) and the European Regional Development Fund (FEDER) under grant AYA2017-88254-P. DK acknowledges the support of the Australian Research Council (ARC) Decra grant (95213534). EM is indebted 
to the anonymous referee for the detailed and careful reading of the manuscript,
which helped to improve significantly the quality of this work.

\end{acknowledgements}

%
%

\end{document}